\documentclass[aps, pre, twocolumn, a4paper, floatfix, nofootinbib, longbibliography]{revtex4-1}

\usepackage[utf8]{inputenc}
\usepackage{epsfig}
\usepackage[T1]{fontenc}
\usepackage[english]{babel}
\usepackage[table]{xcolor}
\usepackage{t1enc}
\usepackage{graphicx}
\usepackage{amssymb}
\usepackage{amsmath}
\usepackage{relsize}
\usepackage{overpic}
\usepackage{bm}

\usepackage[normalem]{ulem}

\newcommand{\avg}[1]{{\left<#1\right>}}

\usepackage{mathtools}
\def\multiset#1#2{\ensuremath{\left(\kern-.3em\left(\genfrac{}{}{0pt}{}{#1}{#2}\right)\kern-.3em\right)}}

\usepackage{amsmath}

\usepackage{verbatim}
\usepackage{overpic}
\usepackage{booktabs}

\begin{document}

\title{Model selection and hypothesis testing for large-scale network models with overlapping groups}

\author{Tiago P. Peixoto}
\email{tiago@itp.uni-bremen.de}
\affiliation{Institut f\"{u}r Theoretische Physik, Universit\"{a}t Bremen, Hochschulring 18, D-28359 Bremen, Germany}


\begin{abstract}
  The effort to understand network systems in increasing detail has
  resulted in a diversity of methods designed to extract their
  large-scale structure from data. Unfortunately, many of these methods
  yield diverging descriptions of the same network, making both the
  comparison and understanding of their results a difficult challenge. A
  possible solution to this outstanding issue is to shift the focus away
  from \emph{ad hoc} methods and move towards more principled
  approaches based on statistical inference of generative models. As a
  result, we face instead the more well-defined task of selecting
  between competing generative processes, which can be done under a
  unified probabilistic framework. Here, we consider the comparison
  between a variety of generative models including features such as
  degree correction, where nodes with arbitrary degrees can belong to
  the same group, and community overlap, where nodes are allowed to
  belong to more than one group. Because such model variants possess an
  increasing number of parameters, they become prone to overfitting. In
  this work, we present a method of model selection based on the minimum
  description length criterion and posterior odds ratios that is capable
  of fully accounting for the increased degrees of freedom of the larger
  models, and selects the best one according to the statistical evidence
  available in the data. In applying this method to many empirical
  unweighted networks from different fields, we observe that community
  overlap is very often not supported by statistical evidence and is
  selected as a better model only for a minority of them. On the other
  hand, we find that degree correction tends to be almost universally
  favored by the available data, implying that intrinsic node
  proprieties (as opposed to group properties) are often an essential
  ingredient of network formation.
\end{abstract}

\maketitle

\section{Introduction}

Many networks possess nontrivial large-scale structures such as
communities~\cite{newman_communities_2011, fortunato_community_2010},
core-peripheries~\cite{holme_core-periphery_2005,
rombach_core-periphery_2012},
bipartitions~\cite{larremore_efficiently_2014} and
hierarchies~\cite{clauset_hierarchical_2008, peixoto_hierarchical_2014}.
These structures presumedly reflect the organizational principles behind
network formation. Furthermore, their detection can be used to predict
missing links~\cite{clauset_hierarchical_2008, guimera_missing_2009} or
detect spurious ones~\cite{guimera_missing_2009}, as well as determine
the robustness of the system to failure or intentional
damage~\cite{buldyrev_catastrophic_2010}, the outcome of the spread of
epidemics~\cite{apolloni_metapopulation_2014} and functional
classification~\cite{guimera_functional_2005}, among many other
applications.  The detail with which such modular features are both
represented and detected reflects directly on the quality of these
tasks. However, the methods of uncovering such structures in empirical
data so far proposed are very different in their suitability to the
aforementioned tasks. Many authors have constructed algorithms which
attempt to divide the network into groups according to some metric
devised specifically for this purpose. Examples of this include
modularity~\cite{newman_modularity_2006},
betweenness~\cite{girvan_community_2002}, link
similarity~\cite{ahn_link_2010}, clique
percolation~\cite{palla_uncovering_2005}, encoding of random
walks~\cite{rosvall_maps_2008}, and many
more~\cite{fortunato_community_2010}. Unfortunately, many of these
methods will result in diverging descriptions for the same
network. Furthermore, the information they obtained cannot be easily
used to generalize the data, and make
predictions~\cite{clauset_hierarchical_2008, guimera_missing_2009}.
Alternatively, other authors have focused on constructing generative
models that encode the large-scale structure as parameters, which can
then be inferred from empirical data
(e.g.~\cite{holland_stochastic_1983, airoldi_mixed_2008,
karrer_stochastic_2011, ball_efficient_2011}). These methods not only
represent a more principled and rigorous stance, but they can also
demonstrably overcome inherent limitations of more \emph{ad hoc}
methods~\cite{peixoto_hierarchical_2014}. Furthermore, they can be used
to generalize the data, and make
predictions~\cite{clauset_hierarchical_2008, guimera_missing_2009}. Both
approaches, however, suffer from a common fundamental problem, namely
the difficulty in deciding which detection method or generative model
provides a more appropriate description of a given network. This issue
tends to escalate as more elaborate models and methods are developed,
including features such as degree
correction~\cite{karrer_stochastic_2011}, community
overlap~\cite{palla_uncovering_2005,airoldi_mixed_2008,ahn_link_2010,ball_efficient_2011},
hierarchical structure~\cite{clauset_hierarchical_2008,
lancichinetti_detecting_2009, peixoto_hierarchical_2014},
self-similarity~\cite{palla_multifractal_2010,leskovec_kronecker_2010},
bipartiteness~\cite{larremore_efficiently_2014}, edge and node
correlates~\cite{mariadassou_uncovering_2010, aicher_learning_2014},
social tiers~\cite{ball_friendship_2013}, multilayer
structure~\cite{kivela_multilayer_2014}, temporal
information~\cite{fu_dynamic_2009}, to name only a few. Although such
developments are essential, they should be made with care, since
increasing the complexity of the network description may lead to
artificial results caused by overfitting. While this is a
well-understood phenomenon when dealing with independent data or time
series, open problems remain when the empirical data are a network, for
which many common assumptions no longer hold and the usual methods
perform very poorly~\cite{yan_model_2014}. This problem is significantly
exacerbated when methods are used which make no attempt to assess the
statistical significance of the results. Unfortunately, most methods
that are not based on generative models fall into this class. Although
for certain specially constructed examples some direct connections
between statistical inference and \emph{ad hoc} methods can be
made~\cite{newman_community_2013, newman_spectral_2013}, and in the case
of some spectral methods a much deeper connection seems to
exist~\cite{nadakuditi_graph_2012,krzakala_spectral_2013}, they still
inherently lack the capacity to reliably distinguish signal from
noise. Furthermore --- what is perhaps even more important --- these
different methods cannot easily be compared to \emph{each other}.  For
example, suppose that for the same network a nonoverlapping partition is
found by compressing random walks, another overlapping partition is
obtained with clique percolation, and yet another with a local method
based on link similarity (all of which are methods not based on
generative models). Most of the time, these three partitions will be
very different, and yet there is no obvious way to decide which one is a
more faithful representation of the network. Although methods such as
network benchmarks~\cite{lancichinetti_benchmark_2008,
lancichinetti_benchmarks_2009, lancichinetti_community_2009} and
perturbation analysis~\cite{karrer_robustness_2007} have been developed
in order to alleviate this issue, they have only limited applicability
to the larger problem. Namely, network benchmarks cannot be used when an
appropriate representation of an empirical network is not known, and if
one wants to decide, for instance, if the network possesses overlapping
groups or not. In a similar vein, perturbation analysis provides
information about the significance of results originating from a single
algorithm, which cannot be directly used to compare two very different
ones.

On the other hand, the situation is different if one focuses on
generative models alone. Since in this context the same problem is posed
in a probabilistic framework, comparison between models is possible,
even if the models are very different. And since models can be designed
to accommodate arbitrary topological features, we lose no explanatory
power when comparing to the \emph{ad hoc} approaches. We show in this
work that this central issue can be tackled in a consistent and
principled manner by performing model selection based on statistical
evidence. In particular, we employ the minimum description length
principle (MDL)~\cite{grunwald_minimum_2007, rissanen_information_2010},
which seeks to minimize the total information necessary to describe the
observed data as well as the model parameters. This can be equivalently
formulated as the maximization of a Bayesian posterior likelihood which
includes noninformative priors on the parameters, from which a posterior
odds ratio between different hypotheses can be computed, yielding a
\emph{degree of confidence} for a model to be rejected in favor of
another.  We focus on the stochastic block model as the underlying
generative model, as well as variants that include degree correction and
mixed memberships. We show that with these models MDL can be used to
produce a very efficient algorithm that scales well for very large
networks and with an arbitrarily large number of groups. Furthermore, we
employ the method to a wide variety of empirical network data sets, and
we show that community overlaps are seldom selected as the most
appropriate model. This casts doubt on the claimed pervasiveness of
group overlaps~\cite{palla_uncovering_2005, ahn_link_2010}, obtained
predominantly with nonstatistical methods, which should perhaps be
interpreted as an artifact of using methods with more degrees of
freedom, instead of an underlying property of many systems --- at least
as long as there is a lack of corroborating evidence supporting the
overlap (such as, potentially, edge
weights~\cite{aicher_learning_2014,rosvall_memory_2014} or multilayer
structure~\cite{kivela_multilayer_2014}, which we do not consider
here). On the other hand, we find that degree correction tends to be
selected for a significant majority of systems, implying that individual
node ``fitness'' that is not uniformly inherited by group membership is
a fundamental aspect of network formation.

This paper is divided as follows. In Sec.~\ref{sec:generative} we
present the generative models considered, and in
Sec.~\ref{sec:model_selection} we describe the model selection procedure
based on MDL. In Sec.~\ref{sec:empirical} we present the results for a
variety of empirical networks. In Sec.~\ref{sec:ident} we analyze the
general identifiability limits of the overlapping models, and in
Sec.~\ref{sec:algorithm} we describe in detail the inference algorithm
used. In Sec.~\ref{sec:conclusion} we finalize with a discussion.

\section{Generative models for network structure}\label{sec:generative}

A generative model is one which attributes to each possible graph $G$ a
probability $P(G|\{\theta\})$ for it to be observed, conditioned on some
set of parameters $\{\theta\}$. Here we will be restricted to discrete
uniform models, where specific choices of $\{\theta\}$ prohibit some
graphs from occurring, but those which are allowed to occur have the
same probability. For these models we can write
$P(G|\{\theta\})=1/\Omega(\{\theta\})=e^{-\mathcal{S}(G|\{\theta\})}$,
with $\Omega(\{\theta\})$ being the total number of possible graphs
compatible with a given choice of parameters,
 and
$\mathcal{S}(G|\{\theta\})=\ln\Omega(\{\theta\})$ is the entropy of this
constrained
ensemble~\cite{bianconi_entropy_2009,peixoto_entropy_2012}. In order to
infer the parameters $\{\theta\}$ via maximum likelihood, we need to
maximize $P(G|\{\theta\})$, or equivalently, minimize
$\mathcal{S}(G|\{\theta\})$. This approach, however, cannot be used if
the \emph{order} of the model is unknown,
i.e. the number of degrees of freedom in the parameter set $\{\theta\}$,
since choices with higher order will almost always increase
$P(G|\{\theta\})$, resulting in overfitting. For the same reason,
maximum likelihood cannot be used to distinguish between models
belonging to different classes, since models with larger degrees of
freedom will inherently lead to larger likelihoods. In order to avoid
overfitting, one needs to maximize instead the Bayesian posterior
probability $P(\{\theta\}|G)= P(G|\{\theta\})P(\{\theta\})/P(G)$, with
$P(G)$ being a normalizing constant. The prior probability
$P(\{\theta\})$, which encodes our \emph{a priori} knowledge of the parameters
(if any) should inherently become smaller if the number of degrees of
freedom increases. We will also be restricted to discrete parameters
with uniform prior probabilities, so that $P(\{\theta\}) =
e^{-\mathcal{L}(\{\theta\})}$, with $\mathcal{L}(\{\theta\})$ being the
entropy of the ensemble of possible parameter choices. We can thus write
the total posterior likelihood as $P(\{\theta\}|G) = e^{-\Sigma}/P(G)$,
with $\Sigma = \mathcal{L}(\{\theta\}) + \mathcal{S}(G|\{\theta\})$. The
value $\Sigma$ is the \emph{description length} of the
data~\cite{grunwald_minimum_2007, rissanen_information_2010},
i.e. the total amount of information required to describe the observed
data conditioned on a set of parameters as well as the parameter set
itself~\cite{rosvall_information-theoretic_2007}. Hence, if we maximize
$P(\{\theta\}|G)$ we are automatically finding the parameter choice that
\emph{compresses} the data most, since it will also minimize its
description length $\Sigma$. Because of this, there is no difference
between specifying probabilistic models for both $G$ and $\{\theta\}$,
or encoding schemes that quantify the amount of information necessary to
describe both. In the following, we will make use of both terminologies
interchangeably, whenever most appropriate.

\subsection{Overlapping model without degree correction}

The main feature we want to consider in our generative model is the
existence of well-defined groups of nodes, which are connected to other
groups with arbitrary probabilities, such that nodes belonging to the
same group play a similar role in the large-scale network structure. We
also want to include the possibility of nodes belonging to more than one
group, and in so doing inherit the topological properties of all groups
to which they belong. In order to implement this, we consider a simple
variation of the stochastic block
model~\cite{holland_stochastic_1983,fienberg_statistical_1985,
faust_blockmodels:_1992, anderson_building_1992} with $N$ nodes and $E$
edges, where the nodes can belong to different groups. Hence, to each
node $i$ we attribute a binary mixture vector $\vec{b}_i$ with $B$ entries,
where a given entry $b_i^r \in \{0,1\}$ specifies whether or not the
node belongs to block $r \in [1, B]$. In addition to this overlapping
partition, we simply define the edge-count matrix $\{e_{rs}\}$, which
specifies how many edges are placed between nodes belonging to blocks
$r$ and $s$ (or twice that number for $r=s$, for convenience of
notation), where we have $\sum_{rs}e_{rs}=2E$. This simple definition
allows one to generate a broad variety of overlapping patterns, which
are not confined to purely assortative structures, and the
nonoverlapping model can be recovered as a special case, simply by
putting each node in a single group.

The likelihood of observing a given graph with the above constraints is
simply
$P(G|\{\vec{b}_i\},\{e_{rs}\})=1/\Omega(\{\vec{b}_i\},\{e_{rs}\})$,
where $\Omega(\{\vec{b}_i\},\{e_{rs}\})$ is the number of possible
graphs, and $\mathcal{S}_t=\ln \Omega(\{\vec{b}_i\},\{e_{rs}\})$ is the
associated ensemble entropy. In this construction, the existence of
multiple edges is allowed. However, the placement of multiple edges
between nodes of blocks $r$ and $s$ should occur with a probability
proportional to $O(e_{rs}/n_rn_s)$, where $n_r$ is the number of nodes
which belong to block $r$, i.e. $n_r = \sum_ib_i^r$ (note that
$\sum_rn_r\ge N$). Since here we are predominantly interested in the
sparse situation where $e_{rs} \sim O(N/B^2)$ and $n_r \sim O(N/B)$, the
probability of observing parallel edges will decay as $O(1/N)$, and
hence can be neglected in the large network limit. Making use of this
simplification, we may approximately count all possible graphs generated
by the parameters $\{\vec{b}_i\},\{e_{rs}\}$ as the number of graphs
where each distinct membership of a single node is considered to be a
different node with a single membership. This corresponds to an
augmented graph generated via a nonoverlapping block model with
$N'=\sum_rn_r$ nodes, where $N'\ge N$, but with the same matrix
$\{e_{rs}\}$, for which the entropy is~\cite{peixoto_entropy_2012}
\begin{equation}\label{eq:st}
  \mathcal{S}_t \simeq E - \frac{1}{2}\sum_{rs}e_{rs}\ln\left(\frac{e_{rs}}{n_rn_s}\right),
\end{equation}
where $n_rn_s \gg e_{rs}$ was assumed. Under this formulation, we
recover trivially the single-membership case simply by assigning each
node to a single group, since Eq.~\ref{eq:st} remains the same in that
special case.  It is possible to remove the approximation that no
parallel edges occur, by defining the model somewhat differently, as in
shown in Appendix~\ref{app:poisson-trad}, in which case the
Eq.~\ref{eq:st} holds exactly as long as no parallel edges are observed.

Like its nonoverlapping counterpart, the block model without degree
correction assumes that nodes belonging to the same group will receive
approximately the same number of edges of that type. Hence, when applied
to empirical data, the modules discovered will also tend to have this
property. This means that if the graph possesses large degree
variability, the groups inferred will tend to correspond to different
degree classes~\cite{karrer_stochastic_2011}. On a similar vein, if a
node belongs to more than one group, it will also tend to have a total
degree that is larger than nodes that belong to either group alone,
since it will receive edges of each type in an independent fashion. In
other words, the group intersections are expected to be strictly
\emph{denser} than the nonoverlapping portions of each group. Note that,
in this respect, this model differs from other popular ones, such as the
mixed membership stochastic block model
(MMSBM)~\cite{airoldi_mixed_2008}, where the density at the
intersections is the weighted average of the groups (see
Appendix~\ref{app:poisson-trad}).

\subsection{Overlapping model with degree correction}

In the preceding model, nodes that belong to the same group mixture
receive, on average, the same number of connections. This means that the
group membership is the only factor regulating the propensity of a given
node to receive links. An alternative possibility, formulated by Karrer
et al~\cite{karrer_stochastic_2011}, is to consider that the nodes have
individual propensities to connect themselves, which are not necessarily
correlated with their group memberships. Therefore, in this
``degree-corrected'' model, nodes of the same group are allowed to
possess very different degrees. It has been demonstrated in
Ref.~\cite{karrer_stochastic_2011} that this model yields more intuitive
partitions for many empirical networks, suggesting that these intrinsic
propensities may be a better model for these systems.  In an analogous
manner, a multiple membership version of the stochastic block model with
degree correction can be defined. This can be achieved simply by
specifying, in addition to the overlapping partition $\{\vec{b}_i\}$,
the number of half-edges incident on a given node $i$ which belong to
group $r$, i.e. $k_i^r$. The combined \emph{labeled degree} of a node
$i$ is denoted $\vec{k}_i = \{k_i^r\}$. Given this \emph{labeled degree
sequence}, one can simply use the same edge count matrix $\{e_{rs}\}$ as
before to generate the graph. If we again make the assumption that the
occurrence of parallel edges can be neglected, the total number of
graphs fulfilling these constraints is approximately equal to the
nonoverlapping ensemble where each set of half-edges incident on any
given node $i$ that belongs to the same group $r$ is considered as an
individual node with degree $k^r_i$, for which the ensemble entropy
is~\cite{peixoto_entropy_2012}
\begin{equation}\label{eq:sd}
  \mathcal{S}_d \simeq -E - \frac{1}{2}\sum_{rs}e_{rs}\ln\left(\frac{e_{rs}}{e_re_s}\right) - \sum_{ir}\ln k^r_i!,
\end{equation}
where $e_{rs}(\avg{k^2}_r - \avg{k}_r)(\avg{k^2}_s -
\avg{k}_s)/\avg{k}^2_r\avg{k}^2_s \ll n_rn_s$ has been assumed.
Similarly to the non-degree-corrected case, it is possible to remove the
approximation that no parallel edges occur, by using a ``Poisson''
version of the model, as is shown in
Appendix~\ref{app:poisson-deg-corr}. Under this formulation, it can be
shown that this model is equivalent to the one proposed by Ball et
al~\cite{ball_efficient_2011}, although here we keep track of the
individual labels on the half-edges as latent variables, instead of
their probabilities.

Since we incorporate the labeled degree sequence as model parameters,
nodes that belong to the same group can have arbitrary
degrees. Furthermore, since the same applies to nodes that belong
simultaneously to more than one group, the overlaps between groups are
neither preferably dense nor sparse; it all depends on the parameters
$\{\vec{k}_i\}$.

\section{Model Selection}\label{sec:model_selection}

As discussed previously, in order to perform model selection, it is
necessary to include the information needed to describe the model
parameters, in addition to the data. The parameters which need to be
described are the overlapping partition $\{\vec{b}_i\}$, the edge counts
$\{e_{rs}\}$, and in the case of the degree-corrected model we also need
to the describe the labeled degree sequence $\{\vec{k}_i\}$.

When choosing an encoding for the parameters (via a particular
generative process) we need to avoid redundancy, and describe them as
parsimoniously as possible, while at the same time averting biases by
being noninformative. In the following, we systematically employ
two-level Bayesian hierarchies, where discrete prior distributions are
parametrized via generic counts, which are themselves sampled from
uniform nonparametric hyperpriors.

\subsection{Overlapping partition, $\{\vec{b}_i\}$}

In order to specify the partition $\{\vec{b}_i\}$, we assume that all
different $2^{B}-1$ mixtures are not necessarily equally likely, and
furthermore the sizes $d_i = \sum_rb_i^r$ of the mixtures are also not a
priori assumed to follow any specific distribution. More specifically,
we consider the mixtures to be the outcome of a generative process with
two steps. We first generate the local mixture sizes $\{d_i\}$, from a
nonparametric distribution. Then, given the mixture sizes, we generate
the actual mixtures $\{\vec{b}_i\}$ for each corresponding subset of the
nodes, again using a nonparametric distribution, conditioned on the
mixture size size.

The mixture sizes $\{d_i\}$ are sampled uniformly from the distribution
with fixed counts $\{n_d\}$, where $n_d$ is the number of nodes with a
mixture of size $d_i=d$, with a likelihood
\begin{equation}
P(\{d_i\}|\{n_d\}) = \frac{\prod_d n_d!}{N!}.
\end{equation}
For the counts $\{n_d\}$ we assume a flat prior $P(\{n_d\}) = 1 /
\multiset{D}{N}$, where $D$ is the maximum value of $d$, and the
denominator is the total number of different choices of $\{n_d\}$, with
$\multiset{n}{m} = {n + m - 1 \choose m}$ being the total number of
$m$-combinations with repetitions from a set of size $n$.

Then, for all $n_d$ nodes with the same value of $d_i=d$, we sample a
sequence of $\{\vec{b}_i\}_d$ from a distribution with support
$|\vec{b}_i|_1 \equiv \sum_rb^r_i = d$ and with fixed counts
$\{n_{\vec{b}}\}_d$, where $n_{\vec{b}}$ is the number of nodes
belonging to a specific mixture $\vec{b}_i=\vec{b}$ of size $d$,
\begin{equation}
P(\{\vec b_i\}_d|\{n_{\vec{b}}\}_d) = \frac{\prod_{|\vec{b}|_1=d} n_{\vec{b}}!}{n_d!}.
\end{equation}
For the counts themselves, we again assume a flat prior
$P(\{n_{\vec{b}}\}_d|n_d)=1/\multiset{{B \choose d}}{n_d}$, where the
denominator enumerates the total number of $\{n_{\vec{b}}\}$ counts with
$|\vec{b}|_1=d$.

The full posterior for the overlapping partition then becomes
\begin{multline}
  P(\{\vec b_i\}) = \left[\prod_dP(\{\vec b_i\}_d|\{n_{\vec{b}}\}_d)P(\{n_{\vec{b}}\}_d|n_d)\right]\times \\
      P(\{d_i\}|\{n_d\})P(\{n_d\}),
\end{multline}
which corresponds to a description length $\mathcal{L}_p = -\ln P(\{\vec b_i\})$,
\begin{align}\label{eq:lp}
  \mathcal{L}_p = \ln{\textstyle\multiset{D}{N}} + \sum_d \ln {\textstyle\multiset{{B \choose d}}{n_d}} + \ln N! - \sum_{\vec{b}}\ln n_{\vec{b}}!.
\end{align}

Although it is possible to encode the partition in different ways
(e.g. by sampling the membership to each group
independently~\cite{latouche_model_2014}), this choice makes no
assumptions regarding the types of overlaps that are more likely to
occur, either according to the number of groups to which a node may
belong, or the actual combination of groups --- it is all left to be
learned from data.  In particular, it is not \emph{a priori} assumed that if
many nodes belong to two specific groups then the overlap between these
same groups will also contain many nodes. As desired, if the observed
partition deviates from this pattern, this will be used to compress it
further. Only if the observed partition falls squarely into this pattern
will further compression be impossible, and we would have an overhead
describing it using Eq.~\ref{eq:lp}, when compared to an encoding that
expects it \emph{a priori}. However, one can also see that in the limit
$n_{\vec{b}} \gg 1$, as the first two terms in Eq.~\ref{eq:lp} grow
asymptotically only with $\ln N$ and $\ln n_d$, respectively, the whole
description length becomes $\mathcal{L}_p \simeq NH(\{n_{\vec{b}}/N\})$,
where $H(\{p_x\})$ is the entropy of the distribution $\{p_x\}$, which
is the optimal limit. Hence if we have a prior that better matches the
observed overlap, the difference in description length compared to
Eq.~\ref{eq:lp} will disappear asymptotically for large systems. Another
advantage of this encoding is that it incurs no overhead when there are
no overlaps at all (i.e. $D=1$), and in this case, the description length
is identical to the nonoverlapping case,
\begin{align}
  \mathcal{L}_p(D=1) = \ln{\textstyle\multiset{B}{N}} + \ln N! - \sum_r\ln n_r!,
\end{align}
as defined in Ref.~\cite{peixoto_hierarchical_2014}.

\subsection{Labeled degree sequence, $\{\vec{k}_i\}$}

For the degree-corrected model, we need to describe the labeled degree
sequence $\{\vec{k}_i\}$. We need to do so in a way which is compatible
with the partition $\{\vec{b}_i\}$ described so far, and with the edge
counts $\{e_{rs}\}$, which will restrict the average degrees of each
type.

In order to fully utilize the partition $\{\vec{b}_i\}$, we describe for
each distinct value of $\vec{b}_i = \vec{b}$ its individual degree
sequence $\{\vec{k}_i\}_{\vec{b}}=\{\vec{k}_i | \vec{b}_i = \vec{b}\}$,
via the counts $n_{\vec{k}}^{\vec{b}}$, i.e. the number of nodes with
mixture $\vec{b}_i=\vec{b}$ which possess labeled degree
$\vec{k}_i=\vec{k}$. We do so in order to preserve the lack of
preference for patterns involving the degrees in the overlaps between
groups. Since the model itself is agnostic with respect to the density
of the overlaps, not only does this choice remain consistent with this
indifference, but also any existing pattern in the degree sequence in
the overlaps will be used to construct a shorter description.

In addition, we must also consider the total number of half-edges of a
given type $r$ incident on a partition $\vec{b}$,
$e^r_{\vec{b}}=\sum_{\vec{k}}k_rn_{\vec{k}}^{\vec{b}}$, where $k_r$ is
the element of $\vec{k}$ corresponding to group $r$, which must be
compatible with the edge counts $\{e_{rs}\}$ via $e_r = \sum_se_{rs} =
\sum_{\vec{b}} e^r_{\vec{b}}$.

An overview of the generative process is as follows: We first consider
the $e_r$ half-edges of each type $r$ and the nonempty ($n_{\vec{b}} >
0$) mixtures $\vec{b}$ which contain the same group $r$. We then
distribute the labeled half-edges among these mixtures, obtaining the
total number of labeled edges incident on each mixture,
$\{e^r_{\vec{b}}\}$. This placement constrains the average degree of
each type inside each mixture. Finally, given $\{e^r_{\vec{b}}\}$, we sample
the actual labeled degree sequence on the nodes of each mixture.

We begin by first distributing all $e_r$ half-edges of type $r$ among
all $m_r$ bins corresponding to each nonempty mixture $\vec{b}$ that
contains the label $r$, i.e. $m_r =
\sum_{\vec{b}}b_r[n_{\vec{b}}>0]$. The total number of such partitions
is simply $\multiset{m_r}{e_r}$, and hence the likelihood for
$\{e^r_{\vec{b}}\}$ becomes
\begin{equation}
  P(\{e^r_{\vec{b}}\}|\{e_{rs}\},\{\vec{b}_i\}) = \multiset{m_r}{e_r}^{-1}.
\end{equation}
Given $\{e^r_{\vec{b}}\}$, we need to distribute the labeled half-edges
inside each partition to obtain each degree sequence. If we sample
uniformly from all possible degree sequences fulfilling all necessary
constraints, we have a likelihood for the degree sequence inside a
mixture $\vec{b}$ given by
\begin{equation}
  P_{\vec{b}}^{(1)}(\{\vec{k}_i\}_{\vec{b}}|\{e^r_{\vec{b}}\},\{\vec{b}_i\}) = \prod_r \multiset{n_{\vec{b}}}{e^r_{\vec{b}}}^{-1},
\end{equation}
where $\multiset{n}{e}$ is the total number of (unlabeled) degree
sequences with a total of $e$ half-edges incident on $n$ nodes. The
corresponding description length would then be
\begin{equation}\label{eq:deg_simple}
  \mathcal{L}^{(1)}_{\vec{b}} = \sum_r\ln{\multiset{n_{\vec{b}}}{e^r_{\vec{b}}}}.
\end{equation}
However, most degree sequences sampled this way will result in nodes
with very similar degrees. Since we want to profit from degree
variability, it is better to condition the description on the degree
counts $\{n_{\vec{k}}^{\vec{b}}\}$, i.e. how many nodes with mixture
$\vec{b}_i=\vec{b}$ possess labeled degree $\vec{k}_i=\vec{k}$. This
alternative distribution is given by
\begin{equation}
  P_{\vec{b}}^{(2)}(\{\vec{k}_i\}_{\vec{b}}|\{n_{\vec{k}}^{\vec{b}}\},\{\vec{b}_i\}) = \frac{\prod_{\vec{k}} n_{\vec{k}}^{\vec{b}}!}{n_{\vec{b}}!}.
\end{equation}
For the degree counts themselves, we choose a uniform prior
$P(\{n_{\vec{k}}^{\vec{b}}\} | \{e^r_{\vec{b}}\}, \{\vec{b}_i\}) = 1 /\Xi_{\vec{b}}$,
where $\Xi_{\vec{b}}$ is the enumeration of all possible
$\{n_{\vec{k}}^{\vec{b}}\}$ counts that fulfill the constraints
$\sum_{\vec{k}}n_{\vec{k}}^{\vec{b}} = n_{\vec{b}}$ and
$\sum_{\vec{k}}k_rn_{\vec{k}}^{\vec{b}} = e^r_{\vec{b}}$.
Unfortunately, this enumeration cannot be done easily in closed
form. However, the maximum entropy ensemble where these constraints are
enforced \emph{on average} is analytically tractable, and as we show in
Appendix~\ref{sec:maxent}, can be well approximated by $\Xi_{\vec{b}} =
\prod_{r\in\vec{b}}\Xi_{\vec{b}}^r$, where
\begin{equation}\label{eq:deg_xi_approx}
  \ln\Xi_{\vec{b}}^r \simeq 2\sqrt{\zeta(2)e_{\vec{b}}^r},
\end{equation}
and $\zeta(x)$ is the Riemann zeta function. The alternative
description length becomes therefore
\begin{equation}\label{eq:deg_xi}
  \mathcal{L}^{(2)}_{\vec{b}}  = \sum_{r\in\vec{b}}\ln\Xi_{\vec{b}}^r + \ln n_{\vec{b}}! - \sum_{\vec{k}} \ln n^{\vec{b}}_{\vec{k}}!.
\end{equation}
This approximation with ``soft'' constraints should become
asymptotically exact as the number of nodes becomes large, but otherwise
will deviate from the actual entropy. On the other hand, if the number
of nodes is very small, describing the degree sequence via
Eq.~\ref{eq:deg_xi} may not provide a shorter description, even if
computed exactly. In this situation, Eq.~\ref{eq:deg_simple} may
actually provide a shorter description of the degree sequence. We
therefore compute both Eq.~\ref{eq:deg_simple} and Eq.~\ref{eq:deg_xi}
and choose whichever is shorter.  Putting it all together, the
complete posterior for the whole labeled degree sequence is
\begin{multline}
  P(\{\vec{k}_i\}|\{e_{rs}\}, \{\vec{b}_i\}) = \left[\prod_{\vec{b}}P_{\vec{b}}(\{\vec{k}_i\}_{\vec{b}}|\{e^r_{\vec{b}}\},\{\vec{b}_i\})\right] \times \\
  P(\{e^r_{\vec{b}}\}|\{e_{rs}\},\{\vec{b}_i\}),
\end{multline}
with $P_{\vec{b}}(\{\vec{k}_i\}_{\vec{b}}|\{e^r_{\vec{b}}\},\{\vec{b}_i\})$ being
the largest choice between
$P_{\vec{b}}^{(1)}(\{\vec{k}_i\}_{\vec{b}}|\{e^r_{\vec{b}}\},\{\vec{b}_i\})$ and
$P_{\vec{b}}^{(2)}(\{\vec{k}_i\}_{\vec{b}}|\{n_{\vec{k}}^{\vec{b}}\})P(\{n_{\vec{k}}^{\vec{b}}\}| \{e^r_{\vec{b}}\}, \{b_i\})$.
Therefore, the description length for the labeled degree sequence
$\mathcal{L}_{\kappa} = -\ln P(\{\vec{k}_i\}|\{e_{rs}\}, \{b_i\})$ becomes
\begin{equation}\label{eq:lkappa}
  \mathcal{L}_{\kappa} = \sum_r\ln{\textstyle\multiset{m_r}{e_r}} + \sum_{\vec{b}}\min\left(\mathcal{L}^{(1)}_{\vec{b}}, \mathcal{L}^{(2)}_{\vec{b}}\right).
\end{equation}
In the limit $n^{\vec{b}}_{\vec{k}} \gg 1$, we have that
$\mathcal{L}_{\kappa} \simeq
\sum_{\vec{b}}n_{\vec{b}}H(\{n^{\vec{b}}_{\vec{k}} / n_{\vec{b}}\})$,
and hence the degree sequences in each partition are described close to
the optimum limit.

For the nonoverlapping case with $D=1$, the description length
simplifies to
\begin{equation}\label{eq:lkappa_D1}
  \mathcal{L}_{\kappa} = \sum_r\min\left(\mathcal{L}^{(1)}_r, \mathcal{L}^{(2)}_r\right),
\end{equation}
with
\begin{align}
  \mathcal{L}^{(1)}_r &= \ln{\multiset{n_r}{e_r}}, \\
  \mathcal{L}^{(2)}_r &= \ln\Xi_r + \ln n_r! - \sum_k \ln n^r_k!,
\end{align}
and $\ln\Xi_r \simeq 2\sqrt{\zeta(2)e_r}$. For $n_r\gg 1$ we obtain
$\mathcal{L}_{\kappa} \simeq \sum_rn_rH(\{n^r_k / n_r\})$. This
approximation was used \emph{a priori} in
Ref.~\cite{peixoto_hierarchical_2014}, but Eq.~\ref{eq:lkappa_D1} is a
more complete description length of the nonoverlapping degree sequence,
and its use should be preferred. Hence, like the description length of
the overlapping partition, the encoding above offers no overhead when
the partition is nonoverlapping.

\subsection{Edge counts, $\{e_{rs}\}$}

The final piece that needs to be described is the matrix of edge counts
$\{e_{rs}\}$. We may view this set as an adjacency matrix of a
multigraph with $B$ nodes and $E=\sum_{rs}e_{rs}/2$ edges. The total
number of such matrices is $\Omega(B,E)=\multiset{{B \choose 2}}{E}$,
and if we assume that they are all equally likely, we have
$P(\{e_{rs}\}) = 1 / \Omega(B,E)$ and $\ln \Omega(B,E)$ can be used as
the description length~\cite{peixoto_parsimonious_2013}. There are,
however, two problems with this approach. First, this uniform
distribution is unlikely to be valid, since most observed networks still
possess structure at the block level. Second, this assumption leads to a
limit in the detection of small groups, with a maximum detectable number
of groups scaling as
$B_{\text{max}}\sim\sqrt{N}$~\cite{peixoto_parsimonious_2013}.
Similarly to what we did for the node partition and the degree sequence,
this can be solved by considering a generative model for the edge counts
themselves, with its own set of hyperparameters.  Since they correspond
to a multigraph, a natural choice is the stochastic block model itself,
which has its own set of edge counts, that can themselves be modeled by
another stochastic block model with fewer nodes, and so on, recursively,
until one has a model only one node and one group at the top. This
nested stochastic block model was proposed in
Ref.~\cite{peixoto_hierarchical_2014}, where it has been shown to reduce
the resolution limit to $B_{\text{max}}\sim N/\log{N}$, making it often
significantly less relevant in practice. Furthermore, since the number
of levels and the topology at each level is obtained by minimizing the
overall description length, it corresponds to a fully nonparametric way
of inferring the multilevel structure of networks. As shown in
Ref.~\cite{peixoto_hierarchical_2014}, if we denote the observed network
to be at the level $l=0$ of the hierarchy, then the total description
length is
\begin{equation}\label{eq}
  \Sigma = \mathcal{S}_{t/c} + \sum_{l=1}^LS_m(\{e^l_{rs}\}, \{n^l_r\}) + \mathcal{L}^{l-1}_t,
\end{equation}
with $\{e^l_{rs}\}$, $\{n^l_r\}$ describing the block model at level
$l$, where
\begin{equation}\label{eq:sm}
  \mathcal{S}_m  = \sum_{r>s} \ln{\textstyle \multiset{n_rn_s}{e_{rs}}} + \sum_r \ln{\textstyle \multiset{\multiset{n_r}{2}}{e_{rr}/2}}
\end{equation}
is the entropy of the corresponding multigraph ensemble and
\begin{equation}\label{eq:dli}
  \mathcal{L}^l_t = \ln{\textstyle \multiset{B_l}{B_{l-1}}} + \ln B_{l-1}! - \sum_r \ln n_r^l!.
\end{equation}
is the description length of the node partition at level $l>0$. For the
level $l=0$ we have $\mathcal{L}^0_t = \mathcal{L}_p$
given by Eq.~\ref{eq:lp}, or $\mathcal{L}^0_t = \mathcal{L}_p +
\mathcal{L}_{\kappa}$ for the degree-corrected model.

Note that here we use the single-membership non-degree-corrected model
in the upper layers. This method could be modified to include arbitrary
mixtures of degree correction and multiple membership, but we stick with
this formulation for simplicity.

\subsection{Significance levels}

By minimizing the description length $\Sigma$, we select the model that
is most favored given the evidence in the data. But in some situations,
one is not merely interested in a binary answer regarding which of two
model choices is best, but instead, one would like to be able to rule out
alternative models with some degree of confidence. In this case, a level
of significance can be obtained by performing a Bayesian hypothesis test
based on the ratio of posterior likelihoods. In this context, there are
different hypotheses which can be tested. For instance, one could ask
whether the entire class of non-degree-corrected overlapping models
(NDCO) is favored in comparison to the class of nonoverlapping
degree-corrected models (DC). This can be done by computing the
posterior distribution for each model class $\mathcal{H}\in
\{\text{NDCO}, \text{DC}\}$,
\begin{equation}
  P(\mathcal{H}|G) = \frac{\sum_{\theta}P(G|\theta,\mathcal{H})P(\theta)P(\mathcal{H})}{P(G)},
\end{equation}
where $\theta$ is shorthand for the entire set of model parameters
[i.e. $\theta = (\{\vec{b}_i\}, \{e_{rs}\})$ for
$\mathcal{H}=\text{NDCO}$, and $\theta = (\{b_i\}, \{e_{rs}\}, \{k_i\})$
  for $\mathcal{H}=\text{DC}$],
with $P(\mathcal{H})$ being the prior belief we have supporting a
given hypothesis, and $P(G)$ is a normalizing constant. The
standard way in Bayesian statistics to evaluate the relative
evidence supporting (or rejecting) hypothesis $\mathcal{H}_1$ over
$\mathcal{H}_2$ is via the posterior odds ratio~\cite{jaynes_probability_2003}
\begin{equation}
  \Lambda = \frac{P(\mathcal{H}_1|G)}{P(\mathcal{H}_2|G)} = \frac{\sum_{\theta}P(G|\theta,\mathcal{H}_1)P(\theta)}{\sum_{\theta}P(G|\theta,\mathcal{H}_2)P(\theta)}\frac{P(\mathcal{H}_1)}{P(\mathcal{H}_2)}.
\end{equation}
However, there are two issues with this approach. First, computing the
sum over all parameter choices is intractable in this context, since it
involves summing over all possible overlapping or nonoverlapping
partitions. Second, and more importantly, this might not be the answer
which is more relevant. If one obtains two model parametrizations by
minimizing the description length as described in the previous section,
with the two results belonging to different model classes, one would be
more interested in selecting or rejecting between these two particular
choices, not necessarily the overall class to which they
belong. Although the description length itself already provides a means
to select the best alternative, one would be interested in obtaining a
confidence level for \emph{this particular decision}. This is a
different sort of hypothesis test than the one above, but which can be
performed analogously. Since the result of the minimization of the
description length is the (possibly overlapping) partition of the
network, our hypothesis is a combination of the model class which we
were using, and the particular partition that was found. The posterior
probability attributed to this hypothesis is therefore
\begin{equation}\label{eq:posterior}
  P(\{\vec{b}_i\},\mathcal{H}|G) = \frac{P(G|\{\vec{b}_i\},\mathcal{H})P(\{\vec{b}_i\}|\mathcal{H})P(\mathcal{H})}{P(G)},
\end{equation}
where again $P(G)$ is a normalization constant. The marginal likelihood
$P(G|\{\vec{b}_i\},\mathcal{H})$ is obtained by summing over the
remaining model parameters. In the case of the overlapping
degree-corrected model ($\mathcal{H} = \text{DCO}$) they are the
$\{e_{rs}\}$ matrix and the labeled degree sequence $\{\vec{k}_i\}$
(which is omitted for the non-degree-corrected model, $\mathcal{H} =
\text{NDCO}$),
\begin{align}
  P(G|\{\vec{b}_i\},\text{DC}) &=
  \begin{aligned}[t]
    \sum_{\{e'_{rs}\},\{\vec{k}_i'\}} & P(G|\{\vec{b}_i\},\{e'_{rs}\},\{\vec{k}_i'\})\times \nonumber\\
    &P(\{e'_{rs}\})P(\{\vec{k}_i'\})
  \end{aligned} \nonumber\\
  &=P(G|\{\vec{b}_i\},\{e_{rs}\},\{\vec{k}_i\})P(\{e_{rs}\})P(\{\vec{k}_i\}),
\end{align}
where the sum trivially contains only one term, since for the same graph
$G$ and partition $\{\vec{b}_i\}$, there is only one possible choice for
the $\{e_{rs}\}$ matrix and degree sequence $\{\vec{k}_i\}$ with nonzero
probability, which is a convenient feature of the microcanonical model
formulation considered here [the same holds for $\mathcal{H} =
\text{NDC}$,
i.e. $P(G|\{\vec{b}_i\},\text{NDC}) =
P(G|\{\vec{b}_i\},\{e_{rs}\})P(\{e_{rs}\})$]. Now if we want to compare
two competing partitions $\{\vec{b}_i\}_a$ and $\{\vec{b}_i\}_b$, this
can be done again via the posterior odds ratio
$\Lambda$,
\begin{align}
  \Lambda &= \frac{P(\{\vec{b}_i\}_a,\mathcal{H}_a|G)}{P(\{\vec{b}_i\}_b,\mathcal{H}_b|G)} \\
          &= \frac{P(G|\{\vec{b}_i\}_a,\mathcal{H}_a)P(\{\vec{b}_i\}_a|\mathcal{H}_a)P(\mathcal{H}_a)}{P(G|\{\vec{b}_i\}_b,\mathcal{H}_b)P(\{\vec{b}_i\}_b|\mathcal{H}_b)P(\mathcal{H}_b)} \label{eq:lambda_or} \\
          &= \exp\left(-\Delta\Sigma\right) \label{eq:lambda_dl},
\end{align}
with $\Delta\Sigma = \Sigma_a-\Sigma_b$ being the difference in the
description length, and in Eq.~\ref{eq:lambda_dl} it was assumed that
$P(\mathcal{H}_a) = P(\mathcal{H}_b)=1/2$, corresponding to a lack of a
priori preference for either model variant (which, in fact, makes
$\Lambda$ identical to the Bayes
factor~\cite{jeffreys_theory_1998}). This is a simple result, which
enables us to use the difference in the description length directly in
the computation of confidence levels. Being a ratio of probabilities,
the value of $\Lambda$ has a straightforward interpretation: For a value
of $\Lambda=1$, both models explain the data equally well, and for
values of $\Lambda<1$ model $a$ is rejected in favor of $b$ with a
confidence increasing as $\Lambda$ diminishes. In order to simplify its
interpretation, the values of $\Lambda$ are usually divided into regions
corresponding to a subjective assessment of the evidence strength. A
common classification is as follows~\cite{jeffreys_theory_1998}: Values
of $\Lambda$ in the intervals $\{[1,1/3], [1/3,1/10],
[1/10,1/30],[1/30,1/100],[1/100,0]\}$ are considered to be very weak,
substantial, strong, very strong and decisive evidence supporting model
$b$, respectively. In the following, when comparing different models, we
will always put the preferred model in the denominator of
Eq.~\ref{eq:lambda_or}, such that $\Lambda \leq 1$.

\begin{figure*}
  \centering
  \begin{minipage}[t]{.49\textwidth}
    \includegraphics[width=1\columnwidth]{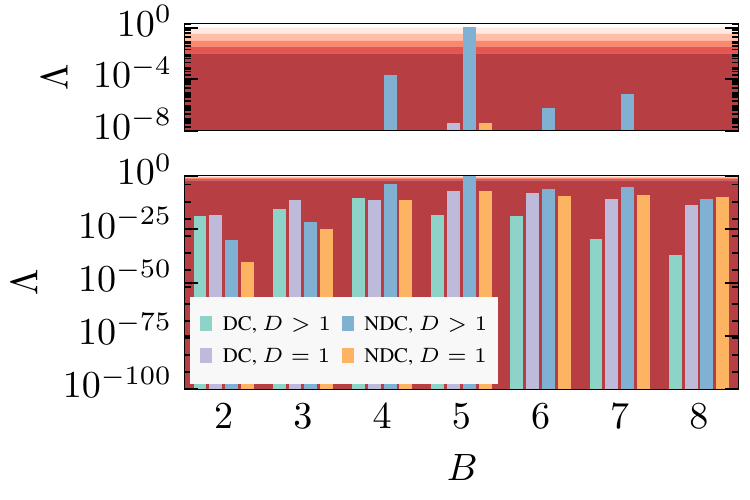}
    \begin{minipage}[t]{.45\columnwidth}
      \centering
      \includegraphics[width=1\columnwidth]{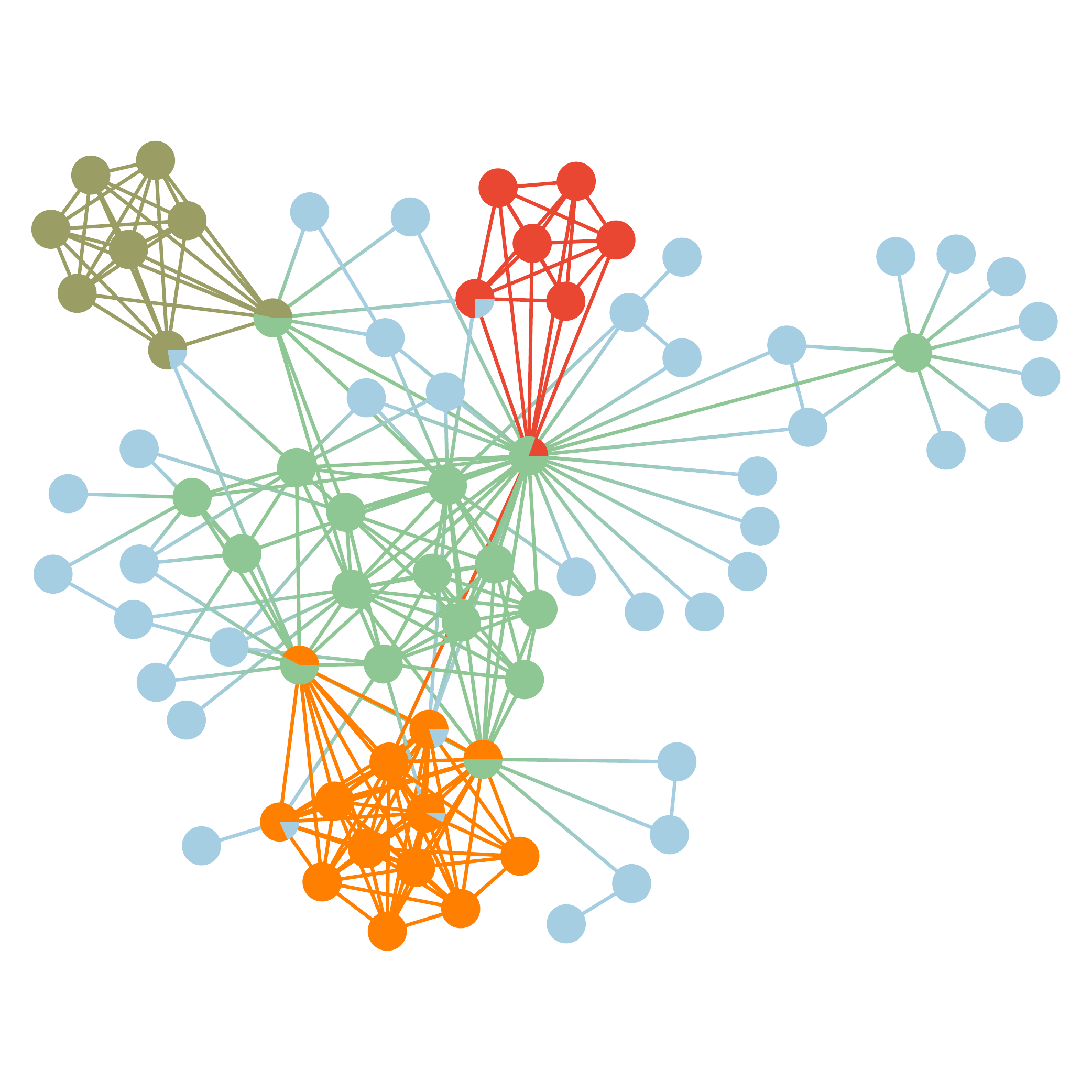}\\
      $B=5$, non-degree-corrected, overlapping, $\Lambda=1$
    \end{minipage}
    \begin{minipage}[t]{.45\columnwidth}
      \centering
      \includegraphics[width=1\columnwidth]{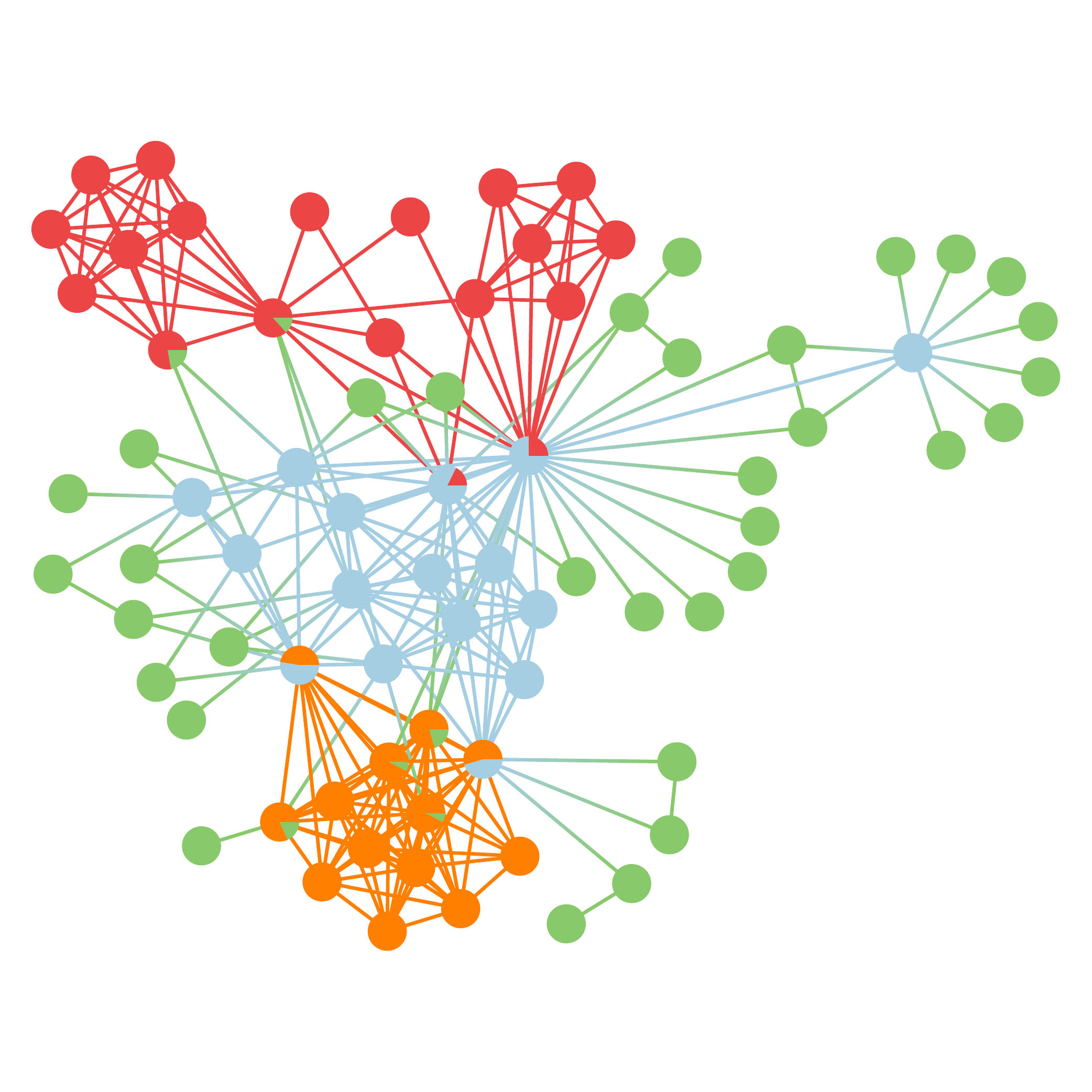}\\
      $B=4$, non-degree-corrected, overlapping, $\Lambda\simeq 2\times 10^{-4}$,
    \end{minipage}
  \end{minipage}
  \begin{minipage}[t]{.49\textwidth}
    \includegraphics[width=1\columnwidth]{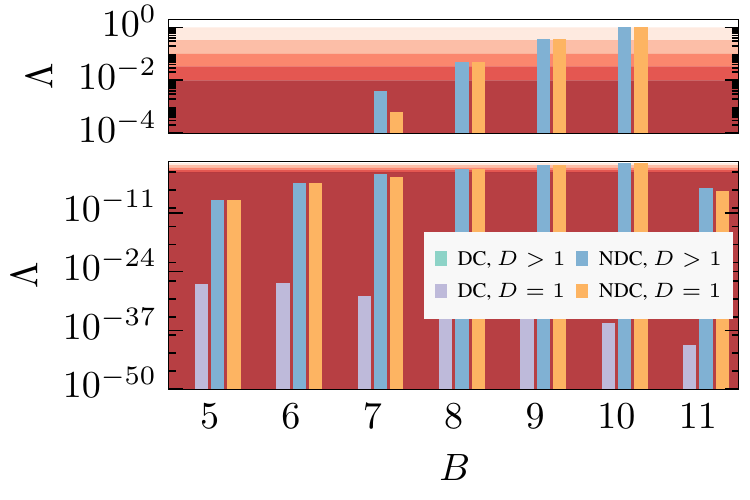}
    \begin{minipage}[t]{.45\columnwidth}
      \centering
      \includegraphics[width=1\columnwidth]{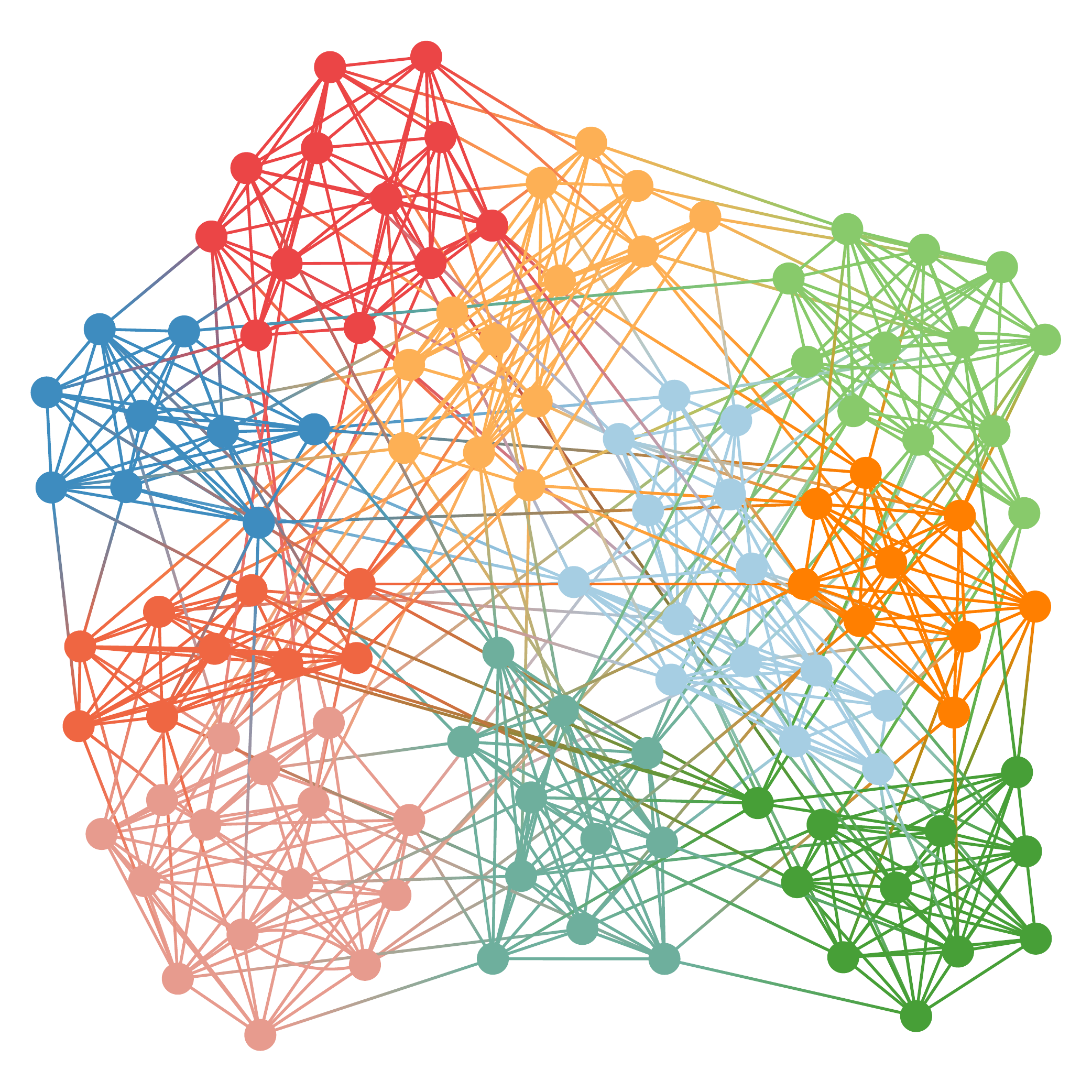}\\
      $B=10$, non-degree-corrected, nonoverlapping, $\Lambda=1$
    \end{minipage}
    \begin{minipage}[t]{.45\columnwidth}
      \centering
      \includegraphics[width=1\columnwidth]{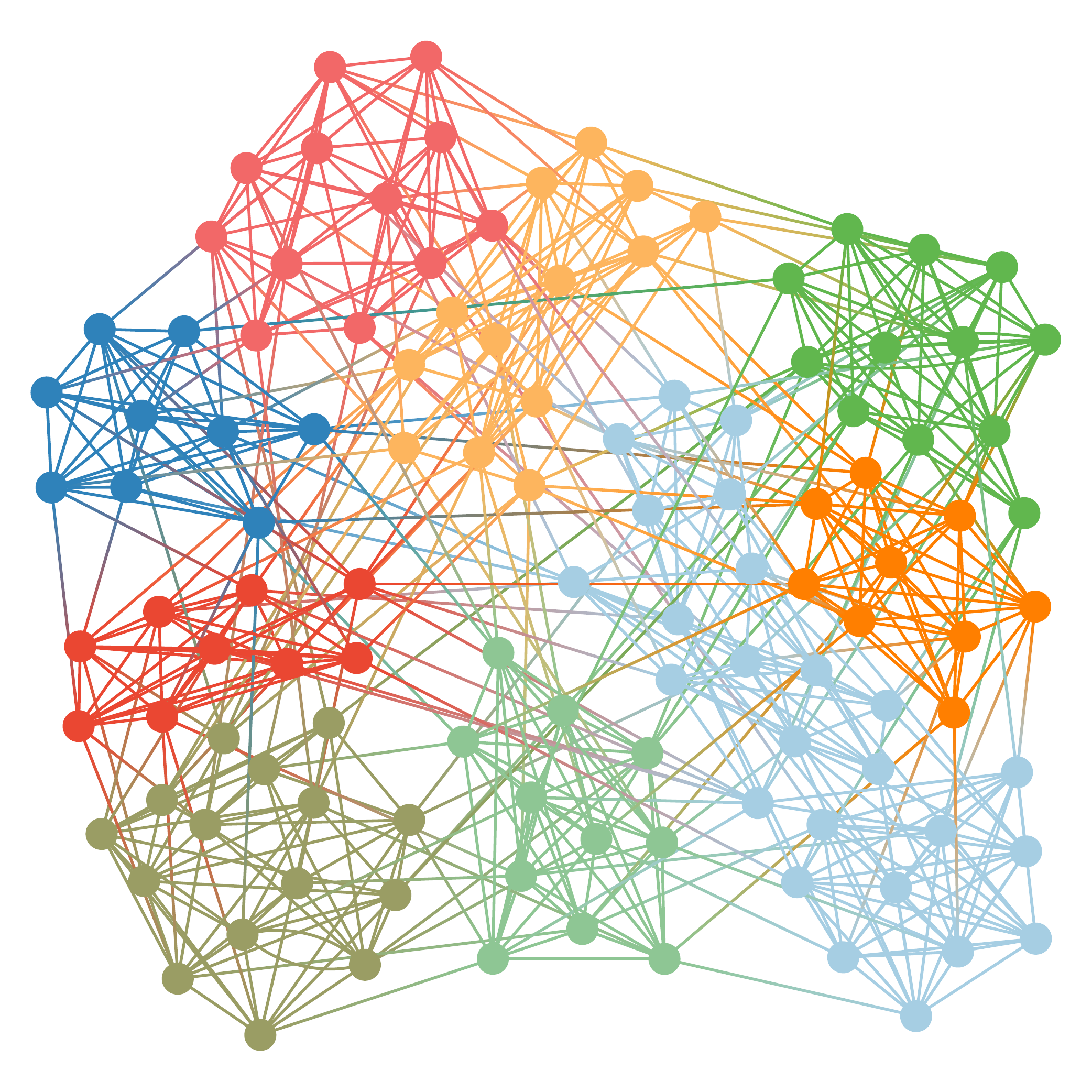}\\
      $B=9$, non-degree-corrected, nonoverlapping, $\Lambda\simeq 0.36$
    \end{minipage}
  \end{minipage}
  \caption{Left: Values for posterior odds ratio $\Lambda$ for the network of co-appearances of
characters in the novel ``Les Misérables'', for all model variations
($D>1$ indicates an overlapping model, ``DC'' a degree-corrected model
and ``NDC'' a non-degree-corrected one). The models with the best and
second-best fits are shown at the bottom. Right: Same as in the left,
but for the American college football
network.\label{fig:lesmis_football}}
\end{figure*}

Using the posterior odds ratio $\Lambda$ is more practical than some
alternative model selection approaches, such as likelihood ratios. As
has been recently shown~\cite{yan_model_2014}, the likelihood
distribution for the stochastic block model does not follow a
$\chi^2$-distribution asymptotically for sparse networks, and hence the
calculation of a $p$-value must be done via an empirical computation of
the likelihood distribution which is computationally costly, and
prohibitively so for very large networks. In contrast, computing
$\Lambda$ can be done easily, and it properly accounts for the increased
complexity of models with larger parameters, and protects against
overfitting. However, it should be emphasized that these different model
selection approaches are designed to answer similar, but not identical
questions. Therefore the most appropriate method should be the one that
more closely matches the questions raised.

\begin{figure*}[t!]
  \centering
  \begin{tabular}{cc}
    \includegraphics[width=1\columnwidth]{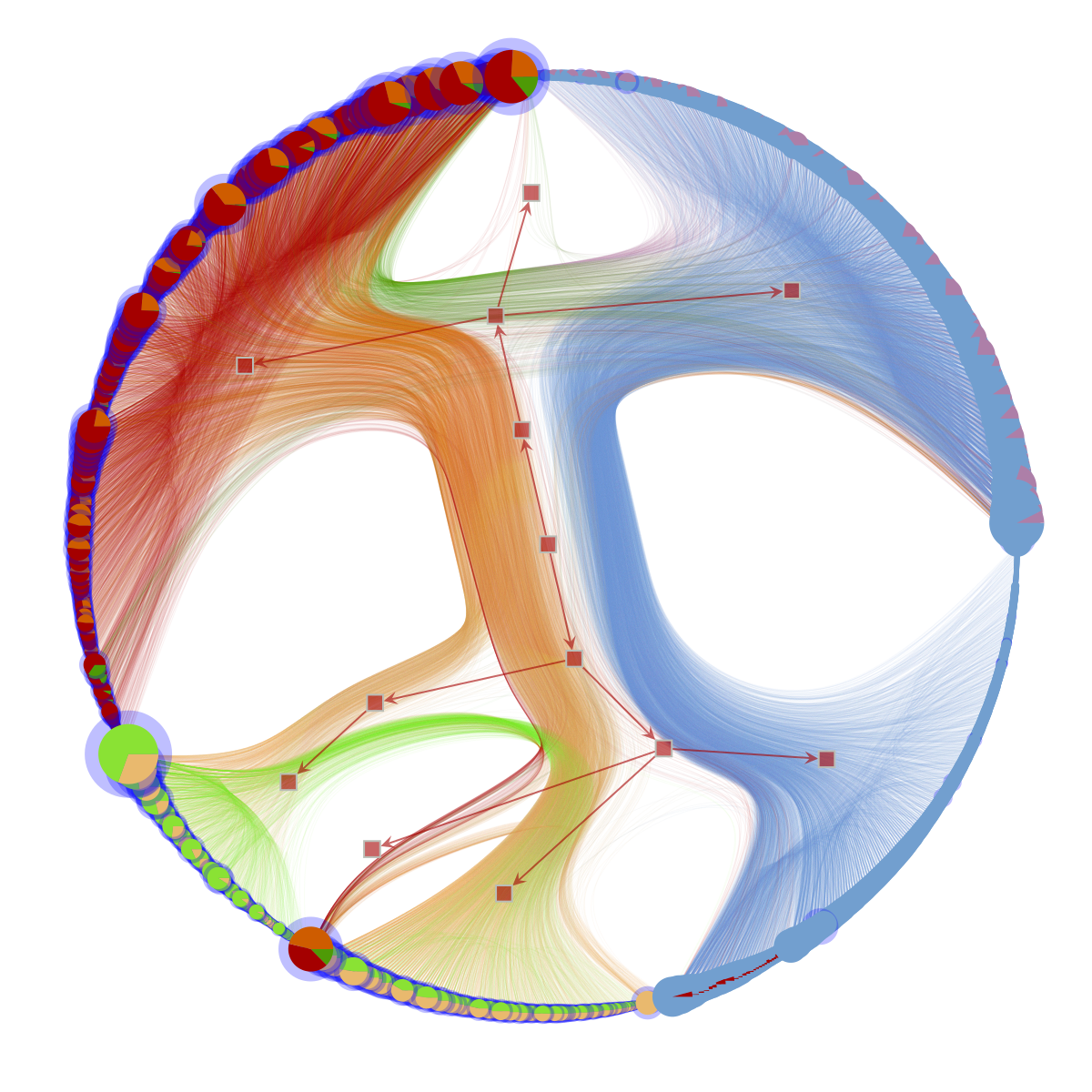} &
    \includegraphics[width=1\columnwidth]{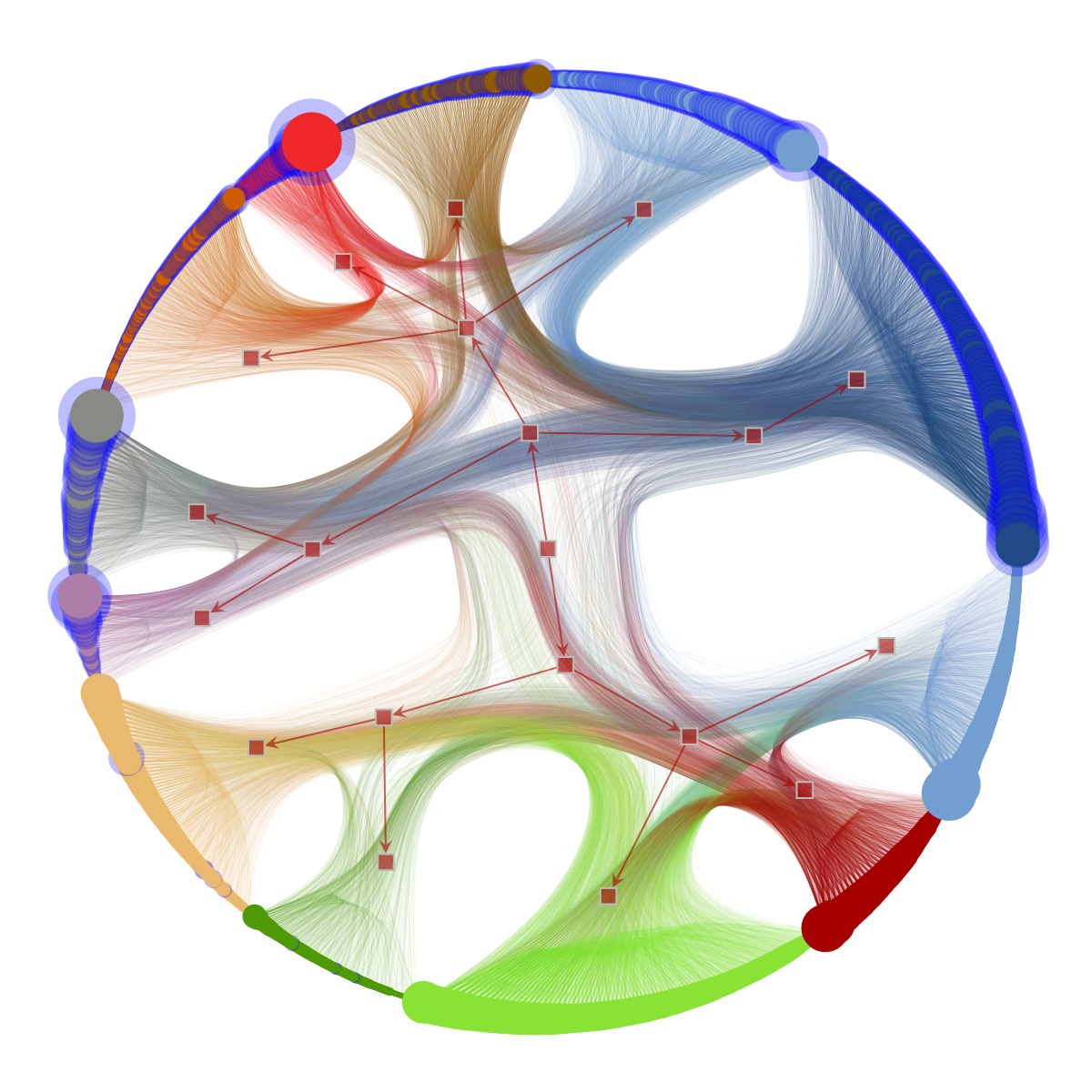} \\
    \parbox[t]{.98\columnwidth}{$B=7$, overlapping, degree-corrected, $\Lambda = 1$} &
    \parbox[t]{.98\columnwidth}{$B=12$, nonoverlapping, degree-corrected, $\log_{10}\Lambda \simeq -747$}
  \end{tabular} \caption{The network of political blogs by Adamic et
  al~\cite{adamic_political_2005}. The left panel shows the best model
  with an overlapping partition, and the right shows the best
  nonoverlapping one. Nodes with a blue halo belong to the Republican
  faction, as determined in Ref.~\cite{adamic_political_2005}. For the
  visualization, the hierarchical edge-bundles
  algorithm~\cite{holten_hierarchical_2006} was used.
  \label{fig:polblogs}}
\end{figure*}

\section{Empirical networks}\label{sec:empirical}

\begin{figure}
  \centering
  \begin{tabular}{cc}
    \includegraphics[width=.49\columnwidth]{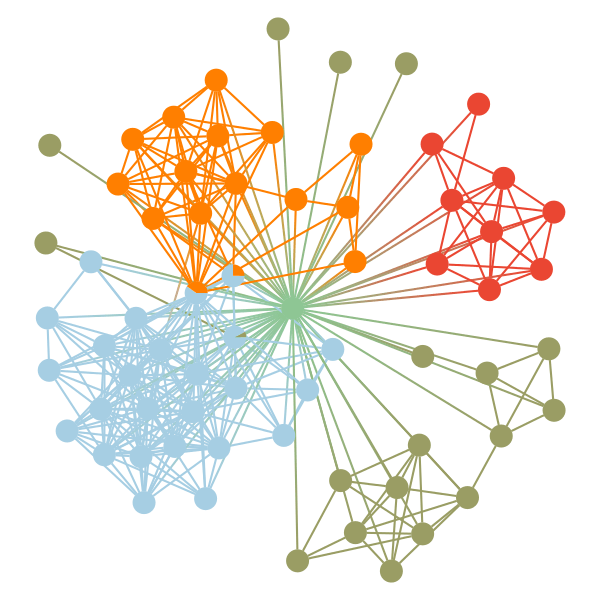} &
    \includegraphics[width=.49\columnwidth]{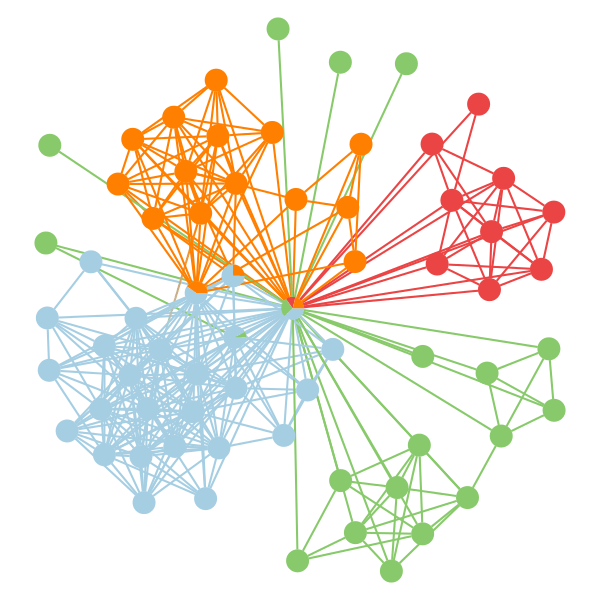} \\
    \begin{minipage}[c]{.4\columnwidth}
      \includegraphics[width=.48\columnwidth]{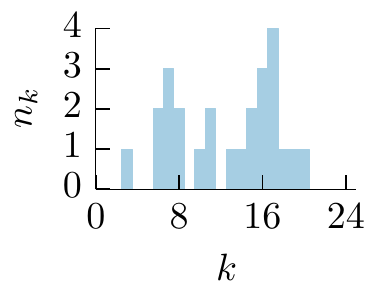}
      \includegraphics[width=.48\columnwidth]{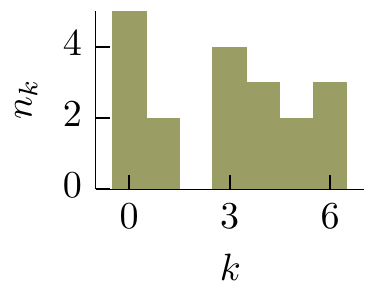}
      \includegraphics[width=.48\columnwidth]{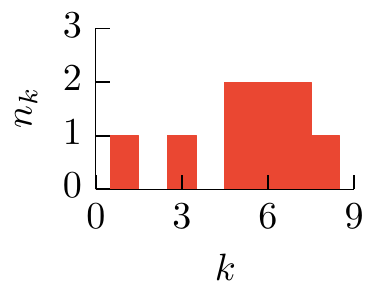}
      \includegraphics[width=.48\columnwidth]{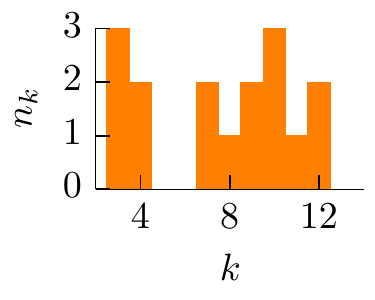}
    \end{minipage} &
    \begin{minipage}[c]{.4\columnwidth}
      \includegraphics[width=.48\columnwidth]{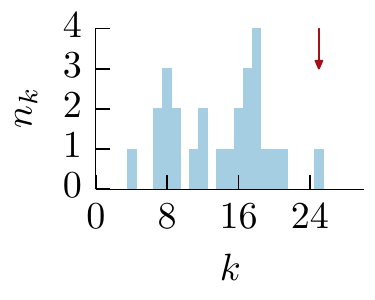}
      \includegraphics[width=.48\columnwidth]{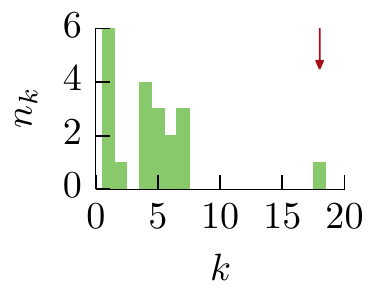}
      \includegraphics[width=.48\columnwidth]{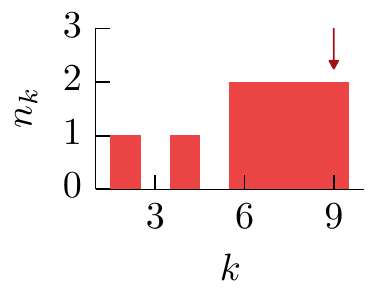}
      \includegraphics[width=.48\columnwidth]{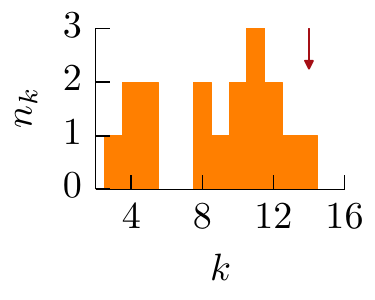}
    \end{minipage} \\
    \parbox[t]{.49\columnwidth}{$B=5$, overlapping, non-degree-corrected, $\Lambda = 1$} &
    \parbox[t]{.49\columnwidth}{$B=4$, overlapping, non-degree-corrected, $\Lambda \simeq 0.053$}
  \end{tabular}

  \caption{\label{fig:ego}Ego network of Facebook
  contacts~\cite{mcauley_discovering_2014}. Left: The best model
  fit across all model variations, which puts the ego node in its own
  group. Right: The alternative hypothesis where the node is split in
  several groups. Below each network are shown the degree distributions
  inside each group. The arrow marks the degree of the ego node.}
\end{figure}

The method outlined in the previous section allows one to determine the
best model from the various available choices. Here we analyze some
empirical examples, and determine the most appropriate model, and
examine the consequences of the balance struck between model complexity
and quality of fit. We start with two small networks, the co-appearance
of characters in the Victor Hugo novel ``Les
Misérables''~\cite{knuth_stanford_1993}, and a network of American
college football games~\cite{girvan_community_2002,
evans_american_2012}. For both networks, we obtain the best partition
according all model variations and for a different number of groups $B$,
and we compute the value of $\Lambda$ relative to the best model, as
shown in Fig.~\ref{fig:lesmis_football}. For the ``Les Misérables''
network, the best fit is a non-degree-corrected overlapping model that
puts the most central characters in more than one group. All other
partitions for different values of $B$ and model types result in values
significantly below the plausibility line of $\Lambda = 10^{-2}$,
indicating that the overlapping model offers a better explanation for
the data with a large degree of confidence. In particular, it offers a
better description than the nonoverlapping model with degree
correction. For the football network, on the other hand, the preferred
model is nonoverlapping and without degree correction with $B=10$, which
matches very well the assumed correct partition into $10$
conferences. The groups are relatively homogeneous, with most nodes
having similar degrees, such that degree correction becomes an extra
burden, with very little added explanatory power. For this network,
however, there are alternative fits with values of $\Lambda$ within the
plausibility region, which means that the communities are not very
strongly defined, and they admit alternative partitions with $B=9$ and
$B=8$ groups which cannot be confidently discarded given the evidence in
the data.

Degree correction tends to become a better choice for larger data sets,
which display stronger degree variability. One example of this is the
network of political blogs obtained by Adamic et
al~\cite{adamic_political_2005}. For this network, the best model is a
degree-corrected, overlapping partition into $B=7$ groups, shown in
Fig.~\ref{fig:polblogs}. Compared to this partition, the best
alternative model without overlap divides the network into $B=12$
groups\footnote{In Ref.~\cite{peixoto_hierarchical_2014} using the same
nonoverlapping model, a value of $B=15$ was found. This is due the
difference in the description length for the degree sequence, where here
we use a more complete estimation than in
Ref.~\cite{peixoto_hierarchical_2014}, which results in this slight
difference.}, but has a posterior odds ratio significantly below the
plausibility region. It should be observed that the nonoverlapping
version captures well the segregation into two groups (Republicans and
Democrats) at the topmost level of the hierarchy. The overlapping
version, on the other hand, tends to classify half-edges belonging to
different camps into different groups, which is compatible with the
accepted division, but the upper layers of the hierarchy do not reflect
this, and prefers to merge together groups that belong to different
factions, but that have otherwise similar roles in the topology.

Overlapping partitions, however, do not always provide better
descriptions, even in situations where it might be considered more
intuitive. One of the contexts where overlapping communities are often
considered to be better explanations is in social networks, where
different social circles could be represented as different groups
(e.g. family, co-workers, friends, etc.), and one could belong to more
than one of these groups. This is illustrated well by so-called ``ego
networks,'' where one examines only the immediate neighbors of a node,
and their mutual connections. One such network, extracted from the
Facebook online social network~\cite{mcauley_discovering_2014}, is shown
in Fig.~\ref{fig:ego}. The common interpretation of networks such as
these is shown on the right in Fig.~\ref{fig:ego}, and corresponds to a
partition of the central ``ego'' node so that it belongs to all of the
different circles. Under this interpretation, the ego node is only
special in the sense that it belongs to all groups, but inside each
group it is just a common member. However, among all model variants, the
best fit turns out to be the one where the ego node is put separately in
its own group, as shown in the left in Fig.~\ref{fig:ego}. In this
example it is easy to see why this is the case: If we observe the degree
distribution inside each group for the network on the left, we see that
there is no strong degree variation. On the right, as the ego is
included in each group, it becomes systematically the most connected
node. This is simply by construction, since the ego must connect to
every other node. The only situation where the ego would not stand out
inside each group, would be if the communities were cliques. Hence,
since the ego is not a typical member of any group, it is simpler to
classify it separately in its own group, which is selected by the method
as a being a more plausible hypothesis. Note that degree correction is
not selected as the most plausible solution, since it is burdened with
the individual description of every degree in the network, which is
fairly uniform with the exception of the ego. One can imagine a
different situation where there would be other very well connected nodes
inside each group, so that the ego could be described as a common member
of each group, but this not observed in any other network obtained in
Ref.~\cite{mcauley_discovering_2014}. Naturally, if one considers the
complete network, of which the ego neighbourhood is only a small part,
the situation may change, since there may be members of each group to
which the ego does not have a direct connection.

\begin{table}
  \centering
 \resizebox{\columnwidth}{!}
            {\smaller[2]
              \setlength{\tabcolsep}{2pt}
              \rowcolors{2}{gray!25}{white}
              \begin{tabular}{lll cccc lll}\toprule
No. & $N$ & $\avg{k}$ & $\log_{10}\Lambda_{\text{DCO}}$ & $\log_{10}\Lambda_{\text{DC}}$ & $\log_{10}\Lambda_{\text{NDCO}}$ & $\log_{10}\Lambda_{\text{NDC}}$ & $B$ & $\avg{d}$ & $\Sigma/E$\\ \midrule

1 & $34$ & $4.6$ & $-2.1$ & $-2.1$ & \multicolumn{1}{c}{---} & $0$ & $2$ & $1$ & $4$ \\
2 & $62$ & $5.1$ & $-4.6$ & $-1.4$ & \multicolumn{1}{c}{---} & $0$ & $2$ & $1$ & $4.8$ \\
3 & $77$ & $6.6$ & $-17$ & $-7.7$ & $0$ & $-7.3$ & $5$ & $1.1$ & $4$ \\
4 & $105$ & $8.4$ & $-12$ & $-2.8$ & $-6.6$ & $0$ & $5$ & $1$ & $4.4$ \\
5 & $115$ & $10.7$ & $-79$ & $-27$ & \multicolumn{1}{c}{---} & $0$ & $10$ & $1$ & $4.3$ \\
6 & $297$ & $15.9$ & $0$ & $-61$ & $-2.0\times10^{2}$ & $-2.1\times10^{2}$ & $5$ & $1.8$ & $5.1$ \\
7 & $379$ & $4.8$ & $-47$ & $-6.6$ & $0$ & $-8.9$ & $20$ & $1.1$ & $6.2$ \\
8 & $903$ & $15.0$ & $-3.8\times10^{2}$ & $-3.7\times10^{2}$ & $0$ & $-3.7\times10^{2}$ & $60$ & $1.2$ & $3.1$ \\
9 & $1,278$ & $2.8$ & $-8.1$ & $0$ & $-1.5\times10^{2}$ & $-89$ & $2$ & $1$ & $7.4$ \\
10 & $1,490$ & $25.6$ & $0$ & $-5.2\times10^{2}$ & $-2.3\times10^{3}$ & $-2.3\times10^{3}$ & $7$ & $1.8$ & $4.4$ \\
11 & $1,536$ & $3.8$ & $-2.5\times10^{2}$ & $0$ & $-65$ & $-62$ & $38$ & $1$ & $6.7$ \\
12 & $1,622$ & $11.2$ & $-4.3\times10^{2}$ & $0$ & $-12$ & $-82$ & $48$ & $1$ & $3.3$ \\
13 & $1,756$ & $4.5$ & $-43$ & $0$ & $-4.0\times10^{2}$ & $-2.8\times10^{2}$ & $7$ & $1$ & $5.9$ \\
14 & $2,018$ & $2.9$ & $-9.2$ & $0$ & $-2.9\times10^{2}$ & $-2.1\times10^{2}$ & $2$ & $1$ & $8.5$ \\
15 & $4,039$ & $43.7$ & $-1.5\times10^{3}$ & $0$ & $-8.1\times10^{2}$ & $-9.5\times10^{2}$ & $158$ & $1$ & $3.2$ \\
16 & $4,941$ & $2.7$ & $-2.2\times10^{2}$ & $0$ & $-21$ & $-25$ & $25$ & $1$ & $11$ \\
17 & $7,663$ & $17.8$ & $0$ & $-1.1\times10^{4}$ & $-5.3\times10^{3}$ & $-1.6\times10^{4}$ & $85$ & $1$ & $3.2$ \\
18 & $7,663$ & $5.3$ & $-1.8\times10^{3}$ & $0$ & $-9.3\times10^{2}$ & $-7.3\times10^{2}$ & $63$ & $1$ & $5$ \\
19 & $8,298$ & $25.0$ & $-9.1\times10^{3}$ & $0$ & $-1.4\times10^{4}$ & $-1.4\times10^{4}$ & $34$ & $1$ & $5.4$ \\
20 & $9,617$ & $7.7$ & $-4.2\times10^{3}$ & $0$ & $-2.3\times10^{3}$ & $-2.5\times10^{3}$ & $34$ & $1$ & $9.3$ \\
21 & $26,197$ & $2.2$ & $-2.4\times10^{3}$ & $-1.2\times10^{3}$ & $0$ & $-2.7\times10^{3}$ & $363$ & $1.3$ & $4.5$ \\
22 & $36,692$ & $20.0$ & $-4.1\times10^{4}$ & $0$ & $-8.5\times10^{4}$ & $-2.8\times10^{4}$ & $1812$ & $1$ & $5.5$ \\
23 & $39,796$ & $15.2$ & $-6.1\times10^{4}$ & $0$ & $-8.8\times10^{4}$ & $-4.5\times10^{4}$ & $1323$ & $1$ & $6.3$ \\
24 & $52,104$ & $15.3$ & $-1.5\times10^{5}$ & $0$ & $-3.7\times10^{4}$ & $-4.0\times10^{4}$ & $172$ & $1$ & $6.4$ \\
25 & $58,228$ & $14.7$ & $0$ & $-5.8\times10^{4}$ & $-1.8\times10^{5}$ & $-1.4\times10^{5}$ & $1995$ & $3.2$ & $7.3$ \\
26 & $65,888$ & $305.2$ & $-4.4\times10^{4}$ & $0$ & $-4.6\times10^{5}$ & $-4.6\times10^{5}$ & $384$ & $1$ & $4.1$ \\
27 & $68,746$ & $1.5$ & $-4.8\times10^{3}$ & $-1.4\times10^{3}$ & $0$ & $-7.0\times10^{3}$ & $719$ & $1.4$ & $6.4$ \\
28 & $75,888$ & $13.4$ & $-1.1\times10^{5}$ & $0$ & $-8.2\times10^{4}$ & $-9.0\times10^{4}$ & $143$ & $1$ & $8.9$ \\
29 & $89,209$ & $5.3$ & $-1.0\times10^{4}$ & $0$ & $-9.7\times10^{3}$ & $-1.1\times10^{4}$ & $848$ & $1$ & $3.2$ \\
30 & $108,300$ & $3.5$ & $-3.3\times10^{3}$ & $-5.2\times10^{3}$ & $0$ & $-2.4\times10^{4}$ & $1660$ & $1.8$ & $5.7$ \\
31 & $133,280$ & $5.9$ & $0$ & $-4.4\times10^{3}$ & $-7.4\times10^{4}$ & $-3.8\times10^{4}$ & $1944$ & $5.3$ & $4.4$ \\
32 & $196,591$ & $19.3$ & $0$ & $-1.9\times10^{5}$ & $-7.1\times10^{5}$ & $-6.6\times10^{5}$ & $6856$ & $3.7$ & $7.8$ \\
33 & $265,214$ & $3.2$ & $-1.4\times10^{4}$ & $0$ & $-9.2\times10^{4}$ & $-8.5\times10^{4}$ & $549$ & $1$ & $8.6$ \\
34 & $273,957$ & $16.8$ & $-5.4\times10^{5}$ & $0$ & $-4.6\times10^{4}$ & $-7.2\times10^{4}$ & $727$ & $1$ & $5.8$ \\
35 & $281,904$ & $16.4$ & $-1.2\times10^{6}$ & $0$ & $-2.8\times10^{5}$ & $-1.5\times10^{5}$ & $6655$ & $1$ & $4.3$ \\
36 & $317,080$ & $6.6$ & $-1.7\times10^{5}$ & $0$ & $-3.9\times10^{5}$ & $-4.2\times10^{5}$ & $8766$ & $1$ & $11$ \\
37 & $325,729$ & $9.2$ & $-5.8\times10^{5}$ & $0$ & $-1.1\times10^{6}$ & $-2.3\times10^{5}$ & $4293$ & $1$ & $5.8$ \\
38 & $325,729$ & $9.2$ & $-5.6\times10^{5}$ & $0$ & $-1.2\times10^{6}$ & $-2.5\times10^{5}$ & $3995$ & $1$ & $5.8$ \\
39 & $334,863$ & $5.5$ & $-3.3\times10^{5}$ & $0$ & $-3.6\times10^{5}$ & $-3.4\times10^{4}$ & $9118$ & $1$ & $11$ \\
40 & $372,787$ & $9.7$ & $-1.0\times10^{6}$ & $0$ & $-1.3\times10^{5}$ & $-1.4\times10^{5}$ & $965$ & $1$ & $11$ \\
41 & $463,347$ & $20.3$ & $-6.4\times10^{5}$ & $0$ & $-1.8\times10^{6}$ & $-1.5\times10^{6}$ & $9276$ & $1$ & $9.3$ \\
42 & $1,134,890$ & $5.3$ & \multicolumn{1}{c}{---} & $0$ & $-4.5\times10^{5}$ & $-4.9\times10^{5}$ & $264$ & $1$ & $13$ \\

\end{tabular}}

 \resizebox{\columnwidth}{!}
            {\smaller[2]
              \setlength{\tabcolsep}{2pt}
              \rowcolors{1}{gray!25}{white}
              \begin{tabular}{l l @{\hspace{8pt}}  l l @{\hspace{8pt}}  l l}\toprule

$1$ & Karate Club~\cite{zachary_information_1977}  & $22$ & Enron emails~\cite{leskovec_community_2008,klimt_introducing_2004}  \\
$2$ & Dolphins~\cite{lusseau_bottlenose_2003}  & $23$ & PGP~\cite{richters_trust_2011} (directed) \\
$3$ & Les Misérables~\cite{knuth_stanford_1993}  & $24$ & Internet AS (Caida)\footnote{Retrieved from \url{http://www.caida.org}.} (directed) \\
$4$ & Political Books\footnote{V. Krebs, retrieved from~\url{http://www-personal.umich.edu/~mejn/netdata/}}  & $25$ & Brightkite social network~\cite{cho_friendship_2011}  \\
$5$ & American football~\cite{girvan_community_2002, evans_american_2012}  & $26$ & netflix-pruned-smaller-u \\
$6$ & \textit{C. elegans} Neurons~\cite{watts_collective_1998} (directed) & $27$ & arXiv Co-Authors (hep-th)~\cite{leskovec_graph_2007}  \\
$7$ & Coauthorships in network science~\cite{newman_finding_2006}  & $28$ & Epinions.com trust network~\cite{richardson_trust_2003} (directed) \\
$8$ & Disease Genes~\cite{goh_human_2007}  & $29$ & arXiv Co-Authors (hep-ph)~\cite{leskovec_graph_2007}  \\
$9$ & Yeast protein interactions (CCSB-YI11)~\cite{yu_high-quality_2008}  & $30$ & arXiv Co-Authors (cond-mat)~\cite{leskovec_graph_2007}  \\
$10$ & Political Blogs~\cite{adamic_political_2005} (directed) & $31$ & arXiv Co-Authors (astro-ph)~\cite{leskovec_graph_2007}  \\
$11$ & Yeast protein interactions (LC)~\cite{reguly_comprehensive_2006}  & $32$ & Gowalla social network~\cite{cho_friendship_2011}  \\
$12$ & Yeast protein interactions (Combined AP/MS)~\cite{collins_toward_2007}  & $33$ & EU email~\cite{leskovec_graph_2007} (directed) \\
$13$ & \textit{E. coli} gene regulation~\cite{salgado_regulondb_2013} (directed) & $34$ & Flickr~\cite{mcauley_image_2012}  \\
$14$ & Yeast protein interactions (Y2H union)~\cite{yu_high-quality_2008}  & $35$ & Web graph of \url{stanford.edu}.~\cite{leskovec_community_2009} (directed) \\
$15$ & Facebook egos~\cite{mcauley_discovering_2014}  & $36$ & DBLP collaboration~\cite{yang_defining_2012}  \\
$16$ & Power Grid~\cite{watts_collective_1998}  & $37$ & Web graph of \url{nd.edu}.~\cite{leskovec_community_2009} (directed) \\
$17$ & Airport routes~\footnote{Retrieved from \url{http://openflights.org/}} (directed) & $38$ & WWW~\cite{albert_internet:_1999} (directed) \\
$18$ & Airport routes  & $39$ & Amazon product network~\cite{yang_defining_2012}  \\
$19$ & Wikipedia Votes~\cite{leskovec_signed_2010, leskovec_predicting_2010} (directed) & $40$ & IMDB film-actor\footnote{Retrieved from \url{http://www.imdb.com/interfaces}.}~\cite{peixoto_parsimonious_2013}  \\
$20$ & Human protein interactions (HPRD r9)~\cite{prasad_human_2009}  & $41$ & APS citations\footnote{Retrieved from \url{http://publish.aps.org/dataset}.} (directed) \\
$21$ & arXiv Co-Authors (gr-qc)~\cite{leskovec_graph_2007}  & $42$ & Youtube social network~\cite{yang_defining_2012} 
\end{tabular}}

\caption{Comparison of different models for many empirical networks. The
columns at the top table correspond to the dataset number (with the
name given at the bottom table), the number of nodes $N$, the average
degree $\avg{k}=2E/N$, the posterior odds ratios relative to the best
model for the degree-corrected overlapping ($\Lambda_{\text{DCO}}$), the
degree-corrected nonoverlapping ($\Lambda_{\text{DC}}$),
non-degree-corrected overlapping ($\Lambda_{\text{NDCO}}$) and
non-degree-corrected nonoverlapping ($\Lambda_{\text{NDC}}$)
models. Missing entries correspond to situations where the best
overlapping partition turns out to be nonoverlapping. The last three
columns show some parameters of the best model: The number of groups
$B$, the average mixture size $\avg{d}$, and the description length per
edge (in bits per edge).\label{tab:empirical}}
\end{table}

When performing model selection for larger networks, it is often the
case that the overlapping models are not chosen. In
table~\ref{tab:empirical} are shown the results for many empirical
networks belonging to different domains. For the majority of cases, the
nonoverlapping degree-corrected models are selected. The are, however,
many exceptions which include two social networks (Gowalla and
Brightkite~\cite{cho_friendship_2011}), the global airport network
of~\url{openflights.com}, the neuronal network of
\textit{C. elegans}~\cite{watts_collective_1998}, the political blog
network already mentioned, the arXiv co-authorship
networks~\cite{leskovec_graph_2007} [in the fields of general relativity
and quantum cosmology (gr-qc), high-energy physics (hep-th), condensed
matter (cond-mat), and astronomy (astro-ph)], co-authorship in network
science~\cite{newman_finding_2006}, and the network of genes implicated
in diseases~\cite{goh_human_2007}, for which some version of the
overlapping model is chosen. Interestingly, for the arXiv co-authorship
network in high-energy physics/phenomenology (hep-ph) a
nonoverlapping model is selected instead. For only one of the remaining
four arXiv networks (astro-ph), the degree-corrected version of the
overlapping model is selected, whereas for the other three the
non-degree-corrected version is preferred. Hence, for co-authorship
networks the model selection procedure seems to correspond to the
intuition that they are composed predominantly of overlapping
groups~\cite{palla_uncovering_2005}.

\begin{figure}
  \centering
  \begin{tabular}{cc}
    \includegraphics[width=.49\columnwidth]{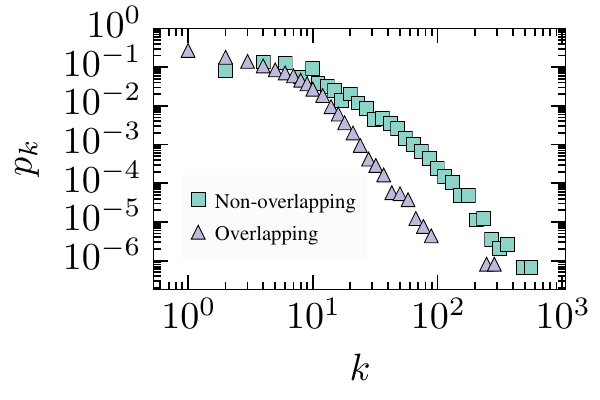} &
    \includegraphics[width=.49\columnwidth]{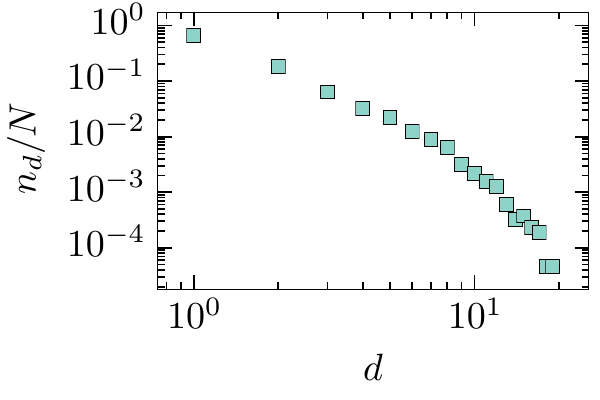} \\
    \includegraphics[width=.49\columnwidth]{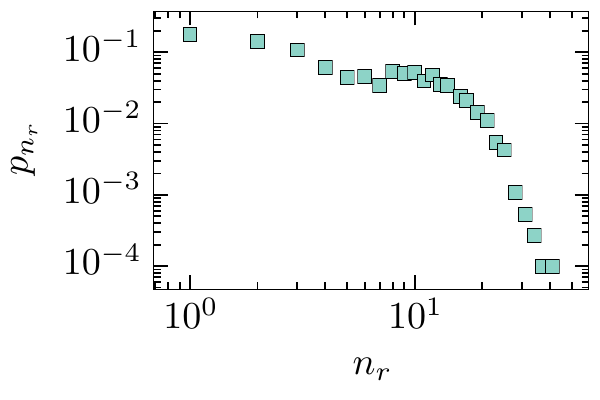} &
    \includegraphics[width=.49\columnwidth]{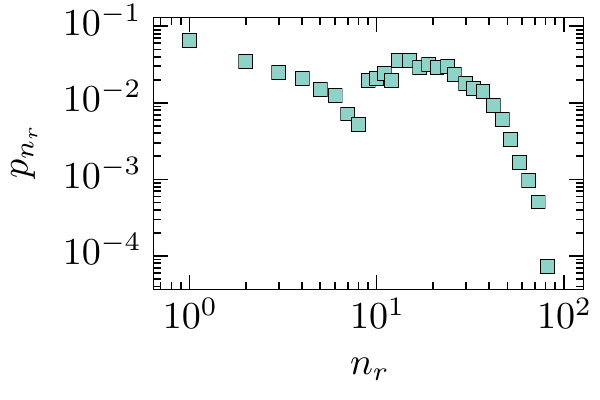}
  \end{tabular} \caption{Statistical properties of the best model
  inferred for the network of arXiv co-authors in the field of
  condensed matter (cond-mat). Top left: Degree distribution of the
  original network and of the overlapping model (where the labeled
  degree sequence $\{\vec{k}_i\}$ is flattened into a single histogram
  for all labeled degrees $\{k_i^r\}$). Top right: Distribution of
  mixture sizes, $n_d$. Bottom left: Distribution of group sizes for the
  best-fitting \emph{nonoverlapping}, non-degree-corrected
  model. Bottom right: Distribution of group sizes for the best-fitting
  \emph{overlapping}, non-degree-corrected model.\label{fig:cond-mat}}
\end{figure}

\begin{figure}
  \centering
  \begin{tabular}{cc}
    \includegraphics[width=.49\columnwidth]{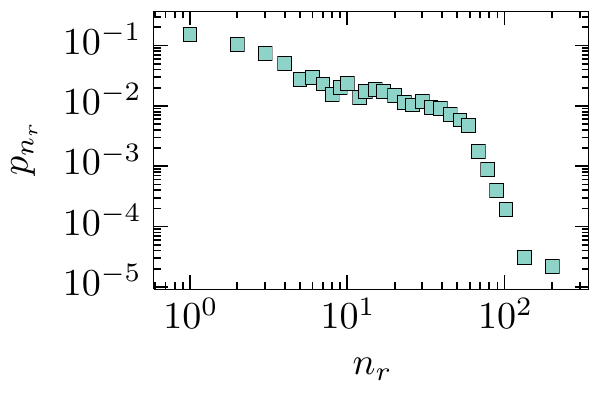} &
    \includegraphics[width=.49\columnwidth]{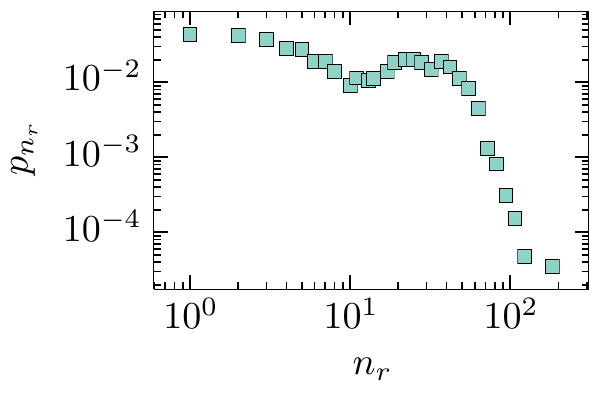} \\
    \includegraphics[width=.49\columnwidth]{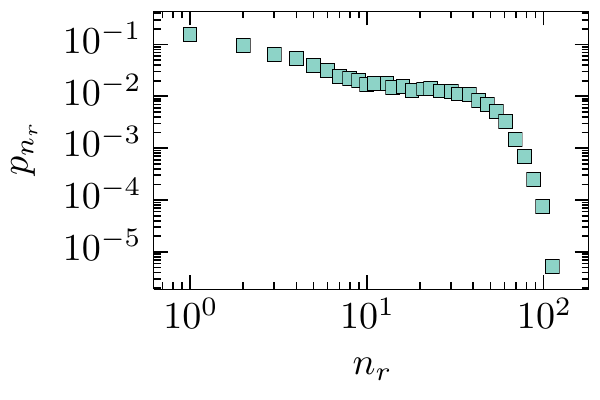} &
    \includegraphics[width=.49\columnwidth]{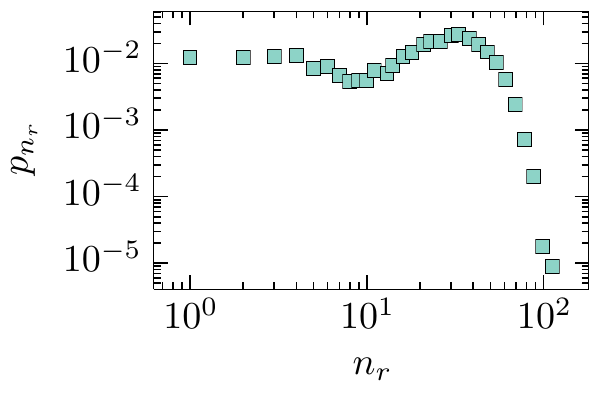}
  \end{tabular} \caption{Distribution of group sizes for the best
  fitting non-degree-corrected nonoverlapping model (left) and the
  degree-corrected nonoverlapping model (right), for the
  PGP~\cite{richters_trust_2011} (top) and DBLP
  collaboration~\cite{yang_defining_2012} (bottom) networks. In both
  cases the degree-corrected model provides a better fit, as shown in
  table~\ref{tab:empirical}.\label{fig:group-sizes}}
\end{figure}

We take the arXiv cond-mat network as a representative example of the
differences between the inferred models. As can be seen in
Fig.~\ref{fig:cond-mat}, although the degree distribution is very broad,
the inferred labeled degree distribution is narrower, meaning that many
large-degree nodes can be well explained as having a smaller degree of
any single type, but belonging simultaneously to many groups (in the
specific context of this network, prolific authors tend to be the ones
which belong to many different types of collaborations). The
distribution of mixture sizes $n_d$ has almost always a maximum at
$d=1$, meaning that most nodes belong to one group, but with a tail
which is comparatively broad (this seems to be a general feature which
is observed in the majority of networks analyzed). The distribution of
group sizes can be very different, depending on which model is
used. nonoverlapping models without degree correction tend to find
groups which are strongly correlated with
degrees~\cite{karrer_stochastic_2011}, and hence lead to a broad
distribution of group sizes when the degree distribution is also
broad. On the other hand, both degree correction and group overlap tend
to change the distribution considerably. In the literature there are
often claims of community sizes following power-law
distributions~\cite{newman_fast_2004,clauset_finding_2004,arenas_community_2004,lancichinetti_detecting_2009}
with figures similar to the lower left panel of
Fig.~\ref{fig:group-sizes}. Regardless to the validity of this hypothesis
for the various methods used in the literature, this is certainly not
the case for the overlapping model as shown in the lower right panel of
the same figure. Indeed, for most networks analyzed, the model which
best fits the data (which tends to be degree-corrected and
nonoverlapping) shows no vestige of group sizes following a scale-free
distribution. Some further examples of this are shown in
Fig.~\ref{fig:group-sizes}, where characteristic size scales can be
clearly identified.

\section{Model identifiability: Overlapping vs. nonoverlapping}\label{sec:ident}

A central issue when selecting between nonoverlapping and overlapping
models is to decide when a group of nodes should belong simultaneously
to two or more groups, of if these nodes should be better represented by
a single membership to a different unique group. The choice is not
always immediately obvious, since we can always generate very similar
networks with either model. If we generate a network with the
overlapping model, but treat it as if it were generated by the
nonoverlapping model, with each distinct mixture $\vec{b}$
corresponding to a separate nonoverlapping group, the associated
entropy will be
\begin{equation}
  \mathcal{S}'_t \simeq E - \frac{1}{2}\sum_{\vec{b}_1\vec{b}_2}e_{\vec{b}_1\vec{b}_2}\ln\left(\frac{e_{\vec{b}_1\vec{b}_2}}{n_{\vec{b}_1}n_{\vec{b}_2}}\right),
\end{equation}
where
\begin{equation}
  e_{\vec{b}_1\vec{b}_2} = \sum_{rs}b_1^rb_2^s\frac{e_{rs}}{n_rn_s}n_{\vec{b}_1}n_{\vec{b}_2}
\end{equation}
is the expected number of edges between mixtures $\vec{b}_1$ and
$\vec{b}_2$.  By exchanging the sums and using Jensen's inequality we
observe directly that
\begin{equation}
  \mathcal{S}'_t \le E - \frac{1}{2}\sum_{rs}e_{rs}\ln\left(\frac{e_{rs}}{n_rn_s}\right),
\end{equation}
with the right-hand side being the entropy of original overlapping model
$\mathcal{S}_t$, and with the equality holding only if the original
model happens to be nonoverlapping to begin with. Thus, the
nonoverlapping model will invariably possess a lower
entropy. Nevertheless, the overlapping hypothesis may still be preferred
if the number of groups $B$ is sufficiently smaller than the number of
individual $\vec{b}$ mixtures, so that the total description length is
shorter. It should be observed, however, that since one model is
contained inside the other, the difference in the description length can
be interpreted simply as the difference in the prior probabilities for
the model parameters. As the amount of available data increases, the
effect of the priors should ``wash out'', and the description length
should be increasingly dominated by the model entropy alone. In these
cases one should expect the nonoverlapping model to be preferred,
regardless of the specific model which was used to generate the
data. However, differently from models that generate independent data
points, the ``amount of available data'' for network models is a finer
issue. In the case of the stochastic block model it involves the
simultaneous scaling of the number of edges $E$, the number of nodes $N$
and the number of groups $B$.

\begin{figure}
  \centering
  \includegraphics[width=\columnwidth]{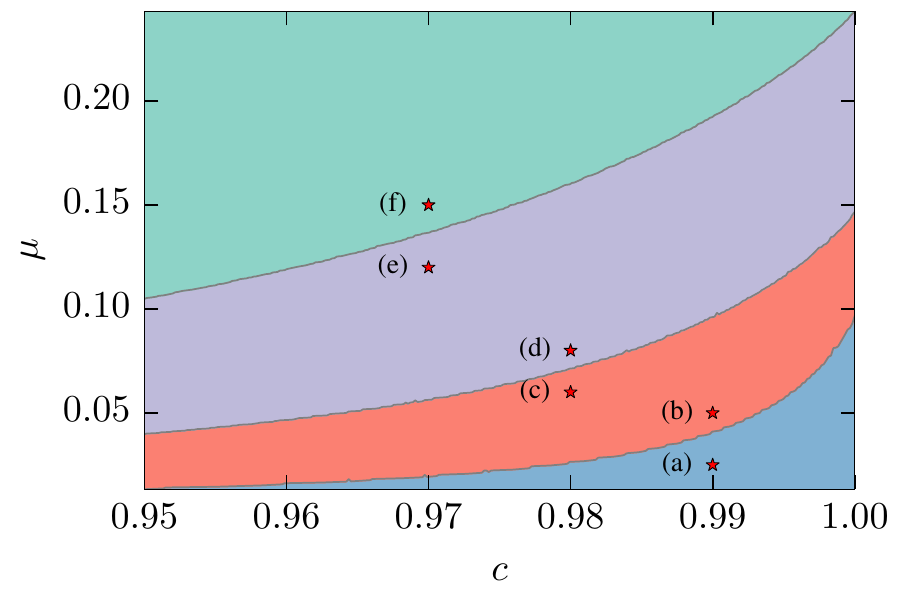}
  \begin{tabular}{ccc}
    \begin{minipage}{.32\columnwidth}
      \centering\smaller
      \includegraphics[width=\columnwidth,trim=0 0cm 0 0cm, clip]{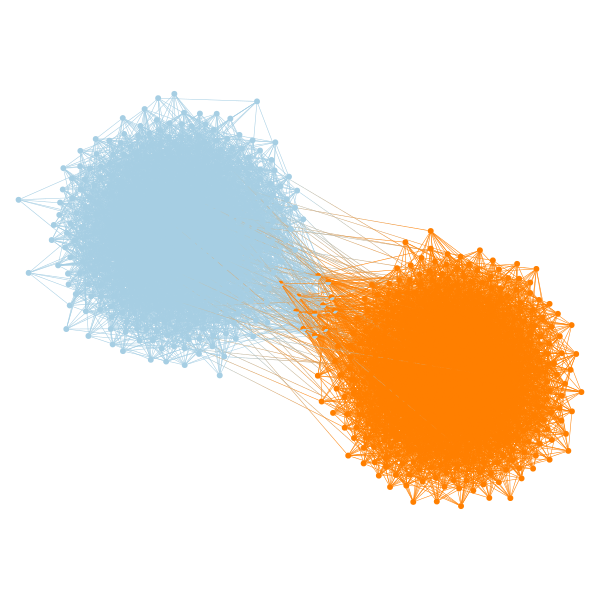}\\
      (a) $B=2$, $c=0.99$, $\mu=0.025$
    \end{minipage} &
    \begin{minipage}{.32\columnwidth}
      \centering\smaller
      \includegraphics[width=\columnwidth,trim=0 0cm 0 0cm, clip]{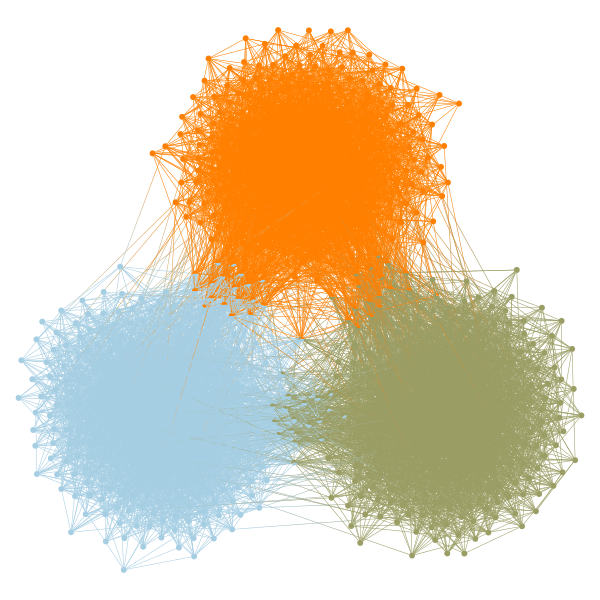}\\
      (c) $B=3$, $c=0.98$, $\mu=0.06$
    \end{minipage} &
    \begin{minipage}{.32\columnwidth}
      \centering\smaller
      \includegraphics[width=\columnwidth,trim=0 0cm 0 0cm, clip]{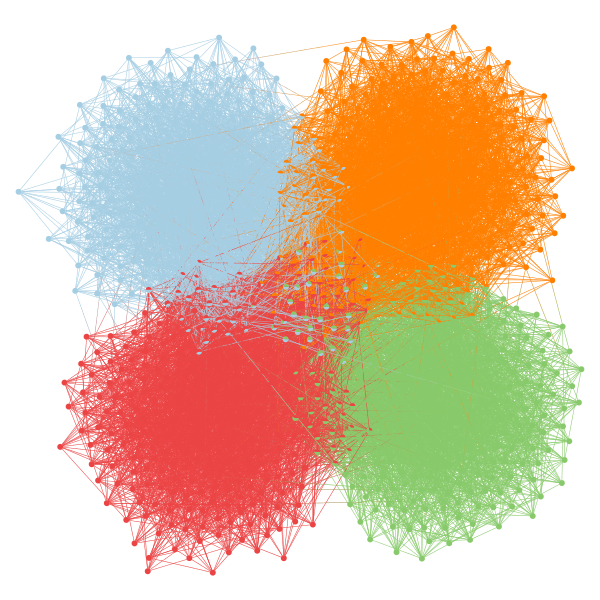}\\
      (e) $B=4$, $c=0.97$, $\mu=0.12$
    \end{minipage}\\
    \begin{minipage}{.32\columnwidth}
      \centering\smaller
      \includegraphics[width=\columnwidth,trim=0 0cm 0 0cm, clip]{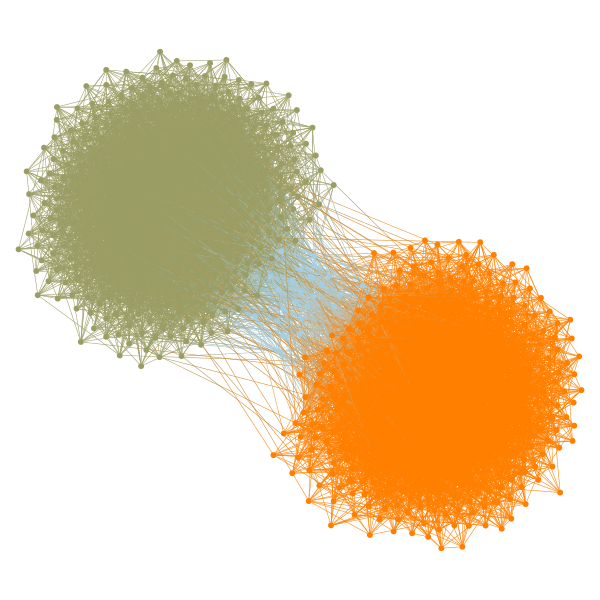}\\
      (b) $B=3$, $c=0.99$, $\mu=0.05$
    \end{minipage} &
    \begin{minipage}{.32\columnwidth}
      \centering\smaller
      \includegraphics[width=\columnwidth,trim=0 0cm 0 0cm, clip]{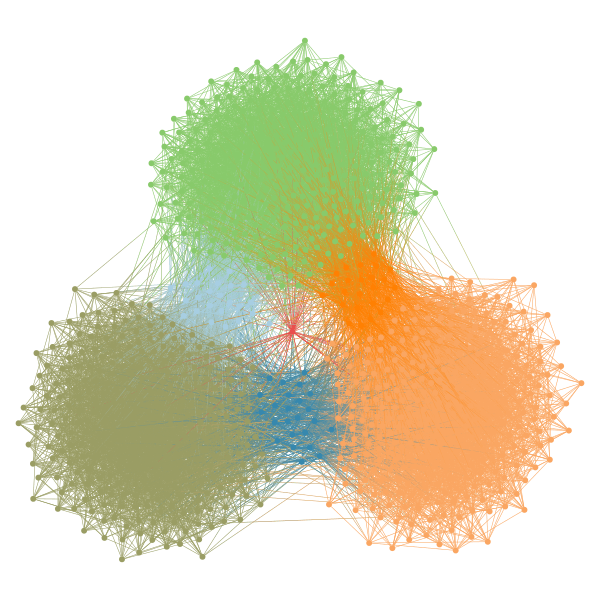}\\
      (d) $B=7$, $c=0.98$, $\mu=0.08$
    \end{minipage} &
    \begin{minipage}{.32\columnwidth}
      \centering\smaller
      \includegraphics[width=\columnwidth,trim=0 0cm 0 0cm, clip]{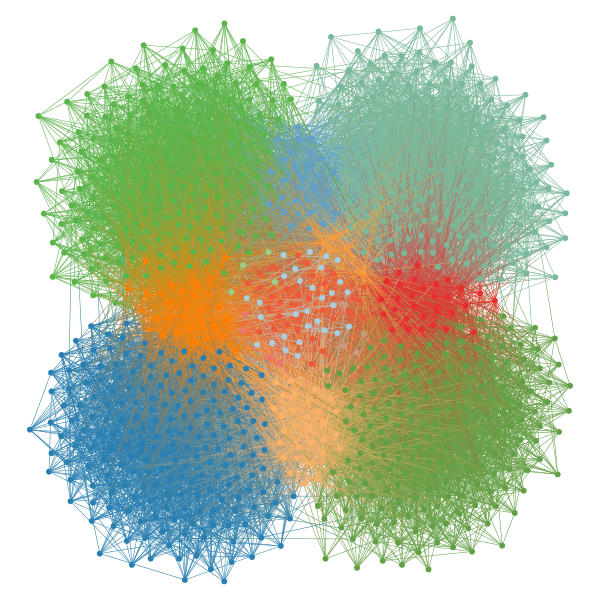}\\
      (f) $B=15$, $c=0.97$, $\mu=0.15$
    \end{minipage}
  \end{tabular} \caption{Top: Parameter regions for the model considered
  in the main text, with $N=10^3$ and $\avg{k} = 2E/N = 20$. Each curve
  corresponds to one value of $B$ and separates a region above where the
  nonoverlapping model is preferred from a region below where the
  overlapping model is chosen. Bottom: Networks and their preferred
  partitions, corresponding to parameter values indicated in the top
  panel. \label{fig:overlap-vs-nonoverlap}}
\end{figure}

As a case example, here we consider a simple overlapping assortative
model, with $e_{rs}=2E[\delta_{rs}c / B + (1-\delta_{rs})(1-c) /
B(B-1)]$, with $c\in[0,1]$ controlling the degree of assortativity. The
mixtures are parameterized as $n_{\vec{b}} = C \prod_r\mu^{b_r}$, with
$C$ being a normalization constant, and $\mu \in [0,1]$ controlling the
degree of overlap. For $\mu \to 0$ we obtain asymptotically a
nonoverlapping partition with $n_r = N/B$, and for $\mu=1$ all mixtures
$\vec{b}$ have the same size. We compare the difference in description
length between this model and its equivalent parametrization with each
mixture as a separate group. As can be seen in
Fig.~\ref{fig:overlap-vs-nonoverlap}, for any given value of $c$, there
is a value of $\mu$ above which the nonoverlapping model is
preferred. In this parameter region, the group intersections are
sufficiently well populated with nodes, so that their representation as
individual groups is chosen. For values of $\mu$ below this value, the
intersections are significantly smaller than the nonoverlapping
portion. In this case, the data are better explained as larger groups of
almost nonoverlapping nodes, with few nodes at the intersections. The
boundary separating the two regions recedes upwards as the number of
groups $B$ is increased, meaning that a larger number of distinct
intersections can compensate for a smaller number of nonoverlapping
nodes. It should also be pointed out that the boundaries move downwards
as the number of nodes and edges is increased, such that the average
degree in the network remains the same (not shown), so it is not only
the relative sizes of the intersections that are the relevant
properties, but also their absolute sizes. The same occurs if the
average degree increases and everything else remains constant. Hence, in
the limit of sufficient data, either with the number of nodes inside
each group and intersections becoming sufficiently large, or with each
part becoming sufficiently dense, the nonoverlapping model is the one
which will be selected. For empirical networks, this may not be the most
representative scaling scenario, since the most appropriate number of
groups and degree of overlap may in fact follow any arbitrary scaling,
and hence the overlapping model may still be selected, even for very
large or very dense networks. Nevertheless, this example seems to
suggest that the nonoverlapping model is general enough to accommodate
structures generated by the overlapping model in these limiting cases,
and may serve as a partial explanation to why the overlapping model is
seldom selected in the empirical systems analyzed in
Sec.~\ref{sec:empirical}.

\section{Inference algorithm}\label{sec:algorithm}

The inference procedure consists in finding the labeling of the
half-edges of the graph such that the description length is
minimized. Such global optimization problems are often NP-hard, and
require heuristics to be solvable in practical time. One possibility is
to use the Markov chain Monte Carlo (MCMC) method, which consists in
modifying the block membership of each half-edge in a random fashion,
and accepting or rejecting each move with a probability given as a
function of the description length difference $\Delta \Sigma$. By
choosing the acceptance probabilities in the appropriate manner, i.e. by
enforcing ergodicity and detailed balance, one can guarantee that the
labelings will be sampled with the correct probability after a
sufficiently long equilibration time is reached. However, naive
formulations of the Markov chain will lead to very long equilibration
times, which become unpractical for large networks. Here we adapt the
algorithm developed in Ref.~\cite{peixoto_efficient_2014} for the
nonoverlapping case which implements a fast Markov chain. It consists in
the move proposal of each half-edge incident on node $i$ of type $r$ to
type $s$ with a probability given by
\begin{equation}\label{eq:move}
  p(r\to s|t) = \frac{e_{ts} + \epsilon}{e_t + \epsilon B},
\end{equation}
where $t$ is the block labeling the half-edge opposing a randomly chosen
half-edge incident to the same node as the half-edge being moved, and
$\epsilon \ge 0$ is a free parameter. Eq.~\ref{eq:move} means that we
attempt to guess the label of a given half-edge by inspecting the group
membership the neighbors of the node to which it belongs, and using the
currently inferred model parameters to choose the most likely group to
which it should be moved. It should be emphasized that this move
proposal does not result in a preference for either assortative or
dissortative networks, since it depends only on the matrix $\{e_{rs}\}$
currently inferred. For any choice of $\epsilon > 0$, this move proposal
preserves ergodicity, but not detailed balance. This last characteristic
can be enforced via the Metropolis-Hastings
criterion~\cite{metropolis_equation_1953,hastings_monte_1970} by
accepting each move with a probability $a$ given by
\begin{equation}\label{eq:a}
a = \min\left\{e^{-\beta\Delta \Sigma} \frac{\sum_tp_t^ip(s\to r|t)}{\sum_tp^i_tp(r \to s|t)}, 1\right\},
\end{equation}
where $p^i_t$ is the fraction of opposing half-edges of node $i$ which
belong to block $t$, and $p(s\to r|t)$ is computed after the proposed
$r\to s$ move (i.e. with the new values of $e_{tr}$), whereas $p(r\to
s|t)$ is computed before. The parameter $\beta$ in Eq.~\ref{eq:a} is an
inverse temperature, which can be used to sample partitions according to
their description length ($\beta=1$) or to find the ground state
($\beta\to\infty$). As explained in Ref.~\cite{peixoto_efficient_2014},
this move proposal as well as the computation of $a$ can be done
efficiently, with minimal book-keeping, so that a sweep of the network
(where each half-edge move is attempted once) is done in time $O(E)$,
independent of the number of groups $B$. This is true even in the
overlapping case, since updating
Eqs.~\ref{eq:st},~\ref{eq:sd},~\ref{eq:lp} and~\ref{eq:lkappa} after
each half-edge move can be done in time $O(1)$.

\begin{figure}
  \centering
  \begin{tabular}{cc}
    \begin{minipage}[c]{.49\columnwidth}\centering
      \includegraphics[width=\columnwidth]{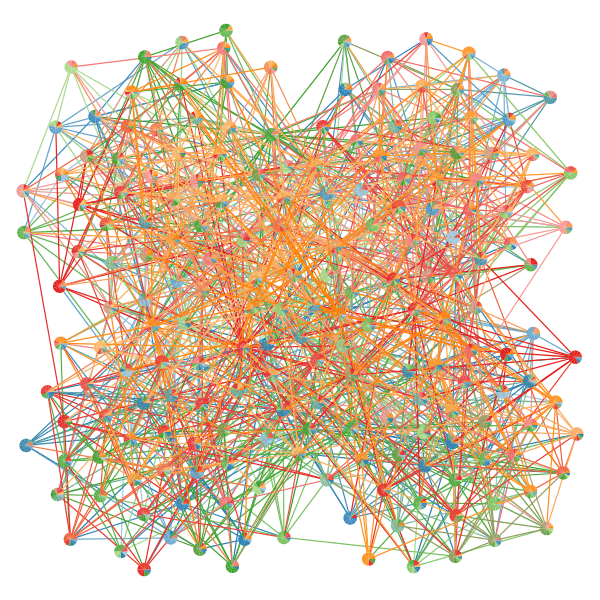}\\
      $B=2E$
    \end{minipage} &
    \begin{minipage}[c]{.49\columnwidth}\centering
      \includegraphics[width=\columnwidth]{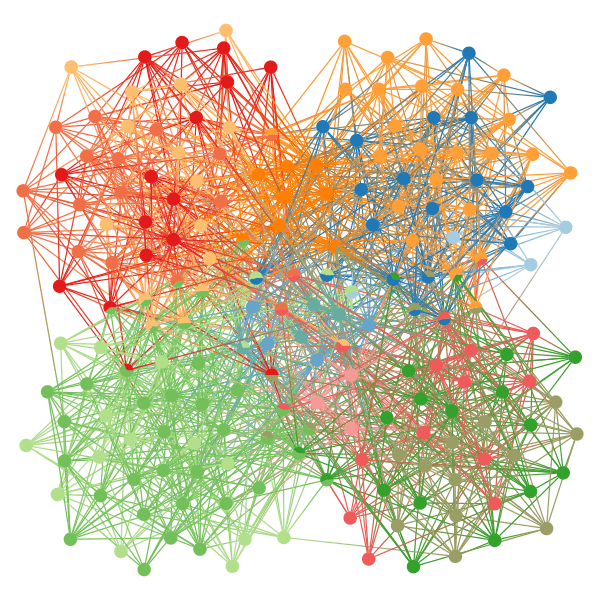}\\
      $B=15$
    \end{minipage} \\
    \begin{minipage}[c]{.49\columnwidth}\centering
      \includegraphics[width=\columnwidth]{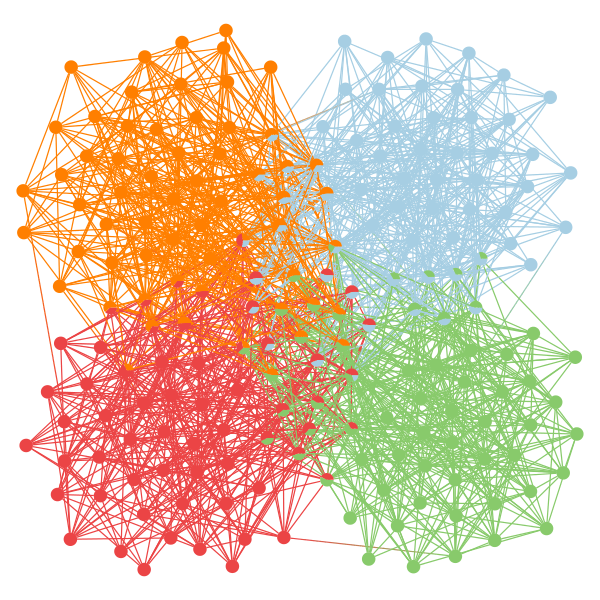}\\
      $B=4$
    \end{minipage} &
    \begin{minipage}[c]{.49\columnwidth}\centering
      \includegraphics[width=\columnwidth]{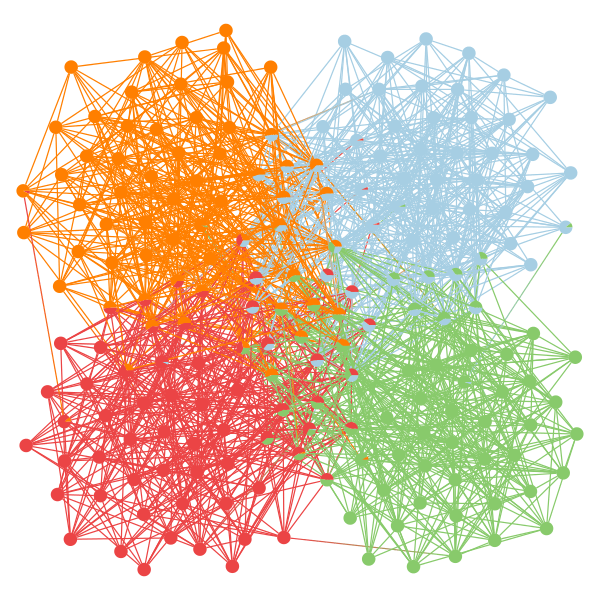}\\
      $B=4$, planted
    \end{minipage}
  \end{tabular} \caption{\label{fig:merging}Typical outcome of the
  greedy multilevel agglomerative algorithm described in the text, for a
  network sampled from the overlapping model with $B=4$. The different
  panels show the progression of the algorithm from $B=2E$ to $B=4$. The
  panel on the lower right shows the planted partition used to generate
  the network.}
\end{figure}

As discussed in Ref.~\cite{peixoto_efficient_2014}, although the MCMC
method above succeeds in equilibrating faster than a naive Markov chain,
it still suffers from a strong dependence on how close one starts from
the global minimum. Usually, starting from a random partition of the
half-edges leads to metastable states where the Markov chain seems to
have equilibrated, but in fact the network structure has only been
partially discovered, and will move from such configurations only after
a very long time. This problem is common to many inference procedures
based on local moves such as expectation
maximization~\cite{ball_efficient_2011} and belief
propagation~\cite{decelle_inference_2011,decelle_asymptotic_2011}. In
Ref.~\cite{peixoto_efficient_2014} a multilevel agglomerative heuristic
was proposed, which significantly alleviates this problem. It consists
in equilibrating the chain for a larger number of groups, and then
merging the groups using the same algorithm used for the block
membership moves. This method, however, cannot be used unmodified in the
overlapping case, since the strict merging of groups will not properly
explore the landscape of possible overlapping partitions. We therefore
modify the approach as follows. Before groups are merged, the half-edges
belonging to each one of them are split into subgroups corresponding to
the different group memberships at the opposing sides. These subgroups
are then treated as separate groups, and are merged together until the
desired number of groups is achieved. All the details of the algorithm
beyond this modification are performed exactly as described in
Ref.~\cite{peixoto_efficient_2014}. Since this algorithm usually does a
good job in finding a partition very close to the final one, it also
tends to perform very well when the algorithm is turned into a greedy
heuristic, by starting with $B=2E$ and each half-edge in its own group,
and by making $\beta\to\infty$. An example of a typical outcome of the
greedy algorithm is shown in Fig.~\ref{fig:merging}. The greedy version
is very fast, with an overall complexity of $O(E\ln^2E)$, which makes it
usable for very large networks. Note that this complexity is independent
on the number of groups, $B$. This is a strong contrast to other methods
proposed for the same problem, such as the stochastic optimization
algorithm of Gopalan et al~\cite{gopalan_efficient_2013}, and the
expectation maximization algorithm of Ball et
al~\cite{ball_efficient_2011}, both of which have a complexity of
$O(EB)$ per sweep, although they only consider strictly assortative
models, and applying the same techniques to the more general models
considered here would lead to an $O(EB^2)$ complexity, similar to belief
propagation algorithms for nonoverlapping
models~\cite{decelle_asymptotic_2011,yan_model_2014}. Although these
approaches can be very efficient if the number of groups is very small,
they quickly become prohibitive if the most appropriate number of groups
scales as some function of the system size (which seems to be generally
the case when model selection is applied, see table~\ref{tab:empirical}
and Ref.~\cite{peixoto_hierarchical_2014}), which is not an issue with
the algorithm described above. It should also be noted that none of the
other algorithms mentioned~\cite{gopalan_efficient_2013,
ball_efficient_2011,decelle_asymptotic_2011,yan_model_2014} is designed
to overcome metastable solutions, like the multilevel approach presented
here.

For most networks analyzed in this work, the fast heuristic version of
the algorithm was used, together with the algorithm described in
Ref.~\cite{peixoto_hierarchical_2014} to infer the upper layers of the
hierarchy (which includes the determination of the number of groups $B$
at the lowest level, in addition to the entire hierarchy, in a
nonparametric fashion)\footnote{A complete implementation of the
algorithm is freely available as part of the {\texttt{graph-tool}}
library~\cite{peixoto_graph-tool_2014}
at~\url{http://graph-tool.skewed.de}.}.

\section{Conclusion}\label{sec:conclusion}

We presented a method of inferring overlapping and degree-corrected
versions of the stochastic block model based on the minimum description
length principle (MDL) that avoids overfitting and allows for the
comparison between model classes. Based on a Bayesian interpretation of
MDL, we derived a posterior odds ratio test that yields a degree of
confidence with which models can be selected or discarded. In applying
this method to a variety of empirical networks, we obtained that for the
majority of them the nonoverlapping degree-corrected model variant is
the one that best fits the data.

The relative success of the degree-corrected model implies that
intrinsic node propensities are an important aspect of the network
formation of many systems, which are not sufficiently well described by
the sole division into node classes. We note, however, that there are
exceptions to this, as there are a few networks that do not show enough
statistical evidence to justify the additional parameters of the
degree-corrected model. In these networks, the groups themselves seem to
be the leading descriptors of the network structure, with the degree
sequence itself providing little additional explanatory power.

Although overlapping structures are often considered to be more
intuitive explanations for some networks, we showed that in many
representative cases the nonoverlapping model can accommodate the same
structure while providing a more parsimonious description of the data.
This contradicts results obtained with nonstatistical
methods~\cite{palla_uncovering_2005, ahn_link_2010}, which claimed that
many or even most networks are better described by overlapping
groups. We believe that this conclusion is most likely a result of
overfitting: Since there are more overlapping structures than
nonoverlapping ones, it is easier to find them in the data. We expect
this fact to bear on tasks that require high-quality fits, such as the
prediction of missing or spurious links~\cite{clauset_hierarchical_2008,
guimera_missing_2009}, or other generalizations of the data.

The models considered in this work generate unlabeled networks, without
any other properties associated with the nodes or edges. However, it is
often the case that either the nodes or edges have
weights~\cite{aicher_learning_2014, viamontes_esquivel_compression_2011,
rosvall_memory_2014} or are of different
types~\cite{mariadassou_uncovering_2010,kivela_multilayer_2014}, or have
temporal information~\cite{fu_dynamic_2009}. This sort of additional
data may corroborate the evidence supporting the generation via a
specific type of model (e.g. with overlaps) and tip the scale towards
it. Therefore, the results presented in this paper should not be
interpreted as a statement on the suitability the abstract notion of
overlapping structures in general, only of the specific formulations
considered. However, the approach presented here is generalizable to
these other cases as well, by augmenting the model to generate
covariates associated with the edges and
nodes~\cite{aicher_learning_2014}. Furthermore, one should be able to
perform a similar comparison with models which belong to very different
classes, such as latent space~\cite{hoff_latent_2002} models, or others.

\appendix
\section{Directed graphs}

The same approach of the main text can be carried over to directed
graphs with no difficulties. In this case the edge counts are in general
asymmetric, $e_{rs}\neq e_{sr}$, which leads to the entropy for the
non-degree-corrected model~\cite{peixoto_entropy_2012}
\begin{equation}
  \mathcal{S}_t \simeq E - \sum_{rs}e_{rs}\ln\left(\frac{e_{rs}}{n_rn_s}\right).
\end{equation}
For the degree-corrected case, there are two degree sequences for the
labeled out- and in-degrees, $\{{k^+}_i^r\}$ and $\{{k^-}_i^r\}$,
respectively. Applying the same argument as for the undirected case, the
entropy becomes~\cite{peixoto_entropy_2012}
\begin{equation}
  \mathcal{S}_d \simeq -E - \sum_{rs}e_{rs}\ln\left(\frac{e_{rs}}{e^+_re^-_s}\right) - \sum_{ir}\ln {k^+}^r_i! - \sum_{ir}\ln {k^-}^r_i!,
\end{equation}
where $e^+_r = \sum_se_{rs}$ and  $e^-_r = \sum_se_{sr}$.

The description length for the overlapping partition is identical to the
undirected case, with $\mathcal{L}_p$ given by Eq.~\ref{eq:lp}. For the
labeled degree sequence, we have instead
\begin{equation}
  \mathcal{L}_{\kappa} = \sum_r\ln{\textstyle\multiset{m_r}{e^+_r}} + \ln{\textstyle\multiset{m_r}{e^-_r}} + \sum_{\vec{b}}\min\left(\mathcal{L}^{(1)}_{\vec{b}}, \mathcal{L}^{(2)}_{\vec{b}}\right).
\end{equation}
with
\begin{equation}
  \mathcal{L}^{(1)}_{\vec{b}} = \sum_r\ln{\multiset{n_{\vec{b}}}{{e^+}^r_{\vec{b}}}} + \ln{\multiset{n_{\vec{b}}}{{e^-}^r_{\vec{b}}}}.
\end{equation}
and
\begin{equation}
  \mathcal{L}^{(2)}_{\vec{b}}  = \sum_rb_r\left(\ln{\Xi_{\vec{b}}^r}^+ + \ln{\Xi_{\vec{b}}^r}^-\right) + \ln n_{\vec{b}}! - \sum_{\vec{k}} \ln n^{\vec{b}}_{\vec{k}^+,\vec{k}^-}!,
\end{equation}
where $\ln{\Xi_{\vec{b}}^r}^+$ and $\ln{\Xi_{\vec{b}}^r}^-$ are computed
as in Eq.~\ref{eq:deg_xi_approx} but using ${e^+}^r_{\vec{b}} =
\sum_{\vec{k}^+,\vec{k}^-}k^+_rn_{\vec{k}^+,\vec{k}^-}^{\vec{b}}$ and
${e^r_{\vec{b}}}^- =
\sum_{\vec{k}^+,\vec{k}^-}k^-_rn_{\vec{k}^+,\vec{k}^-}^{\vec{b}}$,
respectively, which give the total number of out- and in-edges incident
on the mixture $\vec{b}$. In the previous equations the counts
$n_{\vec{k}^+,\vec{k}^-}^{\vec{b}}$ refer to the joint distribution of
labeled in- and out-degrees, so that each vector $\vec{k}^{+/-}$
describes the in- and out-degrees labeled according to degree
membership, i.e. $\vec{k}_i^+ = \{{k^+}_i^r\}$ and $\vec{k}_i^- =
\{{k^-}_i^r\}$.

\section{Poisson Models}

\subsection{Non-degree-corrected} \label{app:poisson-trad}

This approximation of the formulation with ``hard'' constraints of the
multiple membership model discussed in the main text is closely related
to a Poisson variant of the model with ``soft'' constraints, where each
half-edge of the graph is labeled with a latent variable specifying
which group memberships were responsible for its existence, and the
number of edges of type $(r,s)$ between nodes $i$ and $j$,
$A_{ij}^{rs}$, is independently sampled according to a Poisson
distribution (similar to
Refs.~\cite{karrer_stochastic_2011,ball_efficient_2011}), so the
likelihood becomes
\begin{equation}
  P(G|\{\vec{b}_i\},\{p_{rs}\}) = \prod_{i>j}\prod_{r\ge
  s}p_{rs}^{A^{rs}_{ij}}e^{-p_{rs}b_i^rb_j^s}/A^{rs}_{ij}!,
\end{equation}
where $p_{rs}$ is the average number of edges of type $(r,s)$ between
nodes that belong to each group. The log-likelihood can be written as
\begin{equation}
  \ln P = \frac{1}{2}\sum_{rs} e_{rs}\ln p_{rs} - n_rn_sp_{rs} - \sum_{i>j}\sum_{r\ge s} \ln A^{rs}_{ij}!.
\end{equation}
Maximizing $\ln P$ w.r.t. $p_{rs}$, we obtain $\hat{p}_{rs} =
e_{rs}/n_rn_s$, and hence
\begin{equation}
  \ln \hat{P} = -E + \frac{1}{2}\sum_{rs}e_{rs}\ln\left(\frac{e_{rs}}{n_rn_s}\right) - \sum_{i>j}\sum_{r\ge s} \ln A^{rs}_{ij}!.
\end{equation}
For simple graphs with $A^{rs}_{ij} \in \{0, 1\}$, the last term in the
above equation is equal to zero, and we have that the approximation of the
likelihood of the model with ``hard'' constraints in the sparse case is
identical to the exact maximum likelihood of the Poisson model with
``soft'' constraints.

This model is similar to the popular mixed membership stochastic block
model (MMSBM)~\cite{airoldi_mixed_2008}; however it differs in the
important aspect that it generates strictly denser overlaps. In the
MMSBM, the existence of an edge $A_{ij}$ is sampled from a Bernoulli
distribution with parameter $\lambda_{ij} =
\sum_{rs}\theta_i^r\theta_j^sp_{rs}$, where $\theta_i^r$ is the
probability that node $i$ belongs to group $r$, such that
$\sum_r\theta_i^r = 1$, and $p_{rs}\in[0,1]$ is the probability that two
nodes belonging to groups $r$ and $s$ are connected. Although for sparse
graphs the differences between Poisson and Bernoulli models tend to
disappear, with this parametrization the density of the overlaps is
mixed with normalized weights. More specifically, for a node $i$ which
belongs simultaneously to groups $r$ and $s$, its expected degree is
equal to the weighted average of the unmixed degrees, $\avg{k}_i =
\theta_i^r\avg{k}_r + \theta_i^s\avg{k}_s$, where $\avg{k}_r =
\sum_sp_{rs}\sum_i\theta_i^s$ is the expected degree of a node that
belongs only to group $r$.  Thus, in the MMSBM the nodes in the mixture
have an intermediate density between the sparser and the denser groups.
In contrast, in the model considered in the main text, as well as the
Poisson model above, we have simply $\avg{k}_i = \avg{k}_r + \avg{k}_s$,
and therefore the overlaps are always strictly denser than the pure
groups. In this respect, it is equivalent to other formulations of the
MMSBM, see e.g. Refs.~\cite{parkkinen_block_2009,
yang_community-affiliation_2012}.

\subsection{Degree-corrected} \label{app:poisson-deg-corr}

A connection to a version of the model with ``soft'' constraints can
also be made. We may consider each labeled entry $A_{ij}^{rs}$ in the
adjacency matrix to be Poisson distributed with an average given by
$\theta_i^r\theta_j^s\lambda_{rs}$,
\begin{equation}\label{eq:deg_poi}
  P(G|\{\vec{b}_i\},\{\lambda_{rs}\},\{\theta_r\}) =
  \prod_{i>j}\prod_{r\ge s}(\theta_i^r\theta_j^s\lambda_{rs})^{A^{rs}_{ij}}e^{-\theta_i^r\theta_j^s\lambda_{rs}}/A^{rs}_{ij}!,
\end{equation}
where $A^{rs}_{ij}$ is the number of edges of type $(r,s)$ between nodes
$i$ and $j$, and $\theta_i^r$ is the propensity with which a node
receives an edge of type $r$. The log-likelihood can be written as
\begin{multline}
  \ln P = \frac{1}{2}\sum_{rs} e_{rs}\ln \lambda_{rs} + \sum_{ir}k_i^r\ln\theta_i^r - \sum_{r\ge s}\lambda_{rs}\sum_{i>j}\theta_i^r\theta_j^s \\ 
   - \sum_{i>j}\sum_{r\ge s} \ln A^{rs}_{ij}!.
\end{multline}
Maximizing $\ln P$ w.r.t. $\{\lambda_{rs}\}$ and $\{\theta^r_i\}$, we
obtain $\hat{\lambda}_{rs} = e_{rs}/e_re_s$ and $\hat{\theta}^r_i =
k^r_i$, and hence
\begin{multline}
  \ln \hat{P}= -E + \frac{1}{2}\sum_{rs}e_{rs}\ln\left(\frac{e_{rs}}{e_re_s}\right) + \sum_{ir}k_i^r\ln k_i^r \\
  - \sum_{i>j}\sum_{r\ge s} \ln A^{rs}_{ij}!.
\end{multline}
Again, for simple graphs with $A^{rs}_{ij} \in \{0, 1\}$, the last term
in the above equation is equal to zero; however even in that case the
likelihood is not identical to the version with ``hard'' constraints
considered above, as is the case for the single membership version as
well~\cite{peixoto_entropy_2012}. Both likelihoods only become the same
in the limit $k_i^r\gg 1$ such that $\ln k_i^r! \simeq k_i^r\ln k_i^r -
k_i^r$. Nevertheless, for the purpose of this paper, the differences
between these models can be overlooked.

There is a direct connection between this model and the one proposed by
Ball et al~\cite{ball_efficient_2011}. In the not strictly assortative
version of their model, the number of edges $A_{ij}$ is distributed
according to a Poisson with average
$\lambda_{ij}=\sum_{rs}\eta^r_i\eta^s_j\omega_{rs}$, where $\eta_i^r$ is
the propensity with which node $i$ receives edges of type $r$ and
$\omega_{rs}$ regulates the number of edges across groups. The total
likelihood of that model is
\begin{equation}\label{eq:bkn}
  P(G|\{\vec{b}_i\},\{\omega_{rs}\},\{\eta_r\}) = \prod_{i>j}\lambda_{ij}^{A_{ij}}e^{-\lambda_{ij}}/A_{ij}!.
\end{equation}
Since the sum of independent Poisson random variables is also
distributed according to a Poisson, if we generate a graph with the
model of Eq.~\ref{eq:deg_poi} and observe only the total unlabeled edge
counts $A_{ij}=\sum_{rs}A_{ij}^{rs}$, they are distributed exactly like
Eq.~\ref{eq:bkn}, for the same choice of parameters
$\theta_i^r=\eta_i^r$ and $\lambda_{rs}=\omega_{rs}$. Hence, the model
of the main text is an equivalent formulation of the one in
Ref.~\cite{ball_efficient_2011} where one keeps track of the latent
variables specifying the exact type of each half-edge, instead of their
marginal probability. This has the advantage that the maximum likelihood
estimates for the model parameters $\lambda_{rs}$ and $\theta_i^r$ can
be obtained directly by differentiation, and do not require iterations
of an EM algorithm as in Ref.~\cite{ball_efficient_2011}. On the other
hand we are left with the determination of labels in the half-edges,
which is done with the method already described in
Sec.~\ref{sec:algorithm}.

\section{Maximum-entropy ensemble of counts with constrained average}\label{sec:maxent}

Suppose we want to compute the number of all possible non-negative
integer counts $\{n_k\}$, subject to a normalization constraint
$\sum_{k=0}^{\infty}n_k = N$ and a fixed average
$\sum_{k=0}^{\infty}kn_k = E$. This can be obtained approximately, by
relaxing the constraints so that they hold only on average. The maximum
entropy ensemble given these constraints is the one with the
probabilities $P(\{n_k\}) = e^{-H(\{n_k\})} / Z$, with $H(\{n_k\}) =
\lambda \sum_kn_k
+ \mu \sum_kkn_k$, where $\lambda$ and $\mu$ are the Lagrange
multipliers that keep the constraints in place. This ensemble is
mathematically analogous to a simple Bose gas with energy levels given
by $k$. The partition function is given by
\begin{equation}
  Z =\sum_{\{n_k\}}e^{-\lambda\sum_kn_k - \mu \sum_kkn_k}=\prod_kZ_k,
\end{equation}
with
\begin{equation}
  Z_k = \left[1-e^{-\lambda - \mu k}\right]^{-1}.
\end{equation}
The average counts are given by $\left<n_k\right>=-\partial \ln Z_k /
\partial\lambda = \left[\exp(\lambda + \mu k) - 1\right]^{-1}$, and the
parameters $\lambda$ and $\mu$ are determined via the imposed constraints,
\begin{align}
  \sum_{k=0}^{\infty}\left[\exp(\lambda + \mu k) - 1\right]^{-1} = N, \\
  \sum_{k=0}^{\infty}k\left[\exp(\lambda + \mu k) - 1\right]^{-1} = E.
\end{align}
Further analytical progress can be made by replacing the sums with
integrals, and using the polylogarithm function and its connection with
the Bose–Einstein distribution, $\operatorname{Li}_s(z) = \Gamma(s)^{-1}
\int_0^\infty {t^{s-1} \over e^t/z-1}\mathrm{d}t$,
\begin{align}
  \int_0^\infty dk \left[\exp(\lambda + \mu k) - 1\right]^{-1} = \frac{\operatorname{Li}_1(e^{-\lambda})}{\mu} = N, \label{eq:lambda} \\
  \int_0^\infty dk k\left[\exp(\lambda + \mu k) - 1\right]^{-1} = \frac{\operatorname{Li}_2(e^{-\lambda})}{\mu^2} = E. \label{eq:mu}
\end{align}
Eq.~\ref{eq:lambda} can be inverted as $e^{-\lambda}=1-\exp(-N/\mu)$,
but Eq.~\ref{eq:mu} cannot be solved for $\lambda$ in closed
form. However, by assuming a sufficiently ``high temperature'' regime
where $\mu \sim O(1)$, we have that the fugacity simplifies in the
thermodynamic limit, $e^{-\lambda}\to 1$ for $N\gg 1$, and hence we
obtain $\mu\simeq\sqrt{\operatorname{Li}_2(1) / E}$. Using
Eqs.~\ref{eq:lambda} and~\ref{eq:mu}, we can write the entropy of the
ensemble $\ln\Xi = - \sum_k \left[\partial\ln Z_k/\partial\lambda +
\partial\ln Z_k/\partial\mu + \ln Z_k\right]$, as
\begin{equation}\label{eq:xi_lmu}
  \ln\Xi = \lambda N + 2\mu E,
\end{equation}
and for the regime $e^{-\lambda}\to 1$, we have
\begin{equation}\label{eq:bose_ent}
  \ln\Xi \simeq 2\sqrt{\zeta(2) E},
\end{equation}
where the identity $\operatorname{Li}_2(1)=\zeta(2)$ was used, with
$\zeta(x)$ being the Riemann zeta function. Although
Eq.~\ref{eq:bose_ent} becomes asymptotically exact in the thermodynamic
limit with $E\sim N$ and $N\gg 1$, the exact solution can also be
obtained with arbitrary precision simply by iterating
Eqs.~\ref{eq:lambda} and~\ref{eq:mu} as
$\hat{\lambda}(t+1)=1-\exp(-N/\mu(t))$,
$\mu(t+1)=\sqrt{E/\operatorname{Li}_2(\hat{\lambda}(t))}$, where
$\hat{\lambda} \equiv e^{-\lambda}$, with the starting points
$\hat{\lambda}(0)=1$, $\mu(0)=\sqrt{\operatorname{Li}_2(1) / E}$, until
sufficient convergence is reached, and the results are substituted in
Eq.~\ref{eq:xi_lmu}. (We actually use this more precise procedure when
computing Eq.~\ref{eq:deg_xi} in the main text, throughout the
analysis.)

\bibliography{bib}

\begin{thebibliography}{91}%
\makeatletter
\providecommand \@ifxundefined [1]{%
 \@ifx{#1\undefined}
}%
\providecommand \@ifnum [1]{%
 \ifnum #1\expandafter \@firstoftwo
 \else \expandafter \@secondoftwo
 \fi
}%
\providecommand \@ifx [1]{%
 \ifx #1\expandafter \@firstoftwo
 \else \expandafter \@secondoftwo
 \fi
}%
\providecommand \natexlab [1]{#1}%
\providecommand \enquote  [1]{``#1''}%
\providecommand \bibnamefont  [1]{#1}%
\providecommand \bibfnamefont [1]{#1}%
\providecommand \citenamefont [1]{#1}%
\providecommand \href@noop [0]{\@secondoftwo}%
\providecommand \href [0]{\begingroup \@sanitize@url \@href}%
\providecommand \@href[1]{\@@startlink{#1}\@@href}%
\providecommand \@@href[1]{\endgroup#1\@@endlink}%
\providecommand \@sanitize@url [0]{\catcode `\\12\catcode `\$12\catcode
  `\&12\catcode `\#12\catcode `\^12\catcode `\_12\catcode `\%12\relax}%
\providecommand \@@startlink[1]{}%
\providecommand \@@endlink[0]{}%
\providecommand \url  [0]{\begingroup\@sanitize@url \@url }%
\providecommand \@url [1]{\endgroup\@href {#1}{\urlprefix }}%
\providecommand \urlprefix  [0]{URL }%
\providecommand \Eprint [0]{\href }%
\providecommand \doibase [0]{http://dx.doi.org/}%
\providecommand \selectlanguage [0]{\@gobble}%
\providecommand \bibinfo  [0]{\@secondoftwo}%
\providecommand \bibfield  [0]{\@secondoftwo}%
\providecommand \translation [1]{[#1]}%
\providecommand \BibitemOpen [0]{}%
\providecommand \bibitemStop [0]{}%
\providecommand \bibitemNoStop [0]{.\EOS\space}%
\providecommand \EOS [0]{\spacefactor3000\relax}%
\providecommand \BibitemShut  [1]{\csname bibitem#1\endcsname}%
\let\auto@bib@innerbib\@empty
\bibitem [{\citenamefont {Newman}(2011)}]{newman_communities_2011}%
  \BibitemOpen
  \bibfield  {author} {\bibinfo {author} {\bibfnamefont {M.~E.~J.}\
  \bibnamefont {Newman}},\ }\bibfield  {title} {\enquote {\bibinfo {title}
  {Communities, modules and large-scale structure in networks},}\ }\href
  {\doibase 10.1038/nphys2162} {\bibfield  {journal} {\bibinfo  {journal} {Nat
  Phys}\ }\textbf {\bibinfo {volume} {8}},\ \bibinfo {pages} {25--31} (\bibinfo
  {year} {2011})}\BibitemShut {NoStop}%
\bibitem [{\citenamefont {Fortunato}(2010)}]{fortunato_community_2010}%
  \BibitemOpen
  \bibfield  {author} {\bibinfo {author} {\bibfnamefont {Santo}\ \bibnamefont
  {Fortunato}},\ }\bibfield  {title} {\enquote {\bibinfo {title} {Community
  detection in graphs},}\ }\href {\doibase 16/j.physrep.2009.11.002} {\bibfield
   {journal} {\bibinfo  {journal} {Physics Reports}\ }\textbf {\bibinfo
  {volume} {486}},\ \bibinfo {pages} {75--174} (\bibinfo {year}
  {2010})}\BibitemShut {NoStop}%
\bibitem [{\citenamefont {Holme}(2005)}]{holme_core-periphery_2005}%
  \BibitemOpen
  \bibfield  {author} {\bibinfo {author} {\bibfnamefont {Petter}\ \bibnamefont
  {Holme}},\ }\bibfield  {title} {\enquote {\bibinfo {title} {Core-periphery
  organization of complex networks},}\ }\href {\doibase
  10.1103/PhysRevE.72.046111} {\bibfield  {journal} {\bibinfo  {journal}
  {Physical Review E}\ }\textbf {\bibinfo {volume} {72}},\ \bibinfo {pages}
  {046111} (\bibinfo {year} {2005})}\BibitemShut {NoStop}%
\bibitem [{\citenamefont {Rombach}\ \emph {et~al.}(2012)\citenamefont
  {Rombach}, \citenamefont {Porter}, \citenamefont {Fowler},\ and\
  \citenamefont {Mucha}}]{rombach_core-periphery_2012}%
  \BibitemOpen
  \bibfield  {author} {\bibinfo {author} {\bibfnamefont {M.~Puck}\ \bibnamefont
  {Rombach}}, \bibinfo {author} {\bibfnamefont {Mason~A.}\ \bibnamefont
  {Porter}}, \bibinfo {author} {\bibfnamefont {James~H.}\ \bibnamefont
  {Fowler}}, \ and\ \bibinfo {author} {\bibfnamefont {Peter~J.}\ \bibnamefont
  {Mucha}},\ }\bibfield  {title} {\enquote {\bibinfo {title} {Core-periphery
  structure in networks},}\ }\href {http://arxiv.org/abs/1202.2684} {\bibfield
  {journal} {\bibinfo  {journal} {{arXiv}:1202.2684}\ } (\bibinfo {year}
  {2012})}\BibitemShut {NoStop}%
\bibitem [{\citenamefont {Larremore}\ \emph {et~al.}(2014)\citenamefont
  {Larremore}, \citenamefont {Clauset},\ and\ \citenamefont
  {Jacobs}}]{larremore_efficiently_2014}%
  \BibitemOpen
  \bibfield  {author} {\bibinfo {author} {\bibfnamefont {Daniel~B.}\
  \bibnamefont {Larremore}}, \bibinfo {author} {\bibfnamefont {Aaron}\
  \bibnamefont {Clauset}}, \ and\ \bibinfo {author} {\bibfnamefont
  {Abigail~Z.}\ \bibnamefont {Jacobs}},\ }\bibfield  {title} {\enquote
  {\bibinfo {title} {Efficiently inferring community structure in bipartite
  networks},}\ }\href {\doibase 10.1103/PhysRevE.90.012805} {\bibfield
  {journal} {\bibinfo  {journal} {Physical Review E}\ }\textbf {\bibinfo
  {volume} {90}},\ \bibinfo {pages} {012805} (\bibinfo {year}
  {2014})}\BibitemShut {NoStop}%
\bibitem [{\citenamefont {Clauset}\ \emph {et~al.}(2008)\citenamefont
  {Clauset}, \citenamefont {Moore},\ and\ \citenamefont
  {Newman}}]{clauset_hierarchical_2008}%
  \BibitemOpen
  \bibfield  {author} {\bibinfo {author} {\bibfnamefont {Aaron}\ \bibnamefont
  {Clauset}}, \bibinfo {author} {\bibfnamefont {Cristopher}\ \bibnamefont
  {Moore}}, \ and\ \bibinfo {author} {\bibfnamefont {M.~E.~J.}\ \bibnamefont
  {Newman}},\ }\bibfield  {title} {\enquote {\bibinfo {title} {Hierarchical
  structure and the prediction of missing links in networks},}\ }\href
  {\doibase 10.1038/nature06830} {\bibfield  {journal} {\bibinfo  {journal}
  {Nature}\ }\textbf {\bibinfo {volume} {453}},\ \bibinfo {pages} {98--101}
  (\bibinfo {year} {2008})}\BibitemShut {NoStop}%
\bibitem [{\citenamefont
  {Peixoto}(2014{\natexlab{a}})}]{peixoto_hierarchical_2014}%
  \BibitemOpen
  \bibfield  {author} {\bibinfo {author} {\bibfnamefont {Tiago~P.}\
  \bibnamefont {Peixoto}},\ }\bibfield  {title} {\enquote {\bibinfo {title}
  {Hierarchical block structures and high-resolution model selection in large
  networks},}\ }\href {\doibase 10.1103/PhysRevX.4.011047} {\bibfield
  {journal} {\bibinfo  {journal} {Physical Review X}\ }\textbf {\bibinfo
  {volume} {4}},\ \bibinfo {pages} {011047} (\bibinfo {year}
  {2014}{\natexlab{a}})}\BibitemShut {NoStop}%
\bibitem [{\citenamefont {Guimerà}\ and\ \citenamefont
  {Sales-Pardo}(2009)}]{guimera_missing_2009}%
  \BibitemOpen
  \bibfield  {author} {\bibinfo {author} {\bibfnamefont {Roger}\ \bibnamefont
  {Guimerà}}\ and\ \bibinfo {author} {\bibfnamefont {Marta}\ \bibnamefont
  {Sales-Pardo}},\ }\bibfield  {title} {\enquote {\bibinfo {title} {Missing and
  spurious interactions and the reconstruction of complex networks},}\ }\href
  {\doibase 10.1073/pnas.0908366106} {\bibfield  {journal} {\bibinfo  {journal}
  {Proceedings of the National Academy of Sciences}\ }\textbf {\bibinfo
  {volume} {106}},\ \bibinfo {pages} {22073 --22078} (\bibinfo {year}
  {2009})}\BibitemShut {NoStop}%
\bibitem [{\citenamefont {Buldyrev}\ \emph {et~al.}(2010)\citenamefont
  {Buldyrev}, \citenamefont {Parshani}, \citenamefont {Paul}, \citenamefont
  {Stanley},\ and\ \citenamefont {Havlin}}]{buldyrev_catastrophic_2010}%
  \BibitemOpen
  \bibfield  {author} {\bibinfo {author} {\bibfnamefont {Sergey~V.}\
  \bibnamefont {Buldyrev}}, \bibinfo {author} {\bibfnamefont {Roni}\
  \bibnamefont {Parshani}}, \bibinfo {author} {\bibfnamefont {Gerald}\
  \bibnamefont {Paul}}, \bibinfo {author} {\bibfnamefont {H.~Eugene}\
  \bibnamefont {Stanley}}, \ and\ \bibinfo {author} {\bibfnamefont {Shlomo}\
  \bibnamefont {Havlin}},\ }\bibfield  {title} {\enquote {\bibinfo {title}
  {Catastrophic cascade of failures in interdependent networks},}\ }\href
  {\doibase 10.1038/nature08932} {\bibfield  {journal} {\bibinfo  {journal}
  {Nature}\ }\textbf {\bibinfo {volume} {464}},\ \bibinfo {pages} {1025--1028}
  (\bibinfo {year} {2010})}\BibitemShut {NoStop}%
\bibitem [{\citenamefont {Apolloni}\ \emph {et~al.}(2014)\citenamefont
  {Apolloni}, \citenamefont {Poletto}, \citenamefont {Ramasco}, \citenamefont
  {Jensen},\ and\ \citenamefont {Colizza}}]{apolloni_metapopulation_2014}%
  \BibitemOpen
  \bibfield  {author} {\bibinfo {author} {\bibfnamefont {Andrea}\ \bibnamefont
  {Apolloni}}, \bibinfo {author} {\bibfnamefont {Chiara}\ \bibnamefont
  {Poletto}}, \bibinfo {author} {\bibfnamefont {José~J.}\ \bibnamefont
  {Ramasco}}, \bibinfo {author} {\bibfnamefont {Pablo}\ \bibnamefont {Jensen}},
  \ and\ \bibinfo {author} {\bibfnamefont {Vittoria}\ \bibnamefont {Colizza}},\
  }\bibfield  {title} {\enquote {\bibinfo {title} {Metapopulation epidemic
  models with heterogeneous mixing and travel behaviour},}\ }\href {\doibase
  10.1186/1742-4682-11-3} {\bibfield  {journal} {\bibinfo  {journal}
  {Theoretical Biology and Medical Modelling}\ }\textbf {\bibinfo {volume}
  {11}},\ \bibinfo {pages} {3} (\bibinfo {year} {2014})}\BibitemShut {NoStop}%
\bibitem [{\citenamefont {Guimerà}\ and\ \citenamefont
  {Nunes~Amaral}(2005)}]{guimera_functional_2005}%
  \BibitemOpen
  \bibfield  {author} {\bibinfo {author} {\bibfnamefont {Roger}\ \bibnamefont
  {Guimerà}}\ and\ \bibinfo {author} {\bibfnamefont {Luís~A.}\ \bibnamefont
  {Nunes~Amaral}},\ }\bibfield  {title} {{\selectlanguage {english}\enquote
  {\bibinfo {title} {Functional cartography of complex metabolic networks},}\
  }}\href {\doibase 10.1038/nature03288} {\bibfield  {journal} {\bibinfo
  {journal} {Nature}\ }\textbf {\bibinfo {volume} {433}},\ \bibinfo {pages}
  {895--900} (\bibinfo {year} {2005})}\BibitemShut {NoStop}%
\bibitem [{\citenamefont
  {Newman}(2006{\natexlab{a}})}]{newman_modularity_2006}%
  \BibitemOpen
  \bibfield  {author} {\bibinfo {author} {\bibfnamefont {M.~E.~J.}\
  \bibnamefont {Newman}},\ }\bibfield  {title} {{\selectlanguage
  {english}\enquote {\bibinfo {title} {Modularity and community structure in
  networks},}\ }}\href {\doibase 10.1073/pnas.0601602103} {\bibfield  {journal}
  {\bibinfo  {journal} {Proceedings of the National Academy of Sciences}\
  }\textbf {\bibinfo {volume} {103}},\ \bibinfo {pages} {8577--8582} (\bibinfo
  {year} {2006}{\natexlab{a}})}\BibitemShut {NoStop}%
\bibitem [{\citenamefont {Girvan}\ and\ \citenamefont
  {Newman}(2002)}]{girvan_community_2002}%
  \BibitemOpen
  \bibfield  {author} {\bibinfo {author} {\bibfnamefont {M.}~\bibnamefont
  {Girvan}}\ and\ \bibinfo {author} {\bibfnamefont {M.~E.~J.}\ \bibnamefont
  {Newman}},\ }\bibfield  {title} {\enquote {\bibinfo {title} {Community
  structure in social and biological networks},}\ }\href {\doibase
  10.1073/pnas.122653799} {\bibfield  {journal} {\bibinfo  {journal}
  {Proceedings of the National Academy of Sciences}\ }\textbf {\bibinfo
  {volume} {99}},\ \bibinfo {pages} {7821 --7826} (\bibinfo {year}
  {2002})}\BibitemShut {NoStop}%
\bibitem [{\citenamefont {Ahn}\ \emph {et~al.}(2010)\citenamefont {Ahn},
  \citenamefont {Bagrow},\ and\ \citenamefont {Lehmann}}]{ahn_link_2010}%
  \BibitemOpen
  \bibfield  {author} {\bibinfo {author} {\bibfnamefont {Yong-Yeol}\
  \bibnamefont {Ahn}}, \bibinfo {author} {\bibfnamefont {James~P.}\
  \bibnamefont {Bagrow}}, \ and\ \bibinfo {author} {\bibfnamefont {Sune}\
  \bibnamefont {Lehmann}},\ }\bibfield  {title} {{\selectlanguage
  {english}\enquote {\bibinfo {title} {Link communities reveal multiscale
  complexity in networks},}\ }}\href {\doibase 10.1038/nature09182} {\bibfield
  {journal} {\bibinfo  {journal} {Nature}\ }\textbf {\bibinfo {volume} {466}},\
  \bibinfo {pages} {761--764} (\bibinfo {year} {2010})}\BibitemShut {NoStop}%
\bibitem [{\citenamefont {Palla}\ \emph {et~al.}(2005)\citenamefont {Palla},
  \citenamefont {Derényi}, \citenamefont {Farkas},\ and\ \citenamefont
  {Vicsek}}]{palla_uncovering_2005}%
  \BibitemOpen
  \bibfield  {author} {\bibinfo {author} {\bibfnamefont {Gergely}\ \bibnamefont
  {Palla}}, \bibinfo {author} {\bibfnamefont {Imre}\ \bibnamefont {Derényi}},
  \bibinfo {author} {\bibfnamefont {Illés}\ \bibnamefont {Farkas}}, \ and\
  \bibinfo {author} {\bibfnamefont {Tamás}\ \bibnamefont {Vicsek}},\
  }\bibfield  {title} {{\selectlanguage {english}\enquote {\bibinfo {title}
  {Uncovering the overlapping community structure of complex networks in nature
  and society},}\ }}\href {\doibase 10.1038/nature03607} {\bibfield  {journal}
  {\bibinfo  {journal} {Nature}\ }\textbf {\bibinfo {volume} {435}},\ \bibinfo
  {pages} {814--818} (\bibinfo {year} {2005})}\BibitemShut {NoStop}%
\bibitem [{\citenamefont {Rosvall}\ and\ \citenamefont
  {Bergstrom}(2008)}]{rosvall_maps_2008}%
  \BibitemOpen
  \bibfield  {author} {\bibinfo {author} {\bibfnamefont {Martin}\ \bibnamefont
  {Rosvall}}\ and\ \bibinfo {author} {\bibfnamefont {Carl~T.}\ \bibnamefont
  {Bergstrom}},\ }\bibfield  {title} {{\selectlanguage {english}\enquote
  {\bibinfo {title} {Maps of random walks on complex networks reveal community
  structure},}\ }}\href {\doibase 10.1073/pnas.0706851105} {\bibfield
  {journal} {\bibinfo  {journal} {Proceedings of the National Academy of
  Sciences}\ }\textbf {\bibinfo {volume} {105}},\ \bibinfo {pages} {1118--1123}
  (\bibinfo {year} {2008})}\BibitemShut {NoStop}%
\bibitem [{\citenamefont {Holland}\ \emph {et~al.}(1983)\citenamefont
  {Holland}, \citenamefont {Laskey},\ and\ \citenamefont
  {Leinhardt}}]{holland_stochastic_1983}%
  \BibitemOpen
  \bibfield  {author} {\bibinfo {author} {\bibfnamefont {Paul~W.}\ \bibnamefont
  {Holland}}, \bibinfo {author} {\bibfnamefont {Kathryn~Blackmond}\
  \bibnamefont {Laskey}}, \ and\ \bibinfo {author} {\bibfnamefont {Samuel}\
  \bibnamefont {Leinhardt}},\ }\bibfield  {title} {\enquote {\bibinfo {title}
  {Stochastic blockmodels: First steps},}\ }\href {\doibase
  16/0378-8733(83)90021-7} {\bibfield  {journal} {\bibinfo  {journal} {Social
  Networks}\ }\textbf {\bibinfo {volume} {5}},\ \bibinfo {pages} {109--137}
  (\bibinfo {year} {1983})}\BibitemShut {NoStop}%
\bibitem [{\citenamefont {Airoldi}\ \emph {et~al.}(2008)\citenamefont
  {Airoldi}, \citenamefont {Blei}, \citenamefont {Fienberg},\ and\
  \citenamefont {Xing}}]{airoldi_mixed_2008}%
  \BibitemOpen
  \bibfield  {author} {\bibinfo {author} {\bibfnamefont {Edoardo~M.}\
  \bibnamefont {Airoldi}}, \bibinfo {author} {\bibfnamefont {David~M.}\
  \bibnamefont {Blei}}, \bibinfo {author} {\bibfnamefont {Stephen~E.}\
  \bibnamefont {Fienberg}}, \ and\ \bibinfo {author} {\bibfnamefont {Eric~P.}\
  \bibnamefont {Xing}},\ }\bibfield  {title} {\enquote {\bibinfo {title} {Mixed
  membership stochastic blockmodels},}\ }\href
  {http://dl.acm.org/citation.cfm?id=1390681.1442798} {\bibfield  {journal}
  {\bibinfo  {journal} {J. Mach. Learn. Res.}\ }\textbf {\bibinfo {volume}
  {9}},\ \bibinfo {pages} {1981--2014} (\bibinfo {year} {2008})}\BibitemShut
  {NoStop}%
\bibitem [{\citenamefont {Karrer}\ and\ \citenamefont
  {Newman}(2011)}]{karrer_stochastic_2011}%
  \BibitemOpen
  \bibfield  {author} {\bibinfo {author} {\bibfnamefont {Brian}\ \bibnamefont
  {Karrer}}\ and\ \bibinfo {author} {\bibfnamefont {M.~E.~J.}\ \bibnamefont
  {Newman}},\ }\bibfield  {title} {\enquote {\bibinfo {title} {Stochastic
  blockmodels and community structure in networks},}\ }\href {\doibase
  10.1103/PhysRevE.83.016107} {\bibfield  {journal} {\bibinfo  {journal}
  {Physical Review E}\ }\textbf {\bibinfo {volume} {83}},\ \bibinfo {pages}
  {016107} (\bibinfo {year} {2011})}\BibitemShut {NoStop}%
\bibitem [{\citenamefont {Ball}\ \emph {et~al.}(2011)\citenamefont {Ball},
  \citenamefont {Karrer},\ and\ \citenamefont {Newman}}]{ball_efficient_2011}%
  \BibitemOpen
  \bibfield  {author} {\bibinfo {author} {\bibfnamefont {Brian}\ \bibnamefont
  {Ball}}, \bibinfo {author} {\bibfnamefont {Brian}\ \bibnamefont {Karrer}}, \
  and\ \bibinfo {author} {\bibfnamefont {M.~E.~J.}\ \bibnamefont {Newman}},\
  }\bibfield  {title} {\enquote {\bibinfo {title} {Efficient and principled
  method for detecting communities in networks},}\ }\href {\doibase
  10.1103/PhysRevE.84.036103} {\bibfield  {journal} {\bibinfo  {journal}
  {Physical Review E}\ }\textbf {\bibinfo {volume} {84}},\ \bibinfo {pages}
  {036103} (\bibinfo {year} {2011})}\BibitemShut {NoStop}%
\bibitem [{\citenamefont {Lancichinetti}\ \emph {et~al.}(2009)\citenamefont
  {Lancichinetti}, \citenamefont {Fortunato},\ and\ \citenamefont
  {Kertész}}]{lancichinetti_detecting_2009}%
  \BibitemOpen
  \bibfield  {author} {\bibinfo {author} {\bibfnamefont {Andrea}\ \bibnamefont
  {Lancichinetti}}, \bibinfo {author} {\bibfnamefont {Santo}\ \bibnamefont
  {Fortunato}}, \ and\ \bibinfo {author} {\bibfnamefont {János}\ \bibnamefont
  {Kertész}},\ }\bibfield  {title} {{\selectlanguage {english}\enquote
  {\bibinfo {title} {Detecting the overlapping and hierarchical community
  structure in complex networks},}\ }}\href {\doibase
  10.1088/1367-2630/11/3/033015} {\bibfield  {journal} {\bibinfo  {journal}
  {New Journal of Physics}\ }\textbf {\bibinfo {volume} {11}},\ \bibinfo
  {pages} {033015} (\bibinfo {year} {2009})}\BibitemShut {NoStop}%
\bibitem [{\citenamefont {Palla}\ \emph {et~al.}(2010)\citenamefont {Palla},
  \citenamefont {Lovász},\ and\ \citenamefont
  {Vicsek}}]{palla_multifractal_2010}%
  \BibitemOpen
  \bibfield  {author} {\bibinfo {author} {\bibfnamefont {Gergely}\ \bibnamefont
  {Palla}}, \bibinfo {author} {\bibfnamefont {László}\ \bibnamefont
  {Lovász}}, \ and\ \bibinfo {author} {\bibfnamefont {Tamás}\ \bibnamefont
  {Vicsek}},\ }\bibfield  {title} {\enquote {\bibinfo {title} {Multifractal
  network generator},}\ }\href {\doibase 10.1073/pnas.0912983107} {\bibfield
  {journal} {\bibinfo  {journal} {Proceedings of the National Academy of
  Sciences}\ }\textbf {\bibinfo {volume} {107}},\ \bibinfo {pages} {7640
  --7645} (\bibinfo {year} {2010})}\BibitemShut {NoStop}%
\bibitem [{\citenamefont {Leskovec}\ \emph
  {et~al.}(2010{\natexlab{a}})\citenamefont {Leskovec}, \citenamefont
  {Chakrabarti}, \citenamefont {Kleinberg}, \citenamefont {Faloutsos},\ and\
  \citenamefont {Ghahramani}}]{leskovec_kronecker_2010}%
  \BibitemOpen
  \bibfield  {author} {\bibinfo {author} {\bibfnamefont {Jure}\ \bibnamefont
  {Leskovec}}, \bibinfo {author} {\bibfnamefont {Deepayan}\ \bibnamefont
  {Chakrabarti}}, \bibinfo {author} {\bibfnamefont {Jon}\ \bibnamefont
  {Kleinberg}}, \bibinfo {author} {\bibfnamefont {Christos}\ \bibnamefont
  {Faloutsos}}, \ and\ \bibinfo {author} {\bibfnamefont {Zoubin}\ \bibnamefont
  {Ghahramani}},\ }\bibfield  {title} {\enquote {\bibinfo {title} {Kronecker
  graphs: An approach to modeling networks},}\ }\href
  {http://dl.acm.org/citation.cfm?id=1756006.1756039} {\bibfield  {journal}
  {\bibinfo  {journal} {J. Mach. Learn. Res.}\ }\textbf {\bibinfo {volume}
  {11}},\ \bibinfo {pages} {985--1042} (\bibinfo {year}
  {2010}{\natexlab{a}})}\BibitemShut {NoStop}%
\bibitem [{\citenamefont {Mariadassou}\ \emph {et~al.}(2010)\citenamefont
  {Mariadassou}, \citenamefont {Robin},\ and\ \citenamefont
  {Vacher}}]{mariadassou_uncovering_2010}%
  \BibitemOpen
  \bibfield  {author} {\bibinfo {author} {\bibfnamefont {Mahendra}\
  \bibnamefont {Mariadassou}}, \bibinfo {author} {\bibfnamefont {Stéphane}\
  \bibnamefont {Robin}}, \ and\ \bibinfo {author} {\bibfnamefont {Corinne}\
  \bibnamefont {Vacher}},\ }\bibfield  {title} {{\selectlanguage
  {english}\enquote {\bibinfo {title} {Uncovering latent structure in valued
  graphs: A variational approach},}\ }}\href {\doibase 10.1214/10-AOAS361}
  {\bibfield  {journal} {\bibinfo  {journal} {The Annals of Applied
  Statistics}\ }\textbf {\bibinfo {volume} {4}},\ \bibinfo {pages} {715--742}
  (\bibinfo {year} {2010})},\ \bibinfo {note} {mathematical Reviews number
  ({MathSciNet}): {MR}2758646}\BibitemShut {NoStop}%
\bibitem [{\citenamefont {Aicher}\ \emph {et~al.}(2014)\citenamefont {Aicher},
  \citenamefont {Jacobs},\ and\ \citenamefont
  {Clauset}}]{aicher_learning_2014}%
  \BibitemOpen
  \bibfield  {author} {\bibinfo {author} {\bibfnamefont {Christopher}\
  \bibnamefont {Aicher}}, \bibinfo {author} {\bibfnamefont {Abigail~Z.}\
  \bibnamefont {Jacobs}}, \ and\ \bibinfo {author} {\bibfnamefont {Aaron}\
  \bibnamefont {Clauset}},\ }\bibfield  {title} {{\selectlanguage
  {english}\enquote {\bibinfo {title} {Learning latent block structure in
  weighted networks},}\ }}\href {\doibase 10.1093/comnet/cnu026} {\bibfield
  {journal} {\bibinfo  {journal} {Journal of Complex Networks}\ ,\ \bibinfo
  {pages} {cnu026}} (\bibinfo {year} {2014})}\BibitemShut {NoStop}%
\bibitem [{\citenamefont {Ball}\ and\ \citenamefont
  {Newman}(2013)}]{ball_friendship_2013}%
  \BibitemOpen
  \bibfield  {author} {\bibinfo {author} {\bibfnamefont {Brian}\ \bibnamefont
  {Ball}}\ and\ \bibinfo {author} {\bibfnamefont {M.e.j.}\ \bibnamefont
  {Newman}},\ }\bibfield  {title} {\enquote {\bibinfo {title} {Friendship
  networks and social status},}\ }\href {\doibase 10.1017/nws.2012.4}
  {\bibfield  {journal} {\bibinfo  {journal} {Network Science}\ }\textbf
  {\bibinfo {volume} {1}},\ \bibinfo {pages} {16--30} (\bibinfo {year}
  {2013})}\BibitemShut {NoStop}%
\bibitem [{\citenamefont {Kivelä}\ \emph {et~al.}(2014)\citenamefont
  {Kivelä}, \citenamefont {Arenas}, \citenamefont {Barthelemy}, \citenamefont
  {Gleeson}, \citenamefont {Moreno},\ and\ \citenamefont
  {Porter}}]{kivela_multilayer_2014}%
  \BibitemOpen
  \bibfield  {author} {\bibinfo {author} {\bibfnamefont {Mikko}\ \bibnamefont
  {Kivelä}}, \bibinfo {author} {\bibfnamefont {Alex}\ \bibnamefont {Arenas}},
  \bibinfo {author} {\bibfnamefont {Marc}\ \bibnamefont {Barthelemy}}, \bibinfo
  {author} {\bibfnamefont {James~P.}\ \bibnamefont {Gleeson}}, \bibinfo
  {author} {\bibfnamefont {Yamir}\ \bibnamefont {Moreno}}, \ and\ \bibinfo
  {author} {\bibfnamefont {Mason~A.}\ \bibnamefont {Porter}},\ }\bibfield
  {title} {{\selectlanguage {english}\enquote {\bibinfo {title} {Multilayer
  networks},}\ }}\href {\doibase 10.1093/comnet/cnu016} {\bibfield  {journal}
  {\bibinfo  {journal} {Journal of Complex Networks}\ }\textbf {\bibinfo
  {volume} {2}},\ \bibinfo {pages} {203--271} (\bibinfo {year}
  {2014})}\BibitemShut {NoStop}%
\bibitem [{\citenamefont {Fu}\ \emph {et~al.}(2009)\citenamefont {Fu},
  \citenamefont {Song},\ and\ \citenamefont {Xing}}]{fu_dynamic_2009}%
  \BibitemOpen
  \bibfield  {author} {\bibinfo {author} {\bibfnamefont {Wenjie}\ \bibnamefont
  {Fu}}, \bibinfo {author} {\bibfnamefont {Le}~\bibnamefont {Song}}, \ and\
  \bibinfo {author} {\bibfnamefont {Eric~P.}\ \bibnamefont {Xing}},\ }\bibfield
   {title} {\enquote {\bibinfo {title} {Dynamic mixed membership blockmodel for
  evolving networks},}\ }in\ \href {\doibase 10.1145/1553374.1553416} {\emph
  {\bibinfo {booktitle} {Proceedings of the 26th Annual International
  Conference on Machine Learning}}},\ \bibinfo {series and number} {{ICML}
  '09}\ (\bibinfo  {publisher} {{ACM}},\ \bibinfo {address} {New York, {NY},
  {USA}},\ \bibinfo {year} {2009})\ pp.\ \bibinfo {pages}
  {329--336}\BibitemShut {NoStop}%
\bibitem [{\citenamefont {Yan}\ \emph {et~al.}(2014)\citenamefont {Yan},
  \citenamefont {Shalizi}, \citenamefont {Jensen}, \citenamefont {Krzakala},
  \citenamefont {Moore}, \citenamefont {Zdeborová}, \citenamefont {Zhang},\
  and\ \citenamefont {Zhu}}]{yan_model_2014}%
  \BibitemOpen
  \bibfield  {author} {\bibinfo {author} {\bibfnamefont {Xiaoran}\ \bibnamefont
  {Yan}}, \bibinfo {author} {\bibfnamefont {Cosma}\ \bibnamefont {Shalizi}},
  \bibinfo {author} {\bibfnamefont {Jacob~E.}\ \bibnamefont {Jensen}}, \bibinfo
  {author} {\bibfnamefont {Florent}\ \bibnamefont {Krzakala}}, \bibinfo
  {author} {\bibfnamefont {Cristopher}\ \bibnamefont {Moore}}, \bibinfo
  {author} {\bibfnamefont {Lenka}\ \bibnamefont {Zdeborová}}, \bibinfo
  {author} {\bibfnamefont {Pan}\ \bibnamefont {Zhang}}, \ and\ \bibinfo
  {author} {\bibfnamefont {Yaojia}\ \bibnamefont {Zhu}},\ }\bibfield  {title}
  {{\selectlanguage {english}\enquote {\bibinfo {title} {Model selection for
  degree-corrected block models},}\ }}\href {\doibase
  10.1088/1742-5468/2014/05/P05007} {\bibfield  {journal} {\bibinfo  {journal}
  {Journal of Statistical Mechanics: Theory and Experiment}\ }\textbf {\bibinfo
  {volume} {2014}},\ \bibinfo {pages} {P05007} (\bibinfo {year}
  {2014})}\BibitemShut {NoStop}%
\bibitem [{\citenamefont {Newman}(2013{\natexlab{a}})}]{newman_community_2013}%
  \BibitemOpen
  \bibfield  {author} {\bibinfo {author} {\bibfnamefont {M.~E.~J.}\
  \bibnamefont {Newman}},\ }\bibfield  {title} {{\selectlanguage
  {english}\enquote {\bibinfo {title} {Community detection and graph
  partitioning},}\ }}\href {\doibase 10.1209/0295-5075/103/28003} {\bibfield
  {journal} {\bibinfo  {journal} {{EPL} (Europhysics Letters)}\ }\textbf
  {\bibinfo {volume} {103}},\ \bibinfo {pages} {28003} (\bibinfo {year}
  {2013}{\natexlab{a}})}\BibitemShut {NoStop}%
\bibitem [{\citenamefont {Newman}(2013{\natexlab{b}})}]{newman_spectral_2013}%
  \BibitemOpen
  \bibfield  {author} {\bibinfo {author} {\bibfnamefont {M.~E.~J.}\
  \bibnamefont {Newman}},\ }\bibfield  {title} {\enquote {\bibinfo {title}
  {Spectral methods for community detection and graph partitioning},}\ }\href
  {\doibase 10.1103/PhysRevE.88.042822} {\bibfield  {journal} {\bibinfo
  {journal} {Physical Review E}\ }\textbf {\bibinfo {volume} {88}},\ \bibinfo
  {pages} {042822} (\bibinfo {year} {2013}{\natexlab{b}})}\BibitemShut
  {NoStop}%
\bibitem [{\citenamefont {Nadakuditi}\ and\ \citenamefont
  {Newman}(2012)}]{nadakuditi_graph_2012}%
  \BibitemOpen
  \bibfield  {author} {\bibinfo {author} {\bibfnamefont {Raj~Rao}\ \bibnamefont
  {Nadakuditi}}\ and\ \bibinfo {author} {\bibfnamefont {M.~E.~J.}\ \bibnamefont
  {Newman}},\ }\bibfield  {title} {\enquote {\bibinfo {title} {Graph spectra
  and the detectability of community structure in networks},}\ }\href {\doibase
  10.1103/PhysRevLett.108.188701} {\bibfield  {journal} {\bibinfo  {journal}
  {Physical Review Letters}\ }\textbf {\bibinfo {volume} {108}},\ \bibinfo
  {pages} {188701} (\bibinfo {year} {2012})}\BibitemShut {NoStop}%
\bibitem [{\citenamefont {Krzakala}\ \emph {et~al.}(2013)\citenamefont
  {Krzakala}, \citenamefont {Moore}, \citenamefont {Mossel}, \citenamefont
  {Neeman}, \citenamefont {Sly}, \citenamefont {Zdeborová},\ and\
  \citenamefont {Zhang}}]{krzakala_spectral_2013}%
  \BibitemOpen
  \bibfield  {author} {\bibinfo {author} {\bibfnamefont {Florent}\ \bibnamefont
  {Krzakala}}, \bibinfo {author} {\bibfnamefont {Cristopher}\ \bibnamefont
  {Moore}}, \bibinfo {author} {\bibfnamefont {Elchanan}\ \bibnamefont
  {Mossel}}, \bibinfo {author} {\bibfnamefont {Joe}\ \bibnamefont {Neeman}},
  \bibinfo {author} {\bibfnamefont {Allan}\ \bibnamefont {Sly}}, \bibinfo
  {author} {\bibfnamefont {Lenka}\ \bibnamefont {Zdeborová}}, \ and\ \bibinfo
  {author} {\bibfnamefont {Pan}\ \bibnamefont {Zhang}},\ }\bibfield  {title}
  {{\selectlanguage {english}\enquote {\bibinfo {title} {Spectral redemption in
  clustering sparse networks},}\ }}\href {\doibase 10.1073/pnas.1312486110}
  {\bibfield  {journal} {\bibinfo  {journal} {Proceedings of the National
  Academy of Sciences}\ ,\ \bibinfo {pages} {201312486}} (\bibinfo {year}
  {2013})}\BibitemShut {NoStop}%
\bibitem [{\citenamefont {Lancichinetti}\ \emph {et~al.}(2008)\citenamefont
  {Lancichinetti}, \citenamefont {Fortunato},\ and\ \citenamefont
  {Radicchi}}]{lancichinetti_benchmark_2008}%
  \BibitemOpen
  \bibfield  {author} {\bibinfo {author} {\bibfnamefont {Andrea}\ \bibnamefont
  {Lancichinetti}}, \bibinfo {author} {\bibfnamefont {Santo}\ \bibnamefont
  {Fortunato}}, \ and\ \bibinfo {author} {\bibfnamefont {Filippo}\ \bibnamefont
  {Radicchi}},\ }\bibfield  {title} {\enquote {\bibinfo {title} {Benchmark
  graphs for testing community detection algorithms},}\ }\href {\doibase
  10.1103/PhysRevE.78.046110} {\bibfield  {journal} {\bibinfo  {journal}
  {Physical Review E}\ }\textbf {\bibinfo {volume} {78}},\ \bibinfo {pages}
  {046110} (\bibinfo {year} {2008})}\BibitemShut {NoStop}%
\bibitem [{\citenamefont {Lancichinetti}\ and\ \citenamefont
  {Fortunato}(2009{\natexlab{a}})}]{lancichinetti_benchmarks_2009}%
  \BibitemOpen
  \bibfield  {author} {\bibinfo {author} {\bibfnamefont {Andrea}\ \bibnamefont
  {Lancichinetti}}\ and\ \bibinfo {author} {\bibfnamefont {Santo}\ \bibnamefont
  {Fortunato}},\ }\bibfield  {title} {\enquote {\bibinfo {title} {Benchmarks
  for testing community detection algorithms on directed and weighted graphs
  with overlapping communities},}\ }\href {\doibase 10.1103/PhysRevE.80.016118}
  {\bibfield  {journal} {\bibinfo  {journal} {Physical Review E}\ }\textbf
  {\bibinfo {volume} {80}},\ \bibinfo {pages} {016118} (\bibinfo {year}
  {2009}{\natexlab{a}})}\BibitemShut {NoStop}%
\bibitem [{\citenamefont {Lancichinetti}\ and\ \citenamefont
  {Fortunato}(2009{\natexlab{b}})}]{lancichinetti_community_2009}%
  \BibitemOpen
  \bibfield  {author} {\bibinfo {author} {\bibfnamefont {Andrea}\ \bibnamefont
  {Lancichinetti}}\ and\ \bibinfo {author} {\bibfnamefont {Santo}\ \bibnamefont
  {Fortunato}},\ }\bibfield  {title} {\enquote {\bibinfo {title} {Community
  detection algorithms: A comparative analysis},}\ }\href {\doibase
  10.1103/PhysRevE.80.056117} {\bibfield  {journal} {\bibinfo  {journal}
  {Physical Review E}\ }\textbf {\bibinfo {volume} {80}},\ \bibinfo {pages}
  {056117} (\bibinfo {year} {2009}{\natexlab{b}})}\BibitemShut {NoStop}%
\bibitem [{\citenamefont {Karrer}\ \emph {et~al.}(2007)\citenamefont {Karrer},
  \citenamefont {Levina},\ and\ \citenamefont
  {Newman}}]{karrer_robustness_2007}%
  \BibitemOpen
  \bibfield  {author} {\bibinfo {author} {\bibfnamefont {Brian}\ \bibnamefont
  {Karrer}}, \bibinfo {author} {\bibfnamefont {Elizaveta}\ \bibnamefont
  {Levina}}, \ and\ \bibinfo {author} {\bibfnamefont {M.~E.~J}\ \bibnamefont
  {Newman}},\ }\bibfield  {title} {\enquote {\bibinfo {title} {Robustness of
  community structure in networks},}\ }\href {http://arxiv.org/abs/0709.2108}
  {\bibfield  {journal} {\bibinfo  {journal} {0709.2108}\ } (\bibinfo {year}
  {2007})}\BibitemShut {NoStop}%
\bibitem [{\citenamefont {Grünwald}(2007)}]{grunwald_minimum_2007}%
  \BibitemOpen
  \bibfield  {author} {\bibinfo {author} {\bibfnamefont {Peter~D.}\
  \bibnamefont {Grünwald}},\ }\href@noop {} {\emph {\bibinfo {title} {The
  Minimum Description Length Principle}}}\ (\bibinfo  {publisher} {The {MIT}
  Press},\ \bibinfo {year} {2007})\BibitemShut {NoStop}%
\bibitem [{\citenamefont {Rissanen}(2010)}]{rissanen_information_2010}%
  \BibitemOpen
  \bibfield  {author} {\bibinfo {author} {\bibfnamefont {Jorma}\ \bibnamefont
  {Rissanen}},\ }\href@noop {} {\emph {\bibinfo {title} {Information and
  Complexity in Statistical Modeling}}},\ \bibinfo {edition} {1st}\ ed.\
  (\bibinfo  {publisher} {Springer},\ \bibinfo {year} {2010})\BibitemShut
  {NoStop}%
\bibitem [{\citenamefont {Rosvall}\ \emph {et~al.}(2014)\citenamefont
  {Rosvall}, \citenamefont {Esquivel}, \citenamefont {Lancichinetti},
  \citenamefont {West},\ and\ \citenamefont {Lambiotte}}]{rosvall_memory_2014}%
  \BibitemOpen
  \bibfield  {author} {\bibinfo {author} {\bibfnamefont {Martin}\ \bibnamefont
  {Rosvall}}, \bibinfo {author} {\bibfnamefont {Alcides~V.}\ \bibnamefont
  {Esquivel}}, \bibinfo {author} {\bibfnamefont {Andrea}\ \bibnamefont
  {Lancichinetti}}, \bibinfo {author} {\bibfnamefont {Jevin~D.}\ \bibnamefont
  {West}}, \ and\ \bibinfo {author} {\bibfnamefont {Renaud}\ \bibnamefont
  {Lambiotte}},\ }\bibfield  {title} {{\selectlanguage {english}\enquote
  {\bibinfo {title} {Memory in network flows and its effects on spreading
  dynamics and community detection},}\ }}\href {\doibase 10.1038/ncomms5630}
  {\bibfield  {journal} {\bibinfo  {journal} {Nature Communications}\ }\textbf
  {\bibinfo {volume} {5}} (\bibinfo {year} {2014}),\
  10.1038/ncomms5630}\BibitemShut {NoStop}%
\bibitem [{\citenamefont {Bianconi}(2009)}]{bianconi_entropy_2009}%
  \BibitemOpen
  \bibfield  {author} {\bibinfo {author} {\bibfnamefont {Ginestra}\
  \bibnamefont {Bianconi}},\ }\bibfield  {title} {\enquote {\bibinfo {title}
  {Entropy of network ensembles},}\ }\href {\doibase
  10.1103/PhysRevE.79.036114} {\bibfield  {journal} {\bibinfo  {journal}
  {Physical Review E}\ }\textbf {\bibinfo {volume} {79}},\ \bibinfo {pages}
  {036114} (\bibinfo {year} {2009})}\BibitemShut {NoStop}%
\bibitem [{\citenamefont {Peixoto}(2012)}]{peixoto_entropy_2012}%
  \BibitemOpen
  \bibfield  {author} {\bibinfo {author} {\bibfnamefont {Tiago~P.}\
  \bibnamefont {Peixoto}},\ }\bibfield  {title} {\enquote {\bibinfo {title}
  {Entropy of stochastic blockmodel ensembles},}\ }\href {\doibase
  10.1103/PhysRevE.85.056122} {\bibfield  {journal} {\bibinfo  {journal}
  {Physical Review E}\ }\textbf {\bibinfo {volume} {85}},\ \bibinfo {pages}
  {056122} (\bibinfo {year} {2012})}\BibitemShut {NoStop}%
\bibitem [{\citenamefont {Rosvall}\ and\ \citenamefont
  {Bergstrom}(2007)}]{rosvall_information-theoretic_2007}%
  \BibitemOpen
  \bibfield  {author} {\bibinfo {author} {\bibfnamefont {Martin}\ \bibnamefont
  {Rosvall}}\ and\ \bibinfo {author} {\bibfnamefont {Carl~T.}\ \bibnamefont
  {Bergstrom}},\ }\bibfield  {title} {{\selectlanguage {english}\enquote
  {\bibinfo {title} {An information-theoretic framework for resolving community
  structure in complex networks},}\ }}\href {\doibase 10.1073/pnas.0611034104}
  {\bibfield  {journal} {\bibinfo  {journal} {Proceedings of the National
  Academy of Sciences}\ }\textbf {\bibinfo {volume} {104}},\ \bibinfo {pages}
  {7327--7331} (\bibinfo {year} {2007})}\BibitemShut {NoStop}%
\bibitem [{\citenamefont {Fienberg}\ \emph {et~al.}(1985)\citenamefont
  {Fienberg}, \citenamefont {Meyer},\ and\ \citenamefont
  {Wasserman}}]{fienberg_statistical_1985}%
  \BibitemOpen
  \bibfield  {author} {\bibinfo {author} {\bibfnamefont {Stephen~E.}\
  \bibnamefont {Fienberg}}, \bibinfo {author} {\bibfnamefont {Michael~M.}\
  \bibnamefont {Meyer}}, \ and\ \bibinfo {author} {\bibfnamefont {Stanley~S.}\
  \bibnamefont {Wasserman}},\ }\bibfield  {title} {\enquote {\bibinfo {title}
  {Statistical analysis of multiple sociometric relations},}\ }\href {\doibase
  10.2307/2288040} {\bibfield  {journal} {\bibinfo  {journal} {Journal of the
  American Statistical Association}\ }\textbf {\bibinfo {volume} {80}},\
  \bibinfo {pages} {51--67} (\bibinfo {year} {1985})}\BibitemShut {NoStop}%
\bibitem [{\citenamefont {Faust}\ and\ \citenamefont
  {Wasserman}(1992)}]{faust_blockmodels:_1992}%
  \BibitemOpen
  \bibfield  {author} {\bibinfo {author} {\bibfnamefont {Katherine}\
  \bibnamefont {Faust}}\ and\ \bibinfo {author} {\bibfnamefont {Stanley}\
  \bibnamefont {Wasserman}},\ }\bibfield  {title} {\enquote {\bibinfo {title}
  {Blockmodels: Interpretation and evaluation},}\ }\href {\doibase
  16/0378-8733(92)90013-W} {\bibfield  {journal} {\bibinfo  {journal} {Social
  Networks}\ }\textbf {\bibinfo {volume} {14}},\ \bibinfo {pages} {5--61}
  (\bibinfo {year} {1992})}\BibitemShut {NoStop}%
\bibitem [{\citenamefont {Anderson}\ \emph {et~al.}(1992)\citenamefont
  {Anderson}, \citenamefont {Wasserman},\ and\ \citenamefont
  {Faust}}]{anderson_building_1992}%
  \BibitemOpen
  \bibfield  {author} {\bibinfo {author} {\bibfnamefont {Carolyn~J.}\
  \bibnamefont {Anderson}}, \bibinfo {author} {\bibfnamefont {Stanley}\
  \bibnamefont {Wasserman}}, \ and\ \bibinfo {author} {\bibfnamefont
  {Katherine}\ \bibnamefont {Faust}},\ }\bibfield  {title} {\enquote {\bibinfo
  {title} {Building stochastic blockmodels},}\ }\href {\doibase
  16/0378-8733(92)90017-2} {\bibfield  {journal} {\bibinfo  {journal} {Social
  Networks}\ }\textbf {\bibinfo {volume} {14}},\ \bibinfo {pages} {137--161}
  (\bibinfo {year} {1992})}\BibitemShut {NoStop}%
\bibitem [{\citenamefont {Latouche}\ \emph {et~al.}(2014)\citenamefont
  {Latouche}, \citenamefont {Birmelé},\ and\ \citenamefont
  {Ambroise}}]{latouche_model_2014}%
  \BibitemOpen
  \bibfield  {author} {\bibinfo {author} {\bibfnamefont {Pierre}\ \bibnamefont
  {Latouche}}, \bibinfo {author} {\bibfnamefont {Etienne}\ \bibnamefont
  {Birmelé}}, \ and\ \bibinfo {author} {\bibfnamefont {Christophe}\
  \bibnamefont {Ambroise}},\ }\bibfield  {title} {{\selectlanguage
  {english}\enquote {\bibinfo {title} {Model selection in overlapping
  stochastic block models},}\ }}\href {\doibase 10.1214/14-EJS903} {\bibfield
  {journal} {\bibinfo  {journal} {Electronic Journal of Statistics}\ }\textbf
  {\bibinfo {volume} {8}},\ \bibinfo {pages} {762--794} (\bibinfo {year}
  {2014})}\BibitemShut {NoStop}%
\bibitem [{\citenamefont {Peixoto}(2013)}]{peixoto_parsimonious_2013}%
  \BibitemOpen
  \bibfield  {author} {\bibinfo {author} {\bibfnamefont {Tiago~P.}\
  \bibnamefont {Peixoto}},\ }\bibfield  {title} {\enquote {\bibinfo {title}
  {Parsimonious module inference in large networks},}\ }\href {\doibase
  10.1103/PhysRevLett.110.148701} {\bibfield  {journal} {\bibinfo  {journal}
  {Physical Review Letters}\ }\textbf {\bibinfo {volume} {110}},\ \bibinfo
  {pages} {148701} (\bibinfo {year} {2013})}\BibitemShut {NoStop}%
\bibitem [{\citenamefont {Jaynes}(2003)}]{jaynes_probability_2003}%
  \BibitemOpen
  \bibfield  {author} {\bibinfo {author} {\bibfnamefont {E.~T.}\ \bibnamefont
  {Jaynes}},\ }\href@noop {} {{\selectlanguage {english}\emph {\bibinfo {title}
  {Probability Theory: The Logic of Science}}}},\ edited by\ \bibinfo {editor}
  {\bibfnamefont {G.~Larry}\ \bibnamefont {Bretthorst}}\ (\bibinfo  {publisher}
  {Cambridge University Press},\ \bibinfo {address} {Cambridge, {UK} ; New
  York, {NY}},\ \bibinfo {year} {2003})\BibitemShut {NoStop}%
\bibitem [{\citenamefont {Jeffreys}(1998)}]{jeffreys_theory_1998}%
  \BibitemOpen
  \bibfield  {author} {\bibinfo {author} {\bibfnamefont {Sir~Harold}\
  \bibnamefont {Jeffreys}},\ }\href@noop {} {\emph {\bibinfo {title} {The
  Theory of Probability}}}\ (\bibinfo  {publisher} {Oxford University Press},\
  \bibinfo {year} {1998})\BibitemShut {NoStop}%
\bibitem [{\citenamefont {Adamic}\ and\ \citenamefont
  {Glance}(2005)}]{adamic_political_2005}%
  \BibitemOpen
  \bibfield  {author} {\bibinfo {author} {\bibfnamefont {Lada~A.}\ \bibnamefont
  {Adamic}}\ and\ \bibinfo {author} {\bibfnamefont {Natalie}\ \bibnamefont
  {Glance}},\ }\bibfield  {title} {\enquote {\bibinfo {title} {The political
  blogosphere and the 2004 u.s. election: divided they blog},}\ }in\ \href
  {\doibase 10.1145/1134271.1134277} {\emph {\bibinfo {booktitle} {Proceedings
  of the 3rd international workshop on Link discovery}}},\ \bibinfo {series and
  number} {{LinkKDD} '05}\ (\bibinfo  {publisher} {{ACM}},\ \bibinfo {address}
  {New York, {NY}, {USA}},\ \bibinfo {year} {2005})\ pp.\ \bibinfo {pages}
  {36--43}\BibitemShut {NoStop}%
\bibitem [{\citenamefont {Holten}(2006)}]{holten_hierarchical_2006}%
  \BibitemOpen
  \bibfield  {author} {\bibinfo {author} {\bibfnamefont {D.}~\bibnamefont
  {Holten}},\ }\bibfield  {title} {\enquote {\bibinfo {title} {Hierarchical
  edge bundles: Visualization of adjacency relations in hierarchical data},}\
  }\href {\doibase 10.1109/TVCG.2006.147} {\bibfield  {journal} {\bibinfo
  {journal} {{IEEE} Transactions on Visualization and Computer Graphics}\
  }\textbf {\bibinfo {volume} {12}},\ \bibinfo {pages} {741--748} (\bibinfo
  {year} {2006})}\BibitemShut {NoStop}%
\bibitem [{\citenamefont {Mcauley}\ and\ \citenamefont
  {Leskovec}(2014)}]{mcauley_discovering_2014}%
  \BibitemOpen
  \bibfield  {author} {\bibinfo {author} {\bibfnamefont {Julian}\ \bibnamefont
  {Mcauley}}\ and\ \bibinfo {author} {\bibfnamefont {Jure}\ \bibnamefont
  {Leskovec}},\ }\bibfield  {title} {\enquote {\bibinfo {title} {Discovering
  social circles in ego networks},}\ }\href {\doibase 10.1145/2556612}
  {\bibfield  {journal} {\bibinfo  {journal} {{ACM} Trans. Knowl. Discov.
  Data}\ }\textbf {\bibinfo {volume} {8}},\ \bibinfo {pages} {4:1--4:28}
  (\bibinfo {year} {2014})}\BibitemShut {NoStop}%
\bibitem [{\citenamefont {Knuth}(1993)}]{knuth_stanford_1993}%
  \BibitemOpen
  \bibfield  {author} {\bibinfo {author} {\bibfnamefont {Donald~E.}\
  \bibnamefont {Knuth}},\ }\href@noop {} {\emph {\bibinfo {title} {The Stanford
  {GraphBase}: A Platform for Combinatorial Computing}}},\ \bibinfo {edition}
  {1st}\ ed.\ (\bibinfo  {publisher} {Addison-Wesley Professional},\ \bibinfo
  {address} {New York, N.Y. : Reading, Mass},\ \bibinfo {year}
  {1993})\BibitemShut {NoStop}%
\bibitem [{\citenamefont {Evans}(2012)}]{evans_american_2012}%
  \BibitemOpen
  \bibfield  {author} {\bibinfo {author} {\bibfnamefont {T~S}\ \bibnamefont
  {Evans}},\ }\bibfield  {title} {\enquote {\bibinfo {title} {American college
  football network files},}\ }\href {\doibase 10.6084/m9.figshare.93179}
  {\bibfield  {journal} {\bibinfo  {journal} {{FigShare}}\ } (\bibinfo {year}
  {2012}),\ 10.6084/m9.figshare.93179}\BibitemShut {NoStop}%
\bibitem [{\citenamefont {Zachary}(1977)}]{zachary_information_1977}%
  \BibitemOpen
  \bibfield  {author} {\bibinfo {author} {\bibfnamefont {Wayne~W.}\
  \bibnamefont {Zachary}},\ }\bibfield  {title} {\enquote {\bibinfo {title} {An
  information flow model for conflict and fission in small groups},}\ }\href
  {http://www.jstor.org/stable/3629752} {\bibfield  {journal} {\bibinfo
  {journal} {Journal of Anthropological Research}\ }\textbf {\bibinfo {volume}
  {33}},\ \bibinfo {pages} {452--473} (\bibinfo {year} {1977})}\BibitemShut
  {NoStop}%
\bibitem [{\citenamefont {Leskovec}\ \emph {et~al.}(2008)\citenamefont
  {Leskovec}, \citenamefont {Lang}, \citenamefont {Dasgupta},\ and\
  \citenamefont {Mahoney}}]{leskovec_community_2008}%
  \BibitemOpen
  \bibfield  {author} {\bibinfo {author} {\bibfnamefont {Jure}\ \bibnamefont
  {Leskovec}}, \bibinfo {author} {\bibfnamefont {Kevin~J.}\ \bibnamefont
  {Lang}}, \bibinfo {author} {\bibfnamefont {Anirban}\ \bibnamefont
  {Dasgupta}}, \ and\ \bibinfo {author} {\bibfnamefont {Michael~W.}\
  \bibnamefont {Mahoney}},\ }\bibfield  {title} {\enquote {\bibinfo {title}
  {Community structure in large networks: Natural cluster sizes and the absence
  of large well-defined clusters},}\ }\href {http://arxiv.org/abs/0810.1355}
  {\bibfield  {journal} {\bibinfo  {journal} {{arXiv}:0810.1355}\ } (\bibinfo
  {year} {2008})}\BibitemShut {NoStop}%
\bibitem [{\citenamefont {Klimt}\ and\ \citenamefont
  {Yang}(2004)}]{klimt_introducing_2004}%
  \BibitemOpen
  \bibfield  {author} {\bibinfo {author} {\bibfnamefont {Bryan}\ \bibnamefont
  {Klimt}}\ and\ \bibinfo {author} {\bibfnamefont {Yiming}\ \bibnamefont
  {Yang}},\ }\bibfield  {title} {\enquote {\bibinfo {title} {Introducing the
  enron corpus.}}\ }in\ \href {http://bklimt.com/papers/2004_klimt_ceas.pdf}
  {\emph {\bibinfo {booktitle} {{CEAS}}}}\ (\bibinfo {year} {2004})\BibitemShut
  {NoStop}%
\bibitem [{\citenamefont {Lusseau}\ \emph {et~al.}(2003)\citenamefont
  {Lusseau}, \citenamefont {Schneider}, \citenamefont {Boisseau}, \citenamefont
  {Haase}, \citenamefont {Slooten},\ and\ \citenamefont
  {Dawson}}]{lusseau_bottlenose_2003}%
  \BibitemOpen
  \bibfield  {author} {\bibinfo {author} {\bibfnamefont {David}\ \bibnamefont
  {Lusseau}}, \bibinfo {author} {\bibfnamefont {Karsten}\ \bibnamefont
  {Schneider}}, \bibinfo {author} {\bibfnamefont {Oliver~J.}\ \bibnamefont
  {Boisseau}}, \bibinfo {author} {\bibfnamefont {Patti}\ \bibnamefont {Haase}},
  \bibinfo {author} {\bibfnamefont {Elisabeth}\ \bibnamefont {Slooten}}, \ and\
  \bibinfo {author} {\bibfnamefont {Steve~M.}\ \bibnamefont {Dawson}},\
  }\bibfield  {title} {{\selectlanguage {english}\enquote {\bibinfo {title}
  {The bottlenose dolphin community of doubtful sound features a large
  proportion of long-lasting associations},}\ }}\href {\doibase
  10.1007/s00265-003-0651-y} {\bibfield  {journal} {\bibinfo  {journal}
  {Behavioral Ecology and Sociobiology}\ }\textbf {\bibinfo {volume} {54}},\
  \bibinfo {pages} {396--405} (\bibinfo {year} {2003})}\BibitemShut {NoStop}%
\bibitem [{\citenamefont {Richters}\ and\ \citenamefont
  {Peixoto}(2011)}]{richters_trust_2011}%
  \BibitemOpen
  \bibfield  {author} {\bibinfo {author} {\bibfnamefont {Oliver}\ \bibnamefont
  {Richters}}\ and\ \bibinfo {author} {\bibfnamefont {Tiago~P.}\ \bibnamefont
  {Peixoto}},\ }\bibfield  {title} {\enquote {\bibinfo {title} {Trust
  transitivity in social networks},}\ }\href {\doibase
  10.1371/journal.pone.0018384} {\bibfield  {journal} {\bibinfo  {journal}
  {{PLoS} {ONE}}\ }\textbf {\bibinfo {volume} {6}},\ \bibinfo {pages} {e18384}
  (\bibinfo {year} {2011})}\BibitemShut {NoStop}%
\bibitem [{\citenamefont {Cho}\ \emph {et~al.}(2011)\citenamefont {Cho},
  \citenamefont {Myers},\ and\ \citenamefont {Leskovec}}]{cho_friendship_2011}%
  \BibitemOpen
  \bibfield  {author} {\bibinfo {author} {\bibfnamefont {Eunjoon}\ \bibnamefont
  {Cho}}, \bibinfo {author} {\bibfnamefont {Seth~A.}\ \bibnamefont {Myers}}, \
  and\ \bibinfo {author} {\bibfnamefont {Jure}\ \bibnamefont {Leskovec}},\
  }\bibfield  {title} {\enquote {\bibinfo {title} {Friendship and mobility:
  user movement in location-based social networks},}\ }in\ \href {\doibase
  10.1145/2020408.2020579} {\emph {\bibinfo {booktitle} {Proceedings of the
  17th {ACM} {SIGKDD} international conference on Knowledge discovery and data
  mining}}},\ \bibinfo {series and number} {{KDD} '11}\ (\bibinfo  {publisher}
  {{ACM}},\ \bibinfo {address} {New York, {NY}, {USA}},\ \bibinfo {year}
  {2011})\ pp.\ \bibinfo {pages} {1082--1090}\BibitemShut {NoStop}%
\bibitem [{\citenamefont {Watts}\ and\ \citenamefont
  {Strogatz}(1998)}]{watts_collective_1998}%
  \BibitemOpen
  \bibfield  {author} {\bibinfo {author} {\bibfnamefont {D.~J.}\ \bibnamefont
  {Watts}}\ and\ \bibinfo {author} {\bibfnamefont {S.~H.}\ \bibnamefont
  {Strogatz}},\ }\bibfield  {title} {\enquote {\bibinfo {title} {Collective
  dynamics of 'small-world' networks},}\ }\href@noop {} {\bibfield  {journal}
  {\bibinfo  {journal} {Nature}\ }\textbf {\bibinfo {volume} {393}},\ \bibinfo
  {pages} {409--10} (\bibinfo {year} {1998})}\BibitemShut {NoStop}%
\bibitem [{\citenamefont {Leskovec}\ \emph {et~al.}(2007)\citenamefont
  {Leskovec}, \citenamefont {Kleinberg},\ and\ \citenamefont
  {Faloutsos}}]{leskovec_graph_2007}%
  \BibitemOpen
  \bibfield  {author} {\bibinfo {author} {\bibfnamefont {Jure}\ \bibnamefont
  {Leskovec}}, \bibinfo {author} {\bibfnamefont {Jon}\ \bibnamefont
  {Kleinberg}}, \ and\ \bibinfo {author} {\bibfnamefont {Christos}\
  \bibnamefont {Faloutsos}},\ }\bibfield  {title} {\enquote {\bibinfo {title}
  {Graph evolution: Densification and shrinking diameters},}\ }\href {\doibase
  10.1145/1217299.1217301} {\bibfield  {journal} {\bibinfo  {journal} {{ACM}
  Trans. Knowl. Discov. Data}\ }\textbf {\bibinfo {volume} {1}} (\bibinfo
  {year} {2007}),\ 10.1145/1217299.1217301}\BibitemShut {NoStop}%
\bibitem [{\citenamefont {Newman}(2006{\natexlab{b}})}]{newman_finding_2006}%
  \BibitemOpen
  \bibfield  {author} {\bibinfo {author} {\bibfnamefont {M.~E.~J.}\
  \bibnamefont {Newman}},\ }\bibfield  {title} {\enquote {\bibinfo {title}
  {Finding community structure in networks using the eigenvectors of
  matrices},}\ }\href {\doibase 10.1103/PhysRevE.74.036104} {\bibfield
  {journal} {\bibinfo  {journal} {Physical Review E}\ }\textbf {\bibinfo
  {volume} {74}},\ \bibinfo {pages} {036104} (\bibinfo {year}
  {2006}{\natexlab{b}})}\BibitemShut {NoStop}%
\bibitem [{\citenamefont {Richardson}\ \emph {et~al.}(2003)\citenamefont
  {Richardson}, \citenamefont {Agrawal},\ and\ \citenamefont
  {Domingos}}]{richardson_trust_2003}%
  \BibitemOpen
  \bibfield  {author} {\bibinfo {author} {\bibfnamefont {Matthew}\ \bibnamefont
  {Richardson}}, \bibinfo {author} {\bibfnamefont {Rakesh}\ \bibnamefont
  {Agrawal}}, \ and\ \bibinfo {author} {\bibfnamefont {Pedro}\ \bibnamefont
  {Domingos}},\ }\bibfield  {title} {\enquote {\bibinfo {title} {Trust
  management for the semantic web},}\ }in\ \href
  {http://link.springer.com/chapter/10.1007/978-3-540-39718-2_23} {\emph
  {\bibinfo {booktitle} {The Semantic Web - {ISWC} 2003}}},\ \bibinfo {series
  and number} {\bibinfo {series} {Lecture Notes in Computer Science}\ No.\
  \bibinfo {number} {2870}},\ \bibinfo {editor} {edited by\ \bibinfo {editor}
  {\bibfnamefont {Dieter}\ \bibnamefont {Fensel}}, \bibinfo {editor}
  {\bibfnamefont {Katia}\ \bibnamefont {Sycara}}, \ and\ \bibinfo {editor}
  {\bibfnamefont {John}\ \bibnamefont {Mylopoulos}}}\ (\bibinfo  {publisher}
  {Springer Berlin Heidelberg},\ \bibinfo {year} {2003})\ pp.\ \bibinfo {pages}
  {351--368}\BibitemShut {NoStop}%
\bibitem [{\citenamefont {Goh}\ \emph {et~al.}(2007)\citenamefont {Goh},
  \citenamefont {Cusick}, \citenamefont {Valle}, \citenamefont {Childs},
  \citenamefont {Vidal},\ and\ \citenamefont {Barabási}}]{goh_human_2007}%
  \BibitemOpen
  \bibfield  {author} {\bibinfo {author} {\bibfnamefont {K.~I.}\ \bibnamefont
  {Goh}}, \bibinfo {author} {\bibfnamefont {M.~E.}\ \bibnamefont {Cusick}},
  \bibinfo {author} {\bibfnamefont {D.}~\bibnamefont {Valle}}, \bibinfo
  {author} {\bibfnamefont {B.}~\bibnamefont {Childs}}, \bibinfo {author}
  {\bibfnamefont {M.}~\bibnamefont {Vidal}}, \ and\ \bibinfo {author}
  {\bibfnamefont {A.~L.}\ \bibnamefont {Barabási}},\ }\bibfield  {title}
  {\enquote {\bibinfo {title} {The human disease network},}\ }\href
  {http://www.pnas.org/content/104/21/8685.short} {\bibfield  {journal}
  {\bibinfo  {journal} {Proceedings of the National Academy of Sciences}\
  }\textbf {\bibinfo {volume} {104}},\ \bibinfo {pages} {8685} (\bibinfo {year}
  {2007})}\BibitemShut {NoStop}%
\bibitem [{\citenamefont {Yu}\ \emph {et~al.}(2008)\citenamefont {Yu},
  \citenamefont {Braun}, \citenamefont {Yıldırım}, \citenamefont {Lemmens},
  \citenamefont {Venkatesan}, \citenamefont {Sahalie}, \citenamefont
  {Hirozane-Kishikawa}, \citenamefont {Gebreab}, \citenamefont {Li},
  \citenamefont {Simonis}, \citenamefont {Hao}, \citenamefont {Rual},
  \citenamefont {Dricot}, \citenamefont {Vazquez}, \citenamefont {Murray},
  \citenamefont {Simon}, \citenamefont {Tardivo}, \citenamefont {Tam},
  \citenamefont {Svrzikapa}, \citenamefont {Fan}, \citenamefont {Smet},
  \citenamefont {Motyl}, \citenamefont {Hudson}, \citenamefont {Park},
  \citenamefont {Xin}, \citenamefont {Cusick}, \citenamefont {Moore},
  \citenamefont {Boone}, \citenamefont {Snyder}, \citenamefont {Roth},
  \citenamefont {Barabási}, \citenamefont {Tavernier}, \citenamefont {Hill},\
  and\ \citenamefont {Vidal}}]{yu_high-quality_2008}%
  \BibitemOpen
  \bibfield  {author} {\bibinfo {author} {\bibfnamefont {Haiyuan}\ \bibnamefont
  {Yu}}, \bibinfo {author} {\bibfnamefont {Pascal}\ \bibnamefont {Braun}},
  \bibinfo {author} {\bibfnamefont {Muhammed~A.}\ \bibnamefont {Yıldırım}},
  \bibinfo {author} {\bibfnamefont {Irma}\ \bibnamefont {Lemmens}}, \bibinfo
  {author} {\bibfnamefont {Kavitha}\ \bibnamefont {Venkatesan}}, \bibinfo
  {author} {\bibfnamefont {Julie}\ \bibnamefont {Sahalie}}, \bibinfo {author}
  {\bibfnamefont {Tomoko}\ \bibnamefont {Hirozane-Kishikawa}}, \bibinfo
  {author} {\bibfnamefont {Fana}\ \bibnamefont {Gebreab}}, \bibinfo {author}
  {\bibfnamefont {Na}~\bibnamefont {Li}}, \bibinfo {author} {\bibfnamefont
  {Nicolas}\ \bibnamefont {Simonis}}, \bibinfo {author} {\bibfnamefont {Tong}\
  \bibnamefont {Hao}}, \bibinfo {author} {\bibfnamefont {Jean-François}\
  \bibnamefont {Rual}}, \bibinfo {author} {\bibfnamefont {Amélie}\
  \bibnamefont {Dricot}}, \bibinfo {author} {\bibfnamefont {Alexei}\
  \bibnamefont {Vazquez}}, \bibinfo {author} {\bibfnamefont {Ryan~R.}\
  \bibnamefont {Murray}}, \bibinfo {author} {\bibfnamefont {Christophe}\
  \bibnamefont {Simon}}, \bibinfo {author} {\bibfnamefont {Leah}\ \bibnamefont
  {Tardivo}}, \bibinfo {author} {\bibfnamefont {Stanley}\ \bibnamefont {Tam}},
  \bibinfo {author} {\bibfnamefont {Nenad}\ \bibnamefont {Svrzikapa}}, \bibinfo
  {author} {\bibfnamefont {Changyu}\ \bibnamefont {Fan}}, \bibinfo {author}
  {\bibfnamefont {Anne-Sophie~de}\ \bibnamefont {Smet}}, \bibinfo {author}
  {\bibfnamefont {Adriana}\ \bibnamefont {Motyl}}, \bibinfo {author}
  {\bibfnamefont {Michael~E.}\ \bibnamefont {Hudson}}, \bibinfo {author}
  {\bibfnamefont {Juyong}\ \bibnamefont {Park}}, \bibinfo {author}
  {\bibfnamefont {Xiaofeng}\ \bibnamefont {Xin}}, \bibinfo {author}
  {\bibfnamefont {Michael~E.}\ \bibnamefont {Cusick}}, \bibinfo {author}
  {\bibfnamefont {Troy}\ \bibnamefont {Moore}}, \bibinfo {author}
  {\bibfnamefont {Charlie}\ \bibnamefont {Boone}}, \bibinfo {author}
  {\bibfnamefont {Michael}\ \bibnamefont {Snyder}}, \bibinfo {author}
  {\bibfnamefont {Frederick~P.}\ \bibnamefont {Roth}}, \bibinfo {author}
  {\bibfnamefont {Albert-László}\ \bibnamefont {Barabási}}, \bibinfo
  {author} {\bibfnamefont {Jan}\ \bibnamefont {Tavernier}}, \bibinfo {author}
  {\bibfnamefont {David~E.}\ \bibnamefont {Hill}}, \ and\ \bibinfo {author}
  {\bibfnamefont {Marc}\ \bibnamefont {Vidal}},\ }\bibfield  {title}
  {{\selectlanguage {english}\enquote {\bibinfo {title} {High-quality binary
  protein interaction map of the yeast interactome network},}\ }}\href
  {\doibase 10.1126/science.1158684} {\bibfield  {journal} {\bibinfo  {journal}
  {Science}\ }\textbf {\bibinfo {volume} {322}},\ \bibinfo {pages} {104--110}
  (\bibinfo {year} {2008})}\BibitemShut {NoStop}%
\bibitem [{\citenamefont {Reguly}\ \emph {et~al.}(2006)\citenamefont {Reguly},
  \citenamefont {Breitkreutz}, \citenamefont {Boucher}, \citenamefont
  {Breitkreutz}, \citenamefont {Hon}, \citenamefont {Myers}, \citenamefont
  {Parsons}, \citenamefont {Friesen}, \citenamefont {Oughtred}, \citenamefont
  {Tong}, \citenamefont {Stark}, \citenamefont {Ho}, \citenamefont {Botstein},
  \citenamefont {Andrews}, \citenamefont {Boone}, \citenamefont {Troyanskya},
  \citenamefont {Ideker}, \citenamefont {Dolinski}, \citenamefont {Batada},\
  and\ \citenamefont {Tyers}}]{reguly_comprehensive_2006}%
  \BibitemOpen
  \bibfield  {author} {\bibinfo {author} {\bibfnamefont {Teresa}\ \bibnamefont
  {Reguly}}, \bibinfo {author} {\bibfnamefont {Ashton}\ \bibnamefont
  {Breitkreutz}}, \bibinfo {author} {\bibfnamefont {Lorrie}\ \bibnamefont
  {Boucher}}, \bibinfo {author} {\bibfnamefont {Bobby-Joe}\ \bibnamefont
  {Breitkreutz}}, \bibinfo {author} {\bibfnamefont {Gary~C.}\ \bibnamefont
  {Hon}}, \bibinfo {author} {\bibfnamefont {Chad~L.}\ \bibnamefont {Myers}},
  \bibinfo {author} {\bibfnamefont {Ainslie}\ \bibnamefont {Parsons}}, \bibinfo
  {author} {\bibfnamefont {Helena}\ \bibnamefont {Friesen}}, \bibinfo {author}
  {\bibfnamefont {Rose}\ \bibnamefont {Oughtred}}, \bibinfo {author}
  {\bibfnamefont {Amy}\ \bibnamefont {Tong}}, \bibinfo {author} {\bibfnamefont
  {Chris}\ \bibnamefont {Stark}}, \bibinfo {author} {\bibfnamefont {Yuen}\
  \bibnamefont {Ho}}, \bibinfo {author} {\bibfnamefont {David}\ \bibnamefont
  {Botstein}}, \bibinfo {author} {\bibfnamefont {Brenda}\ \bibnamefont
  {Andrews}}, \bibinfo {author} {\bibfnamefont {Charles}\ \bibnamefont
  {Boone}}, \bibinfo {author} {\bibfnamefont {Olga~G.}\ \bibnamefont
  {Troyanskya}}, \bibinfo {author} {\bibfnamefont {Trey}\ \bibnamefont
  {Ideker}}, \bibinfo {author} {\bibfnamefont {Kara}\ \bibnamefont {Dolinski}},
  \bibinfo {author} {\bibfnamefont {Nizar~N.}\ \bibnamefont {Batada}}, \ and\
  \bibinfo {author} {\bibfnamefont {Mike}\ \bibnamefont {Tyers}},\ }\bibfield
  {title} {{\selectlanguage {english}\enquote {\bibinfo {title} {Comprehensive
  curation and analysis of global interaction networks in saccharomyces
  cerevisiae},}\ }}\href {\doibase 10.1186/jbiol36} {\bibfield  {journal}
  {\bibinfo  {journal} {Journal of Biology}\ }\textbf {\bibinfo {volume} {5}},\
  \bibinfo {pages} {11} (\bibinfo {year} {2006})}\BibitemShut {NoStop}%
\bibitem [{\citenamefont {Collins}\ \emph {et~al.}(2007)\citenamefont
  {Collins}, \citenamefont {Kemmeren}, \citenamefont {Zhao}, \citenamefont
  {Greenblatt}, \citenamefont {Spencer}, \citenamefont {Holstege},
  \citenamefont {Weissman},\ and\ \citenamefont
  {Krogan}}]{collins_toward_2007}%
  \BibitemOpen
  \bibfield  {author} {\bibinfo {author} {\bibfnamefont {Sean~R.}\ \bibnamefont
  {Collins}}, \bibinfo {author} {\bibfnamefont {Patrick}\ \bibnamefont
  {Kemmeren}}, \bibinfo {author} {\bibfnamefont {Xue-Chu}\ \bibnamefont
  {Zhao}}, \bibinfo {author} {\bibfnamefont {Jack~F.}\ \bibnamefont
  {Greenblatt}}, \bibinfo {author} {\bibfnamefont {Forrest}\ \bibnamefont
  {Spencer}}, \bibinfo {author} {\bibfnamefont {Frank C.~P.}\ \bibnamefont
  {Holstege}}, \bibinfo {author} {\bibfnamefont {Jonathan~S.}\ \bibnamefont
  {Weissman}}, \ and\ \bibinfo {author} {\bibfnamefont {Nevan~J.}\ \bibnamefont
  {Krogan}},\ }\bibfield  {title} {{\selectlanguage {english}\enquote {\bibinfo
  {title} {Toward a comprehensive atlas of the physical interactome of
  saccharomyces cerevisiae},}\ }}\href {\doibase 10.1074/mcp.M600381-MCP200}
  {\bibfield  {journal} {\bibinfo  {journal} {Molecular \& cellular proteomics:
  {MCP}}\ }\textbf {\bibinfo {volume} {6}},\ \bibinfo {pages} {439--450}
  (\bibinfo {year} {2007})}\BibitemShut {NoStop}%
\bibitem [{\citenamefont {Salgado}\ \emph {et~al.}(2013)\citenamefont
  {Salgado}, \citenamefont {Peralta-Gil}, \citenamefont {Gama-Castro},
  \citenamefont {Santos-Zavaleta}, \citenamefont {Muñiz-Rascado},
  \citenamefont {García-Sotelo}, \citenamefont {Weiss}, \citenamefont
  {Solano-Lira}, \citenamefont {Martínez-Flores}, \citenamefont
  {Medina-Rivera}, \citenamefont {Salgado-Osorio}, \citenamefont
  {Alquicira-Hernández}, \citenamefont {Alquicira-Hernández}, \citenamefont
  {López-Fuentes}, \citenamefont {Porrón-Sotelo}, \citenamefont {Huerta},
  \citenamefont {Bonavides-Martínez}, \citenamefont {Balderas-Martínez},
  \citenamefont {Pannier}, \citenamefont {Olvera}, \citenamefont {Labastida},
  \citenamefont {Jiménez-Jacinto}, \citenamefont {Vega-Alvarado},
  \citenamefont {Del Moral-Chávez}, \citenamefont {Hernández-Alvarez},
  \citenamefont {Morett},\ and\ \citenamefont
  {Collado-Vides}}]{salgado_regulondb_2013}%
  \BibitemOpen
  \bibfield  {author} {\bibinfo {author} {\bibfnamefont {Heladia}\ \bibnamefont
  {Salgado}}, \bibinfo {author} {\bibfnamefont {Martin}\ \bibnamefont
  {Peralta-Gil}}, \bibinfo {author} {\bibfnamefont {Socorro}\ \bibnamefont
  {Gama-Castro}}, \bibinfo {author} {\bibfnamefont {Alberto}\ \bibnamefont
  {Santos-Zavaleta}}, \bibinfo {author} {\bibfnamefont {Luis}\ \bibnamefont
  {Muñiz-Rascado}}, \bibinfo {author} {\bibfnamefont {Jair~S.}\ \bibnamefont
  {García-Sotelo}}, \bibinfo {author} {\bibfnamefont {Verena}\ \bibnamefont
  {Weiss}}, \bibinfo {author} {\bibfnamefont {Hilda}\ \bibnamefont
  {Solano-Lira}}, \bibinfo {author} {\bibfnamefont {Irma}\ \bibnamefont
  {Martínez-Flores}}, \bibinfo {author} {\bibfnamefont {Alejandra}\
  \bibnamefont {Medina-Rivera}}, \bibinfo {author} {\bibfnamefont {Gerardo}\
  \bibnamefont {Salgado-Osorio}}, \bibinfo {author} {\bibfnamefont {Shirley}\
  \bibnamefont {Alquicira-Hernández}}, \bibinfo {author} {\bibfnamefont
  {Kevin}\ \bibnamefont {Alquicira-Hernández}}, \bibinfo {author}
  {\bibfnamefont {Alejandra}\ \bibnamefont {López-Fuentes}}, \bibinfo {author}
  {\bibfnamefont {Liliana}\ \bibnamefont {Porrón-Sotelo}}, \bibinfo {author}
  {\bibfnamefont {Araceli~M.}\ \bibnamefont {Huerta}}, \bibinfo {author}
  {\bibfnamefont {César}\ \bibnamefont {Bonavides-Martínez}}, \bibinfo
  {author} {\bibfnamefont {Yalbi~I.}\ \bibnamefont {Balderas-Martínez}},
  \bibinfo {author} {\bibfnamefont {Lucia}\ \bibnamefont {Pannier}}, \bibinfo
  {author} {\bibfnamefont {Maricela}\ \bibnamefont {Olvera}}, \bibinfo {author}
  {\bibfnamefont {Aurora}\ \bibnamefont {Labastida}}, \bibinfo {author}
  {\bibfnamefont {Verónica}\ \bibnamefont {Jiménez-Jacinto}}, \bibinfo
  {author} {\bibfnamefont {Leticia}\ \bibnamefont {Vega-Alvarado}}, \bibinfo
  {author} {\bibfnamefont {Victor}\ \bibnamefont {Del Moral-Chávez}}, \bibinfo
  {author} {\bibfnamefont {Alfredo}\ \bibnamefont {Hernández-Alvarez}},
  \bibinfo {author} {\bibfnamefont {Enrique}\ \bibnamefont {Morett}}, \ and\
  \bibinfo {author} {\bibfnamefont {Julio}\ \bibnamefont {Collado-Vides}},\
  }\bibfield  {title} {{\selectlanguage {english}\enquote {\bibinfo {title}
  {{RegulonDB} v8.0: omics data sets, evolutionary conservation, regulatory
  phrases, cross-validated gold standards and more},}\ }}\href {\doibase
  10.1093/nar/gks1201} {\bibfield  {journal} {\bibinfo  {journal} {Nucleic
  Acids Research}\ }\textbf {\bibinfo {volume} {41}},\ \bibinfo {pages}
  {D203--213} (\bibinfo {year} {2013})}\BibitemShut {NoStop}%
\bibitem [{\citenamefont {McAuley}\ and\ \citenamefont
  {Leskovec}(2012)}]{mcauley_image_2012}%
  \BibitemOpen
  \bibfield  {author} {\bibinfo {author} {\bibfnamefont {Julian}\ \bibnamefont
  {McAuley}}\ and\ \bibinfo {author} {\bibfnamefont {Jure}\ \bibnamefont
  {Leskovec}},\ }\bibfield  {title} {\enquote {\bibinfo {title} {Image labeling
  on a network: Using social-network metadata for image classification},}\ }in\
  \href {http://link.springer.com/chapter/10.1007/978-3-642-33765-9_59} {\emph
  {\bibinfo {booktitle} {Computer Vision – {ECCV} 2012}}},\ \bibinfo {series
  and number} {\bibinfo {series} {Lecture Notes in Computer Science}\ No.\
  \bibinfo {number} {7575}},\ \bibinfo {editor} {edited by\ \bibinfo {editor}
  {\bibfnamefont {Andrew}\ \bibnamefont {Fitzgibbon}}, \bibinfo {editor}
  {\bibfnamefont {Svetlana}\ \bibnamefont {Lazebnik}}, \bibinfo {editor}
  {\bibfnamefont {Pietro}\ \bibnamefont {Perona}}, \bibinfo {editor}
  {\bibfnamefont {Yoichi}\ \bibnamefont {Sato}}, \ and\ \bibinfo {editor}
  {\bibfnamefont {Cordelia}\ \bibnamefont {Schmid}}}\ (\bibinfo  {publisher}
  {Springer Berlin Heidelberg},\ \bibinfo {year} {2012})\ pp.\ \bibinfo {pages}
  {828--841}\BibitemShut {NoStop}%
\bibitem [{\citenamefont {Leskovec}\ \emph {et~al.}(2009)\citenamefont
  {Leskovec}, \citenamefont {Lang}, \citenamefont {Dasgupta},\ and\
  \citenamefont {Mahoney}}]{leskovec_community_2009}%
  \BibitemOpen
  \bibfield  {author} {\bibinfo {author} {\bibfnamefont {Jure}\ \bibnamefont
  {Leskovec}}, \bibinfo {author} {\bibfnamefont {Kevin~J.}\ \bibnamefont
  {Lang}}, \bibinfo {author} {\bibfnamefont {Anirban}\ \bibnamefont
  {Dasgupta}}, \ and\ \bibinfo {author} {\bibfnamefont {Michael~W.}\
  \bibnamefont {Mahoney}},\ }\bibfield  {title} {\enquote {\bibinfo {title}
  {Community structure in large networks: Natural cluster sizes and the absence
  of large well-defined clusters},}\ }\href
  {http://www.tandfonline.com/doi/abs/10.1080/15427951.2009.10129177}
  {\bibfield  {journal} {\bibinfo  {journal} {Internet Mathematics}\ }\textbf
  {\bibinfo {volume} {6}},\ \bibinfo {pages} {29--123} (\bibinfo {year}
  {2009})}\BibitemShut {NoStop}%
\bibitem [{\citenamefont {Yang}\ and\ \citenamefont
  {Leskovec}(2012{\natexlab{a}})}]{yang_defining_2012}%
  \BibitemOpen
  \bibfield  {author} {\bibinfo {author} {\bibfnamefont {Jaewon}\ \bibnamefont
  {Yang}}\ and\ \bibinfo {author} {\bibfnamefont {Jure}\ \bibnamefont
  {Leskovec}},\ }\bibfield  {title} {\enquote {\bibinfo {title} {Defining and
  evaluating network communities based on ground-truth},}\ }in\ \href {\doibase
  10.1145/2350190.2350193} {\emph {\bibinfo {booktitle} {Proceedings of the
  {ACM} {SIGKDD} Workshop on Mining Data Semantics}}},\ \bibinfo {series and
  number} {{MDS} '12}\ (\bibinfo  {publisher} {{ACM}},\ \bibinfo {address} {New
  York, {NY}, {USA}},\ \bibinfo {year} {2012})\ pp.\ \bibinfo {pages}
  {3:1--3:8}\BibitemShut {NoStop}%
\bibitem [{\citenamefont {Albert}\ \emph {et~al.}(1999)\citenamefont {Albert},
  \citenamefont {Jeong},\ and\ \citenamefont
  {Barabási}}]{albert_internet:_1999}%
  \BibitemOpen
  \bibfield  {author} {\bibinfo {author} {\bibfnamefont {Réka}\ \bibnamefont
  {Albert}}, \bibinfo {author} {\bibfnamefont {Hawoong}\ \bibnamefont {Jeong}},
  \ and\ \bibinfo {author} {\bibfnamefont {Albert-László}\ \bibnamefont
  {Barabási}},\ }\bibfield  {title} {{\selectlanguage {english}\enquote
  {\bibinfo {title} {Internet: Diameter of the world-wide web},}\ }}\href
  {\doibase 10.1038/43601} {\bibfield  {journal} {\bibinfo  {journal} {Nature}\
  }\textbf {\bibinfo {volume} {401}},\ \bibinfo {pages} {130--131} (\bibinfo
  {year} {1999})}\BibitemShut {NoStop}%
\bibitem [{\citenamefont {Leskovec}\ \emph
  {et~al.}(2010{\natexlab{b}})\citenamefont {Leskovec}, \citenamefont
  {Huttenlocher},\ and\ \citenamefont {Kleinberg}}]{leskovec_signed_2010}%
  \BibitemOpen
  \bibfield  {author} {\bibinfo {author} {\bibfnamefont {Jure}\ \bibnamefont
  {Leskovec}}, \bibinfo {author} {\bibfnamefont {Daniel}\ \bibnamefont
  {Huttenlocher}}, \ and\ \bibinfo {author} {\bibfnamefont {Jon}\ \bibnamefont
  {Kleinberg}},\ }\bibfield  {title} {\enquote {\bibinfo {title} {Signed
  networks in social media},}\ }in\ \href {\doibase 10.1145/1753326.1753532}
  {\emph {\bibinfo {booktitle} {Proceedings of the {SIGCHI} Conference on Human
  Factors in Computing Systems}}},\ \bibinfo {series and number} {{CHI} '10}\
  (\bibinfo  {publisher} {{ACM}},\ \bibinfo {address} {New York, {NY}, {USA}},\
  \bibinfo {year} {2010})\ pp.\ \bibinfo {pages} {1361--1370}\BibitemShut
  {NoStop}%
\bibitem [{\citenamefont {Leskovec}\ \emph
  {et~al.}(2010{\natexlab{c}})\citenamefont {Leskovec}, \citenamefont
  {Huttenlocher},\ and\ \citenamefont {Kleinberg}}]{leskovec_predicting_2010}%
  \BibitemOpen
  \bibfield  {author} {\bibinfo {author} {\bibfnamefont {Jure}\ \bibnamefont
  {Leskovec}}, \bibinfo {author} {\bibfnamefont {Daniel}\ \bibnamefont
  {Huttenlocher}}, \ and\ \bibinfo {author} {\bibfnamefont {Jon}\ \bibnamefont
  {Kleinberg}},\ }\bibfield  {title} {\enquote {\bibinfo {title} {Predicting
  positive and negative links in online social networks},}\ }in\ \href
  {\doibase 10.1145/1772690.1772756} {\emph {\bibinfo {booktitle} {Proceedings
  of the 19th International Conference on World Wide Web}}},\ \bibinfo {series
  and number} {{WWW} '10}\ (\bibinfo  {publisher} {{ACM}},\ \bibinfo {address}
  {New York, {NY}, {USA}},\ \bibinfo {year} {2010})\ pp.\ \bibinfo {pages}
  {641--650}\BibitemShut {NoStop}%
\bibitem [{\citenamefont {Prasad}\ \emph {et~al.}(2009)\citenamefont {Prasad},
  \citenamefont {Goel}, \citenamefont {Kandasamy}, \citenamefont
  {Keerthikumar}, \citenamefont {Kumar}, \citenamefont {Mathivanan},
  \citenamefont {Telikicherla}, \citenamefont {Raju}, \citenamefont {Shafreen},
  \citenamefont {Venugopal}, \citenamefont {Balakrishnan}, \citenamefont
  {Marimuthu}, \citenamefont {Banerjee}, \citenamefont {Somanathan},
  \citenamefont {Sebastian}, \citenamefont {Rani}, \citenamefont {Ray},
  \citenamefont {Kishore}, \citenamefont {Kanth}, \citenamefont {Ahmed},
  \citenamefont {Kashyap}, \citenamefont {Mohmood}, \citenamefont
  {Ramachandra}, \citenamefont {Krishna}, \citenamefont {Rahiman},
  \citenamefont {Mohan}, \citenamefont {Ranganathan}, \citenamefont
  {Ramabadran}, \citenamefont {Chaerkady},\ and\ \citenamefont
  {Pandey}}]{prasad_human_2009}%
  \BibitemOpen
  \bibfield  {author} {\bibinfo {author} {\bibfnamefont {T.~S.~Keshava}\
  \bibnamefont {Prasad}}, \bibinfo {author} {\bibfnamefont {Renu}\ \bibnamefont
  {Goel}}, \bibinfo {author} {\bibfnamefont {Kumaran}\ \bibnamefont
  {Kandasamy}}, \bibinfo {author} {\bibfnamefont {Shivakumar}\ \bibnamefont
  {Keerthikumar}}, \bibinfo {author} {\bibfnamefont {Sameer}\ \bibnamefont
  {Kumar}}, \bibinfo {author} {\bibfnamefont {Suresh}\ \bibnamefont
  {Mathivanan}}, \bibinfo {author} {\bibfnamefont {Deepthi}\ \bibnamefont
  {Telikicherla}}, \bibinfo {author} {\bibfnamefont {Rajesh}\ \bibnamefont
  {Raju}}, \bibinfo {author} {\bibfnamefont {Beema}\ \bibnamefont {Shafreen}},
  \bibinfo {author} {\bibfnamefont {Abhilash}\ \bibnamefont {Venugopal}},
  \bibinfo {author} {\bibfnamefont {Lavanya}\ \bibnamefont {Balakrishnan}},
  \bibinfo {author} {\bibfnamefont {Arivusudar}\ \bibnamefont {Marimuthu}},
  \bibinfo {author} {\bibfnamefont {Sutopa}\ \bibnamefont {Banerjee}}, \bibinfo
  {author} {\bibfnamefont {Devi~S.}\ \bibnamefont {Somanathan}}, \bibinfo
  {author} {\bibfnamefont {Aimy}\ \bibnamefont {Sebastian}}, \bibinfo {author}
  {\bibfnamefont {Sandhya}\ \bibnamefont {Rani}}, \bibinfo {author}
  {\bibfnamefont {Somak}\ \bibnamefont {Ray}}, \bibinfo {author} {\bibfnamefont
  {C.~J.~Harrys}\ \bibnamefont {Kishore}}, \bibinfo {author} {\bibfnamefont
  {Sashi}\ \bibnamefont {Kanth}}, \bibinfo {author} {\bibfnamefont {Mukhtar}\
  \bibnamefont {Ahmed}}, \bibinfo {author} {\bibfnamefont {Manoj~K.}\
  \bibnamefont {Kashyap}}, \bibinfo {author} {\bibfnamefont {Riaz}\
  \bibnamefont {Mohmood}}, \bibinfo {author} {\bibfnamefont {Y.~L.}\
  \bibnamefont {Ramachandra}}, \bibinfo {author} {\bibfnamefont
  {V.}~\bibnamefont {Krishna}}, \bibinfo {author} {\bibfnamefont {B.~Abdul}\
  \bibnamefont {Rahiman}}, \bibinfo {author} {\bibfnamefont {Sujatha}\
  \bibnamefont {Mohan}}, \bibinfo {author} {\bibfnamefont {Prathibha}\
  \bibnamefont {Ranganathan}}, \bibinfo {author} {\bibfnamefont {Subhashri}\
  \bibnamefont {Ramabadran}}, \bibinfo {author} {\bibfnamefont {Raghothama}\
  \bibnamefont {Chaerkady}}, \ and\ \bibinfo {author} {\bibfnamefont
  {Akhilesh}\ \bibnamefont {Pandey}},\ }\bibfield  {title} {{\selectlanguage
  {english}\enquote {\bibinfo {title} {Human protein reference database—2009
  update},}\ }}\href {\doibase 10.1093/nar/gkn892} {\bibfield  {journal}
  {\bibinfo  {journal} {Nucleic Acids Research}\ }\textbf {\bibinfo {volume}
  {37}},\ \bibinfo {pages} {D767--D772} (\bibinfo {year} {2009})}\BibitemShut
  {NoStop}%
\bibitem [{\citenamefont {Newman}(2004)}]{newman_fast_2004}%
  \BibitemOpen
  \bibfield  {author} {\bibinfo {author} {\bibfnamefont {M.~E.~J.}\
  \bibnamefont {Newman}},\ }\bibfield  {title} {\enquote {\bibinfo {title}
  {Fast algorithm for detecting community structure in networks},}\ }\href
  {\doibase 10.1103/PhysRevE.69.066133} {\bibfield  {journal} {\bibinfo
  {journal} {Physical Review E}\ }\textbf {\bibinfo {volume} {69}},\ \bibinfo
  {pages} {066133} (\bibinfo {year} {2004})}\BibitemShut {NoStop}%
\bibitem [{\citenamefont {Clauset}\ \emph {et~al.}(2004)\citenamefont
  {Clauset}, \citenamefont {Newman},\ and\ \citenamefont
  {Moore}}]{clauset_finding_2004}%
  \BibitemOpen
  \bibfield  {author} {\bibinfo {author} {\bibfnamefont {Aaron}\ \bibnamefont
  {Clauset}}, \bibinfo {author} {\bibfnamefont {M.~E.~J.}\ \bibnamefont
  {Newman}}, \ and\ \bibinfo {author} {\bibfnamefont {Cristopher}\ \bibnamefont
  {Moore}},\ }\bibfield  {title} {\enquote {\bibinfo {title} {Finding community
  structure in very large networks},}\ }\href {\doibase
  10.1103/PhysRevE.70.066111} {\bibfield  {journal} {\bibinfo  {journal}
  {Physical Review E}\ }\textbf {\bibinfo {volume} {70}},\ \bibinfo {pages}
  {066111} (\bibinfo {year} {2004})}\BibitemShut {NoStop}%
\bibitem [{\citenamefont {Arenas}\ \emph {et~al.}(2004)\citenamefont {Arenas},
  \citenamefont {Danon}, \citenamefont {Díaz-Guilera}, \citenamefont
  {Gleiser},\ and\ \citenamefont {Guimerá}}]{arenas_community_2004}%
  \BibitemOpen
  \bibfield  {author} {\bibinfo {author} {\bibfnamefont {A.}~\bibnamefont
  {Arenas}}, \bibinfo {author} {\bibfnamefont {L.}~\bibnamefont {Danon}},
  \bibinfo {author} {\bibfnamefont {A.}~\bibnamefont {Díaz-Guilera}}, \bibinfo
  {author} {\bibfnamefont {P.~M.}\ \bibnamefont {Gleiser}}, \ and\ \bibinfo
  {author} {\bibfnamefont {R.}~\bibnamefont {Guimerá}},\ }\bibfield  {title}
  {{\selectlanguage {english}\enquote {\bibinfo {title} {Community analysis in
  social networks},}\ }}\href {\doibase 10.1140/epjb/e2004-00130-1} {\bibfield
  {journal} {\bibinfo  {journal} {The European Physical Journal B - Condensed
  Matter and Complex Systems}\ }\textbf {\bibinfo {volume} {38}},\ \bibinfo
  {pages} {373--380} (\bibinfo {year} {2004})}\BibitemShut {NoStop}%
\bibitem [{\citenamefont
  {Peixoto}(2014{\natexlab{b}})}]{peixoto_efficient_2014}%
  \BibitemOpen
  \bibfield  {author} {\bibinfo {author} {\bibfnamefont {Tiago~P.}\
  \bibnamefont {Peixoto}},\ }\bibfield  {title} {\enquote {\bibinfo {title}
  {Efficient monte carlo and greedy heuristic for the inference of stochastic
  block models},}\ }\href {\doibase 10.1103/PhysRevE.89.012804} {\bibfield
  {journal} {\bibinfo  {journal} {Physical Review E}\ }\textbf {\bibinfo
  {volume} {89}},\ \bibinfo {pages} {012804} (\bibinfo {year}
  {2014}{\natexlab{b}})}\BibitemShut {NoStop}%
\bibitem [{\citenamefont {Metropolis}\ \emph {et~al.}(1953)\citenamefont
  {Metropolis}, \citenamefont {Rosenbluth}, \citenamefont {Rosenbluth},
  \citenamefont {Teller},\ and\ \citenamefont
  {Teller}}]{metropolis_equation_1953}%
  \BibitemOpen
  \bibfield  {author} {\bibinfo {author} {\bibfnamefont {Nicholas}\
  \bibnamefont {Metropolis}}, \bibinfo {author} {\bibfnamefont {Arianna~W.}\
  \bibnamefont {Rosenbluth}}, \bibinfo {author} {\bibfnamefont {Marshall~N.}\
  \bibnamefont {Rosenbluth}}, \bibinfo {author} {\bibfnamefont {Augusta~H.}\
  \bibnamefont {Teller}}, \ and\ \bibinfo {author} {\bibfnamefont {Edward}\
  \bibnamefont {Teller}},\ }\bibfield  {title} {\enquote {\bibinfo {title}
  {Equation of state calculations by fast computing machines},}\ }\href
  {\doibase 10.1063/1.1699114} {\bibfield  {journal} {\bibinfo  {journal} {The
  Journal of Chemical Physics}\ }\textbf {\bibinfo {volume} {21}},\ \bibinfo
  {pages} {1087} (\bibinfo {year} {1953})}\BibitemShut {NoStop}%
\bibitem [{\citenamefont {Hastings}(1970)}]{hastings_monte_1970}%
  \BibitemOpen
  \bibfield  {author} {\bibinfo {author} {\bibfnamefont {W.~K.}\ \bibnamefont
  {Hastings}},\ }\bibfield  {title} {\enquote {\bibinfo {title} {Monte carlo
  sampling methods using markov chains and their applications},}\ }\href
  {\doibase 10.1093/biomet/57.1.97} {\bibfield  {journal} {\bibinfo  {journal}
  {Biometrika}\ }\textbf {\bibinfo {volume} {57}},\ \bibinfo {pages} {97 --109}
  (\bibinfo {year} {1970})}\BibitemShut {NoStop}%
\bibitem [{\citenamefont {Decelle}\ \emph
  {et~al.}(2011{\natexlab{a}})\citenamefont {Decelle}, \citenamefont
  {Krzakala}, \citenamefont {Moore},\ and\ \citenamefont
  {Zdeborová}}]{decelle_inference_2011}%
  \BibitemOpen
  \bibfield  {author} {\bibinfo {author} {\bibfnamefont {Aurelien}\
  \bibnamefont {Decelle}}, \bibinfo {author} {\bibfnamefont {Florent}\
  \bibnamefont {Krzakala}}, \bibinfo {author} {\bibfnamefont {Cristopher}\
  \bibnamefont {Moore}}, \ and\ \bibinfo {author} {\bibfnamefont {Lenka}\
  \bibnamefont {Zdeborová}},\ }\bibfield  {title} {\enquote {\bibinfo {title}
  {Inference and phase transitions in the detection of modules in sparse
  networks},}\ }\href {\doibase 10.1103/PhysRevLett.107.065701} {\bibfield
  {journal} {\bibinfo  {journal} {Physical Review Letters}\ }\textbf {\bibinfo
  {volume} {107}},\ \bibinfo {pages} {065701} (\bibinfo {year}
  {2011}{\natexlab{a}})}\BibitemShut {NoStop}%
\bibitem [{\citenamefont {Decelle}\ \emph
  {et~al.}(2011{\natexlab{b}})\citenamefont {Decelle}, \citenamefont
  {Krzakala}, \citenamefont {Moore},\ and\ \citenamefont
  {Zdeborová}}]{decelle_asymptotic_2011}%
  \BibitemOpen
  \bibfield  {author} {\bibinfo {author} {\bibfnamefont {Aurelien}\
  \bibnamefont {Decelle}}, \bibinfo {author} {\bibfnamefont {Florent}\
  \bibnamefont {Krzakala}}, \bibinfo {author} {\bibfnamefont {Cristopher}\
  \bibnamefont {Moore}}, \ and\ \bibinfo {author} {\bibfnamefont {Lenka}\
  \bibnamefont {Zdeborová}},\ }\bibfield  {title} {\enquote {\bibinfo {title}
  {Asymptotic analysis of the stochastic block model for modular networks and
  its algorithmic applications},}\ }\href {\doibase 10.1103/PhysRevE.84.066106}
  {\bibfield  {journal} {\bibinfo  {journal} {Physical Review E}\ }\textbf
  {\bibinfo {volume} {84}},\ \bibinfo {pages} {066106} (\bibinfo {year}
  {2011}{\natexlab{b}})}\BibitemShut {NoStop}%
\bibitem [{\citenamefont {Gopalan}\ and\ \citenamefont
  {Blei}(2013)}]{gopalan_efficient_2013}%
  \BibitemOpen
  \bibfield  {author} {\bibinfo {author} {\bibfnamefont {Prem~K.}\ \bibnamefont
  {Gopalan}}\ and\ \bibinfo {author} {\bibfnamefont {David~M.}\ \bibnamefont
  {Blei}},\ }\bibfield  {title} {{\selectlanguage {english}\enquote {\bibinfo
  {title} {Efficient discovery of overlapping communities in massive
  networks},}\ }}\href {\doibase 10.1073/pnas.1221839110} {\bibfield  {journal}
  {\bibinfo  {journal} {Proceedings of the National Academy of Sciences}\
  }\textbf {\bibinfo {volume} {110}},\ \bibinfo {pages} {14534--14539}
  (\bibinfo {year} {2013})}\BibitemShut {NoStop}%
\bibitem [{\citenamefont
  {Peixoto}(2014{\natexlab{c}})}]{peixoto_graph-tool_2014}%
  \BibitemOpen
  \bibfield  {author} {\bibinfo {author} {\bibfnamefont {Tiago~P.}\
  \bibnamefont {Peixoto}},\ }\bibfield  {title} {\enquote {\bibinfo {title}
  {The graph-tool python library},}\ }\href {\doibase
  10.6084/m9.figshare.1164194} {\bibfield  {journal} {\bibinfo  {journal}
  {figshare}\ } (\bibinfo {year} {2014}{\natexlab{c}}),\
  10.6084/m9.figshare.1164194}\BibitemShut {NoStop}%
\bibitem [{\citenamefont {Viamontes~Esquivel}\ and\ \citenamefont
  {Rosvall}(2011)}]{viamontes_esquivel_compression_2011}%
  \BibitemOpen
  \bibfield  {author} {\bibinfo {author} {\bibfnamefont {Alcides}\ \bibnamefont
  {Viamontes~Esquivel}}\ and\ \bibinfo {author} {\bibfnamefont {Martin}\
  \bibnamefont {Rosvall}},\ }\bibfield  {title} {\enquote {\bibinfo {title}
  {Compression of flow can reveal overlapping-module organization in
  networks},}\ }\href {\doibase 10.1103/PhysRevX.1.021025} {\bibfield
  {journal} {\bibinfo  {journal} {Physical Review X}\ }\textbf {\bibinfo
  {volume} {1}},\ \bibinfo {pages} {021025} (\bibinfo {year}
  {2011})}\BibitemShut {NoStop}%
\bibitem [{\citenamefont {Hoff}\ \emph {et~al.}(2002)\citenamefont {Hoff},
  \citenamefont {Raftery},\ and\ \citenamefont {Handcock}}]{hoff_latent_2002}%
  \BibitemOpen
  \bibfield  {author} {\bibinfo {author} {\bibfnamefont {Peter~D}\ \bibnamefont
  {Hoff}}, \bibinfo {author} {\bibfnamefont {Adrian~E}\ \bibnamefont
  {Raftery}}, \ and\ \bibinfo {author} {\bibfnamefont {Mark~S}\ \bibnamefont
  {Handcock}},\ }\bibfield  {title} {\enquote {\bibinfo {title} {Latent space
  approaches to social network analysis},}\ }\href {\doibase
  10.1198/016214502388618906} {\bibfield  {journal} {\bibinfo  {journal}
  {Journal of the American Statistical Association}\ }\textbf {\bibinfo
  {volume} {97}},\ \bibinfo {pages} {1090--1098} (\bibinfo {year}
  {2002})}\BibitemShut {NoStop}%
\bibitem [{\citenamefont {Parkkinen}\ \emph {et~al.}(2009)\citenamefont
  {Parkkinen}, \citenamefont {Sinkkonen}, \citenamefont {Gyenge},\ and\
  \citenamefont {Kaski}}]{parkkinen_block_2009}%
  \BibitemOpen
  \bibfield  {author} {\bibinfo {author} {\bibfnamefont {Juuso}\ \bibnamefont
  {Parkkinen}}, \bibinfo {author} {\bibfnamefont {Janne}\ \bibnamefont
  {Sinkkonen}}, \bibinfo {author} {\bibfnamefont {Adam}\ \bibnamefont
  {Gyenge}}, \ and\ \bibinfo {author} {\bibfnamefont {Samuel}\ \bibnamefont
  {Kaski}},\ }\bibfield  {title} {\enquote {\bibinfo {title} {A block model
  suitable for sparse graphs},}\ }in\ \href
  {http://research.ics.aalto.fi/mi/info/online-papers/mlg09sibm.pdf} {\emph
  {\bibinfo {booktitle} {Proceedings of the 7th International Workshop on
  Mining and Learning with Graphs ({MLG} 2009), Leuven}}}\ (\bibinfo {year}
  {2009})\BibitemShut {NoStop}%
\bibitem [{\citenamefont {Yang}\ and\ \citenamefont
  {Leskovec}(2012{\natexlab{b}})}]{yang_community-affiliation_2012}%
  \BibitemOpen
  \bibfield  {author} {\bibinfo {author} {\bibfnamefont {Jaewon}\ \bibnamefont
  {Yang}}\ and\ \bibinfo {author} {\bibfnamefont {J.}~\bibnamefont
  {Leskovec}},\ }\bibfield  {title} {\enquote {\bibinfo {title}
  {Community-affiliation graph model for overlapping network community
  detection},}\ }in\ \href {\doibase 10.1109/ICDM.2012.139} {\emph {\bibinfo
  {booktitle} {2012 {IEEE} 12th International Conference on Data Mining
  ({ICDM})}}}\ (\bibinfo {year} {2012})\ pp.\ \bibinfo {pages}
  {1170--1175}\BibitemShut {NoStop}%
\end{thebibliography}%
\end{document}